\documentclass[12pt]{article}
\usepackage{axodraw}
\usepackage{feynarts2}

\makeatletter

\def\paragraph{\@startsection{paragraph}{4}{\z@}{+2.00ex plus
 +1ex minus +.2ex}{1.5ex plus .2ex}{\it\normalsize}}

\def\section{\@startsection {section}{1}{\z@}{+3.0ex plus +1ex minus
  +.2ex}{2.3ex plus .2ex}{\normalsize\bf\boldmath}}
\def\subsection{\@startsection{subsection}{2}{\z@}{+2.5ex plus +1ex
minus +.2ex}{1.5ex plus .2ex}{\normalsize\bf\boldmath}}
\def\subsubsection{\@startsection{subsubsection}{3}{\z@}{+3.25ex plus
 +1ex minus +.2ex}{1.5ex plus .2ex}{\normalsize\it}}

\expandafter\ifx\csname mathrm\endcsname\relax\def\mathrm#1{{\rm #1}}\fi

\renewcommand{\theequation}{\thesection.\arabic{equation}}
\newcounter{saveeqn}

\@addtoreset{equation}{section}

\newcount\@tempcntc
\def\@citex[#1]#2{\if@filesw\immediate\write\@auxout{\string\citation{#2}}\fi
  \@tempcnta\z@\@tempcntb\m@ne\def\@citea{}\@cite{\@for\@citeb:=#2\do
    {\@ifundefined
       {b@\@citeb}{\@citeo\@tempcntb\m@ne\@citea
        \def\@citea{,\penalty\@m\ }{\bf ?}\@warning
       {Citation `\@citeb' on page \thepage \space undefined}}%
    {\setbox\z@\hbox{\global\@tempcntc0\csname
b@\@citeb\endcsname\relax}%
     \ifnum\@tempcntc=\z@ \@citeo\@tempcntb\m@ne
       \@citea\def\@citea{,\penalty\@m}
       \hbox{\csname b@\@citeb\endcsname}%
     \else
      \advance\@tempcntb\@ne
      \ifnum\@tempcntb=\@tempcntc
      \else\advance\@tempcntb\m@ne\@citeo
      \@tempcnta\@tempcntc\@tempcntb\@tempcntc\fi\fi}}\@citeo}{#1}}

\def\@citeo{\ifnum\@tempcnta>\@tempcntb\else\@citea
  \def\@citea{,\penalty\@m}%
  \ifnum\@tempcnta=\@tempcntb\the\@tempcnta\else
   {\advance\@tempcnta\@ne\ifnum\@tempcnta=\@tempcntb \else
\def\@citea{--}\fi
    \advance\@tempcnta\m@ne\the\@tempcnta\@citea\the\@tempcntb}\fi\fi}

\def\nl{\nonumber\\}

\newcommand{\lsim}
{\mathrel{\raisebox{-.3em}{$\stackrel{\displaystyle <}{\sim}$}}}
\newcommand{\gsim}
{\mathrel{\raisebox{-.3em}{$\stackrel{\displaystyle >}{\sim}$}}}
\def\asymp#1%
{\mathrel{\raisebox{-.4em}{$\widetilde{\scriptstyle #1}$}}}

\def\Nequal#1%
{\mathrel{\raisebox{-.5em}{$\stackrel{=}{\scriptstyle\rm#1}$}}}
\newcommand{\dsl}[1]{\not \hspace{-0.7mm}#1}
\def\dsl{\mathpalette\make@slash}
\def\make@slash#1#2{\setbox\z@\hbox{$#1#2$}%
  \hbox to 0pt{\hss$#1/$\hss\kern-\wd0}\box0}
\def\eqref#1{\stackrel{\mbox{\scriptsize{(\ref{#1})}}}{=}}

\makeatother

\def\beq{\begin{equation}}
\def\eeq{\end{equation}}
\def\beqar{\begin{eqnarray}}
\def\eeqar{\end{eqnarray}}
\def\barr#1{\begin{array}{#1}}
\def\earr{\end{array}}
\def\bfi{\begin{figure}}
\def\efi{\end{figure}}
\def\btab{\begin{table}}
\def\etab{\end{table}}
\def\bce{\begin{center}}
\def\ece{\end{center}}
\def\nn{\nonumber}
\def\disp{\displaystyle}
\def\text{\textstyle}

\def\al{\alpha}
\def\be{\beta}

\def\ga{\gamma}
\def\de{\delta}
\def\De{\Delta}
\def\eps{\epsilon}
\def\veps{\varepsilon}

\def\si{\sigma}
\def\Si{\Sigma}

\def\refeq#1{\mbox{(\ref{#1})}}
\def\refeqs#1{\mbox{(\ref{#1})}}
\def\refeqf#1{\mbox{(\ref{#1})}}
\def\reffi#1{\mbox{Figure~\ref{#1}}}
\def\reffis#1{\mbox{Figures~\ref{#1}}}
\def\refta#1{\mbox{Table~\ref{#1}}}

\def\refse#1{\mbox{Section~\ref{#1}}}
\def\refses#1{\mbox{Sections~\ref{#1}}}

\def\citere#1{\mbox{Ref.~\cite{#1}}}
\def\citeres#1{\mbox{Refs.~\cite{#1}}}

\newcommand{\GeV}{\unskip\,\mathrm{GeV}}
\newcommand{\MeV}{\unskip\,\mathrm{MeV}}

\newcommand{\ri}{{\mathrm{i}}}
\newcommand{\rd}{{\mathrm{d}}}
\newcommand{\rE}{{\mathrm{E}}}

\newcommand{\rT}{{\mathrm{T}}}
\newcommand{\rw}{\mathswitchr w}

\newcommand{\ord}{\mathswitch{{\cal{O}}}}
\newcommand{\Oa}{\mathswitch{{\cal{O}}(\alpha)}}
\newcommand{\Oaa}{\mathswitch{{\cal{O}}(\alpha^2)}}
\newcommand{\Oaaa}{\mathswitch{{\cal{O}}(\alpha^3)}}

\newcommand{\M}{{\cal{M}}}

\def\mathswitchr#1{\relax\ifmmode{\mathrm{#1}}\else$\mathrm{#1}$\fi}

\newcommand{\PW}{\mathswitchr W}
\newcommand{\PZ}{\mathswitchr Z}

\newcommand{\PH}{\mathswitchr H}
\newcommand{\Pe}{\mathswitchr e}

\newcommand{\Pne}{\mathswitch \nu_{\mathrm{e}}}
\newcommand{\Pnebar}{\mathswitch \bar\nu_{\mathrm{e}}}
\newcommand{\Pd}{\mathswitchr d}
\newcommand{\Pdbar}{\bar{\mathswitchr d}}
\newcommand{\Pu}{\mathswitchr u}

\newcommand{\Ps}{\mathswitchr s}

\newcommand{\Pc}{\mathswitchr c}

\newcommand{\Pb}{\mathswitchr b}

\newcommand{\Pt}{\mathswitchr t}

\newcommand{\Pep}{\mathswitchr {e^+}}
\newcommand{\Pem}{\mathswitchr {e^-}}
\newcommand{\PWp}{\mathswitchr {W^+}}
\newcommand{\PWm}{\mathswitchr {W^-}}

\def\mathswitch#1{\relax\ifmmode#1\else$#1$\fi}

\newcommand{\MW}{\mathswitch {M_\PW}}

\newcommand{\MZ}{\mathswitch {M_\PZ}}
\newcommand{\MH}{\mathswitch {M_\PH}}
\newcommand{\Me}{\mathswitch {m_\Pe}}

\newcommand{\Md}{\mathswitch {m_\Pd}}
\newcommand{\Mu}{\mathswitch {m_\Pu}}
\newcommand{\Ms}{\mathswitch {m_\Ps}}
\newcommand{\Mc}{\mathswitch {m_\Pc}}
\newcommand{\Mb}{\mathswitch {m_\Pb}}
\newcommand{\Mt}{\mathswitch {m_\Pt}}
\newcommand{\GH}{\Gamma_{\PH}}
\newcommand{\Gt}{\Gamma_{\Pt}}
\newcommand{\GW}{\Gamma_{\PW}}
\newcommand{\GZ}{\Gamma_{\PZ}}

\newcommand{\alphas}{\mathswitch {\alpha_{\mathrm{s}}}}
\newcommand{\sw}{\mathswitch {s_\rw}}
\newcommand{\cw}{\mathswitch {c_\rw}}

\newcommand{\GF}{\mathswitch {G_\mu}}

\def\solid{\raise.9mm\hbox{\protect\rule{1.1cm}{.2mm}}}
\def\dash{\raise.9mm\hbox{\protect\rule{2mm}{.2mm}}\hspace*{1mm}}

\def\ie{i.e.\ }
\def\eg{e.g.\ }

\newcommand{\DPA}{{\mathrm{DPA}}}

\newcommand{\Born}{{\mathrm{Born}}}
\newcommand{\born}{{\mathrm{Born}}}

\newcommand{\eefourf}{{\mathswitchr{ee4f}}}

\newcommand{\CMS}{{\mathrm{CMS}}}
\newcommand{\FW}{{\mathrm{FW}}}

\newcommand{\virt}{{\mathrm{virt}}}

\newcommand{\recomb}{{\mathrm{rec}}}

\def\Re{\mathop{\mathrm{Re}}\nolimits}
\def\Im{\mathop{\mathrm{Im}}\nolimits}
\def\sgn{\mathop{\mathrm{sgn}}\nolimits}

\newcommand{\eeWWffff}{\Pep\Pem\to\PW\PW\to 4f}
\newcommand{\eeffff}{\Pep\Pem\to 4f}
\newcommand{\cceeffff}{\Pep\Pem\to f_1\bar f_2 f_3 \bar f_4}

\newcommand{\Mhat}{\hat\M}
\newcommand{\Sn}{S}

\newcommand{\sss}{\scriptscriptstyle}

\newtheorem{step}{Step}

\newcommand{\DiracGamma}{A}
\newcommand{\indexsep}{,}

\def\draftdate{\relax}
\def\mda{\relax}
\def\mua{\relax}
\def\mla{\relax}
\def\draft{
\def\thtystars{******************************}
\def\sixtystars{\thtystars\thtystars}
\typeout{}
\typeout{\sixtystars**}
\typeout{* Draft mode!
         For final version remove \protect\draft\space in source file *}
\typeout{\sixtystars**}
\typeout{}
\def\draftdate{\today}
\def\mua{\marginpar[\boldmath\hfil$\uparrow$]%
                   {\boldmath$\uparrow$\hfil}%
                    \typeout{marginpar: $\uparrow$}\ignorespaces}
\def\mda{\marginpar[\boldmath\hfil$\downarrow$]%
                   {\boldmath$\downarrow$\hfil}%
                    \typeout{marginpar: $\downarrow$}\ignorespaces}
\def\mla{\marginpar[\boldmath\hfil$\rightarrow$]%
                   {\boldmath$\leftarrow $\hfil}%
                    \typeout{marginpar: $\leftrightarrow$}\ignorespaces}
\def\Mua{\marginpar[\boldmath\hfil$\Uparrow$]%
                   {\boldmath$\Uparrow$\hfil}%
                    \typeout{marginpar: $\uparrow$}\ignorespaces}
\def\Mda{\marginpar[\boldmath\hfil$\Downarrow$]%
                   {\boldmath$\Downarrow$\hfil}%
                    \typeout{marginpar: $\downarrow$}\ignorespaces}
\def\Mla{\marginpar[\boldmath\hfil$\Rightarrow$]%
                   {\boldmath$\Leftarrow $\hfil}%
                    \typeout{marginpar: $\leftrightarrow$}\ignorespaces}
\overfullrule 5pt
\oddsidemargin -15mm
\marginparwidth 29mm
}

\def\stars{\strut\leaders\hbox{*}\hfill\strut}
\def\starline{\hfil\strut\hfil\hbox to \textwidth {\stars}\hfil}

\begin{document}

\thispagestyle{empty}
\def\thefootnote{\fnsymbol{footnote}}
\setcounter{footnote}{1}
\null
\draftdate\hfill MPP-2005-23 \\
\strut\hfill PSI-PR-05-05\\
\strut\hfill hep-ph/0505042
\vfill
\begin{center}
{\Large \bf\boldmath
Electroweak corrections to \\[.2em]
charged-current $\Pep\Pem\to4\,$fermion processes \\[.2em]
--- \\[.2em]
technical details and further results 
\par} \vskip 1.5em
\vspace{.5cm}
{\large
{\sc A.\ Denner$^1$, S.\ Dittmaier$^2$, M. Roth$^2$ and 
L.H.\ Wieders$^{1,3}$}} \\[1cm]
$^1$ {\it Paul Scherrer Institut, W\"urenlingen und Villigen\\
CH-5232 Villigen PSI, Switzerland} \\[0.5cm]
$^2$ {\it Max-Planck-Institut f\"ur Physik 
(Werner-Heisenberg-Institut) \\
D-80805 M\"unchen, Germany}
\\[0.5cm]
$^3$ {\it Institute for Theoretical Physics\\ University of Z\"urich, CH-8057 
Z\"urich, Switzerland}
\par \vskip 1em
\end{center}\par
\vfill \vskip 1cm {\bf Abstract:} \par The complete electroweak ${\cal
  O}(\alpha)$ corrections have been calculated for the charged-current
four-fermion production processes
$\Pep\Pem\to\nu_\tau\tau^+\mu^-\bar\nu_\mu$,
$\Pu\bar\Pd\mu^-\bar\nu_\mu$, and $\Pu\bar\Pd\Ps\bar\Pc$. Here,
technical details of this calculation are presented. These include the
algebraic reduction of spinor chains to a few standard structures and
the consistent implementation of the finite width of the W~boson.  To
this end, a generalization of the complex-mass scheme to the one-loop
level is proposed, and the practical application of this method is
described.  Finally, the effects of the complete $\Oa$ corrections to
various differential cross sections of physical interest are discussed
and compared to predictions based on the double-pole approximation,
revealing that the latter approximation is not sufficient to fully
exploit the potential of a future linear collider in an analysis of
W-boson pairs at high energies.
\par
\vskip 1cm
\noindent
August 2011
\null
\setcounter{page}{0}
\clearpage
\def\thefootnote{\arabic{footnote}}
\setcounter{footnote}{0}

\section{Introduction}
\label{se:intro}

W-pair production is an important process for testing the Electroweak
Standard Model (SM). On the one hand, it allows to measure the mass of
the W~boson, a fundamental parameter of the SM, precisely. On the
other hand, it is sensitive to the triple non-abelian gauge couplings
and, in particular at high energies, allows to test the non-abelian
structure of the SM accurately, owing to the delicate gauge
cancellations in its lowest-order matrix elements.

Experimentally, W-pair production has been studied intensively at LEP2
with quite high precision \cite{lep2}.  The total cross section was
measured from threshold up to a centre-of-mass (CM) energy of
$207\GeV$ leading to a combined experimental accuracy of $\sim 1\%$.
While the W-boson mass $\MW$ was determined from the threshold cross
section with an error of $\sim 200\MeV$ and by reconstructing the
$\PW$ bosons from their decay products within $\sim 40\MeV$,
deviations from the SM triple gauge-boson couplings were constrained
within a few per cent.  More precise experimental investigations of
W-pair production will be possible at a future International $\Pe^+
\Pe^-$ Linear Collider (ILC)
\cite{Aguilar-Saavedra:2001rg,Abe:2001wn,Abe:2001gc}. Owing to the
high luminosity of such a collider, the accuracy of the cross section
measurement will be at the per-mille level, and the precision of the
$\PW$-mass determination is expected to be $\sim10\MeV$ \cite{talkKM}
by direct reconstruction and \mbox{$\sim 7\MeV$} from a threshold scan
of the total W-pair-production cross section
\cite{Aguilar-Saavedra:2001rg,Abe:2001wn}.  The higher energy of the
ILC will enable much more precise measurements of the triple
gauge-boson couplings, improving the sensitivity of LEP2 by more than
an order of magnitude.

Because of its theoretical importance, W-pair production found early
interest. The lowest-order amplitudes for on-shell W-pair production
were already considered at the end of the 1970's \cite{Alles:1976qv}.
The electroweak corrections to on-shell W-pair production have been
calculated by four different groups \cite{Lemoine:1979pm,Bohm:1987ck}
shortly after and supplemented by hard photon radiation
\cite{Beenakker:1990sf}.  Already at that time, these calculations
were at the forefront of the technical developments in higher-order
calculations. Later, the structure of these corrections has been
investigated by constructing improved Born approximations
\cite{Dittmaier:1991np} and high-energy approximations
\cite{Beenakker:1993tt}.  With the advent of LEP2 it became quickly
clear that the decays of the W~bosons into fermion pairs had to be
included.  The electroweak corrections to the on-shell W-boson decay
were given in \citere{Bardin:1986fi}.  Different types of programs
(ranging from semianalytical codes to Monte Carlo generators) for
lowest-order predictions for $\eeffff$ were developed
\cite{Berends:1994xn} (see also
\citeres{Beenakker:1996kt,Bardin:1997gc,Grunewald:2000ju} and
references therein) and subsequently supplemented by universal
corrections thus reaching an accuracy of about $2\%$.  The universal
corrections included running couplings, initial-state radiation, and
also the effects of the Coulomb singularity for off-shell W-pair
production \cite{Fadin:1993kg}.  Since the accuracy of $2\%$ was not
sufficient for LEP2, the $\Oa$ corrections to $\eeWWffff$ were
calculated in the double-pole approximation (DPA), where only the
leading terms in an expansion about the resonance poles of the two
W-boson propagators were taken into account
\cite{Beenakker:1998gr,Jadach:1998tz,Denner:2000kn,Denner:2000bj,Kurihara:2001um}.
These corrections were implemented into the event generators {\sc
  YFSWW} \cite{Jadach:1998tz} and {\sc RacoonWW}
\cite{Denner:2000kn,Denner:2000bj,Denner:1999gp}. A discussion of the
remaining theoretical uncertainties of these calculations is presented
in \citeres{Grunewald:2000ju,Jadach:2001cz}.  In view of the improved
precision of the ILC, a further reduction of the uncertainties from
missing radiative corrections is necessary.  This requires, in
particular, the calculation of the full one-loop corrections for
W-pair-mediated $\eeffff$ processes.

Such a calculation poses a number of theoretical challenges.%
\footnote{Some of the problems appearing in a first attempt of such a
  calculation were already described in \citere{Vicini:1998iy}.} 
Neglecting diagrams involving couplings of Higgs bosons to light
fermions, which are proportional to the fermion masses, already for
the simplest final states $\Pep\Pem\to\nu_\tau\tau^+\mu^-\bar\nu_\mu$,
$\Pu\bar\Pd\mu^-\bar\nu_\mu$, and $\Pu\bar\Pd\Ps\bar\Pc$ about 1200
Feynman diagrams contribute (counting the contributions of the
three fermion generations in the loops only once), and for the most
complicated final state $\Pne\Pep\Pem\Pnebar$, there are about 6000
diagrams.  The large number and the complexity of the diagrams require
to develop improved reduction algorithms, in order to keep the
expressions manageable and to produce an efficient and numerically
stable computer code.

Because of the complicated multi-particle final state, the appearing
loop integrals in general cannot be evaluated with standard methods.
Using the Passarino--Veltman reduction to calculate the tensor
integrals leads to serious numerical problems when the Gram
determinants that appear in the denominators become small.  This
usually happens near the boundary of phase space but can also occur
within phase space because of the indefinite Minkowski metric.  Thus,
in order to obtain numerically stable results, one has to devise and
implement alternative methods for the calculation of the tensor
integrals, at least, in the critical regions.

The inclusion of the finite gauge-boson decay width constitutes a
further important problem in the calculation of radiative corrections
to W-pair-mediated four-fermion production.  An appropriate
description of resonances in perturbation theory requires a Dyson
summation of self-energy insertions in the resonant propagators.  It
is well known that this procedure in general violates gauge
invariance, \ie destroys Slavnov--Taylor and Ward identities and
disturbs the cancellation of gauge-parameter dependences, because
different perturbative orders are mixed
\cite{Berends:1969nt,Argyres:1995ym}.  Several solutions have been
described for lowest-order predictions.  The early attempts have been
summarized in \citere{Aeppli:1993cb}, and some of the schemes have
been compared in \citere{Beenakker:1996kn}.  More recent approaches
include the ``pole-scheme'' \cite{Stuart:1991xk,Aeppli:1993rs}, the
``fermion-loop scheme''
\cite{Argyres:1995ym,Beenakker:1996kn,Baur:1995aa,Passarino:1999zh},
the use of effective Lagrangians
\cite{Beenakker:1999hi,Beneke:2003xh}%
\footnote{The recently proposed approach \cite{Beneke:2003xh} to
  describe unstable particles within an effective field theory is
  equivalent to a pole expansion.}, and the ``complex-mass scheme''
(CMS) \cite{Denner:1999gp}.  Apart from the pole expansions, none of
these approaches has been elaborated beyond tree level so far.  The
pole scheme provides a gauge-invariant answer in terms of an expansion
about the resonance, but is only applicable sufficiently far above the
W-pair threshold.  However, in the full calculation we are after a
unified description that is valid everywhere in phase space, without
any matching between different treatments for different regions.  Some
problems related to the finite width appearing in a calculation of
radiative corrections to $\Pep\Pem\to\mu^-\bar\nu_\mu\Pu\bar\Pd$ are
illustrated in \citere{Boudjema:2004id}. Here we propose to solve
these problems by using a generalization of the CMS, which was
introduced in \citere{Denner:1999gp} for lowest-order calculations, to
the one-loop level.

In this paper we present details of our solutions to the
above-mentioned problems. We discuss, in particular, our methods for
the algebraic reduction of the one-loop amplitude and describe the use
of the CMS at the one-loop level. In addition, we provide numerical
results for the effects of the complete $\Oa$ corrections to the
processes $\Pep\Pem\to\nu_\tau\tau^+\mu^-\bar\nu_\mu$,
$\Pu\bar\Pd\mu^-\bar\nu_\mu$, and $\Pu\bar\Pd\Ps\bar\Pc$ for various
differential cross sections of physical interest. The effects of the
complete corrections on the total cross section without cuts have
already been discussed in \citere{Denner:2005es}.

The paper is organized as follows: In \refse{se:strategy} we fix our
conventions and sketch the general strategy of our calculation. Our
methods for the reduction of spinor chains to a few standard
structures are described in \refse{se:spinorchains}. Section
\ref{se:complex-masses} is devoted to the description of the
complex-mass scheme for loop calculations with complex masses. In
\refse{se:numres}, we discuss the corrections for various
distributions.  We note that \refses{se:spinorchains},
\ref{se:complex-masses}, and \ref{se:numres} can be read
independently.  Section~\ref{se:concl} contains our conclusions.

\section{Strategy of the calculation}
\label{se:strategy}

The actual calculation builds upon the {\sc RacoonWW} approach
\cite{Denner:2000kn,Denner:2000bj}, where real-photonic corrections
are based on full matrix elements and virtual corrections are treated
in DPA. Real and virtual corrections are combined either using
two-cutoff phase-space slicing or employing the dipole subtraction
method as formulated in \citere{Dittmaier:2000mb} for photon
radiation.  We also include leading-logarithmic initial-state
radiation (ISR) beyond $\Oa$ in the structure-function approach (see
\citere{Beenakker:1996kt} and references therein).  The presented
calculation only differs from {\sc RacoonWW} in the treatment of the
(IR and collinear finite part of the) virtual corrections.
Therefore, we only describe the calculation of the complete $\Oa$
virtual corrections in the following.

\subsection{Notation and conventions}
\label{se:conventions}

We consider the process
\beqar\label{eq:ee4f}
\Pep(p_+,\si_+)+\Pem(p_-,\si_-) &\;\to\;&
f_1(k_1,\si_1)+\bar f_2(k_2,\si_2)+f_3(k_3,\si_3)+\bar f_4(k_4,\si_4).
\eeqar
The arguments label the momenta $p_\pm$, $k_i$ ($i=1,2,3,4$) and
helicities $\si_\pm$, $\si_i=\pm1/2$ of the corresponding particles.
We often use only the signs to denote the helicities.  The particle
momenta obey the mass-shell conditions $p_\pm^2=\Me^2$ and
$k_i^2=m_i^2$.  The masses of the external fermions are neglected
whenever possible, \ie everywhere but in the mass-singular
logarithms. For later use, the following set of kinematical 
invariants is defined:
\beq
s = (p_1+p_2)^2, \qquad
s_{ij} = (k_i+k_j)^2, \qquad 
t_{\pm i} = (p_\pm-k_i)^2, \qquad i,j=1,2,3,4.
\eeq

In this paper we consider only final states where $f_1$ and $f_3$ are
different fermions excluding electrons and electron neutrinos; $f_2$
and $f_4$ are their isospin partners, respectively. This corresponds
to the CC11 family in the classification of \citere{Bardin:1997gc}. It
represents the gauge-invariant subclass of general $\eeffff$ processes
that includes all diagrams with pairs of potentially resonant
W~bosons.  In this class, the lowest-order and one-loop amplitudes
vanish unless $\si=\si_-=-\si_+$, $\si_1=-\si_2$, and $\si_3=-\si_4$.
Moreover, the helicities of the outgoing fermions are fixed,
$\si_{1,3} = -\si_{2,4} = -1/2$, owing to the left-handed coupling of
the \PW~bosons. In general this does not hold for other $\eeffff$
final states and diagrams.  We set the quark-mixing matrix to the unit
matrix, but in the limit of small masses for the external fermions a
non-trivial quark-mixing matrix can be easily taken into account by
rescaling the cross sections for definite flavours accordingly.

The lowest-order cross section reads
\beq
\sigma_{0} = \frac{1}{2s} \,
\int\rd\Phi_4 \sum_{\si=\pm\frac{1}{2}}
\frac{1}{4}(1+2P_-\si)(1-2P_+\si) \,
|\M_0^{\si--}|^2.
\eeq
Here $\M^{\si\si_1\si_3}_0$ denotes the lowest-order matrix element,
$P_\pm$ the polarization degrees of the $\Pe^\pm$ beams, and
$\rd\Phi_{4}$ the 4-particle phase-space volume element
\beq
\rd\Phi_4 =
\left( \prod_{i=1}^4 \frac{\rd^3 {\bf k}_i}{(2\pi)^3 2k_i^0} \right)\,
(2\pi)^4 \delta\Biggl(p_++p_--\sum_{j=1}^4 k_j\Biggr).
\label{eq:dG3}
\eeq

\subsection{Survey of one-loop diagrams}

\begin{figure}
\centerline{\footnotesize  
\input{Tree}
}
\caption{Lowest-order diagrams for  $\cceeffff$}
\label{fi:Born-diagrams}
\end{figure}

The virtual corrections receive contributions from self-energy,
vertex, box, pentagon, and hexagon diagrams.  In this section, we
survey the diagrams contributing to the massless charged-current
processes $\Pe^+\Pe^- \to f_1 \bar{f}_2 f_3 \bar{f}_4$, where $f_1$
and $f_3$ are different fermions, excluding electrons and electron
neutrinos, and $f_2$ and $f_4$ their respective isospin partners.

The contributions from self-energy and vertex corrections are obtained
by inserting self-energies and vertex corrections in all propagators
and vertices of the tree-level diagrams shown in
\reffi{fi:Born-diagrams}.  When inserting a self-energy in a
$\gamma/Z$ line one obtains $\ga\ga$ and $ZZ$ self-energies as well as
$\gamma Z$ and $Z\gamma$ mixing-energies. Since we neglect the masses
of the external fermions, all diagrams that involve Higgs-boson
couplings to these fermions obviously vanish.  Nevertheless there
remain diagrams containing contributions to the $H\ga$, $HZ$, and
$\phi W$ mixing energies and to the $Hee$ and $\phi ff$ vertices with
on-shell fermions, where $\phi$ denotes the would-be Goldstone boson
corresponding to the W boson. One can easily verify that these
contributions also vanish in the limit of vanishing masses for the
external fermions.

The diagrams for the appearing $ee\ga$, $eeZ$, $e\nu_eW$, $\ga WW$,
and $ZWW$ vertex functions are listed in \citere{Bohm:1987ck}, where
the process $\Pep\Pem\to\PWp\PWm$ was treated at one loop. The
diagrams for the other $\ga f\bar{f}$, $Zf\bar{f}$, $Wf\bar{f}'$
vertex functions are simply obtained from the latter by obvious
substitutions and adding some diagrams with internal Z~bosons replaced
by photons, which are absent for the $e\nu_e W$ vertex because of the
vanishing charge of the neutrino.  The diagrams for the gauge-boson
and fermion self-energies can be found in \citere{Hollik:1988ii}.  For
all these vertex functions, the corresponding counterterm diagrams
must be included.

The generic contributions of the different vertex functions with more
than three external legs are shown in \reffi{fi:generic-diagrams}.
\begin{figure}
\centerline{\footnotesize  
\input{Vertex}
}
\caption{Contributions of vertex functions with at least four external
  legs to $\cceeffff$}
\label{fi:generic-diagrams}
\end{figure}
These are all ultraviolet (UV) finite.  There are 40 hexagon diagrams,
112 pentagon diagrams, and 227 (220) box diagrams in the conventional
't~Hooft--Feynman gauge (background-field gauge \cite{Denner:1994xt}).
\begin{figure}
\centerline{\footnotesize  
\input{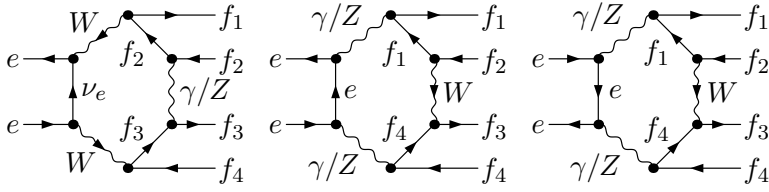}
}
\caption{Ten hexagon diagrams for $\cceeffff$. The remaining 30
  hexagon diagrams are obtained by reversing the fermion flow in one
  or both of the fermion lines of the outgoing fermions and by
  exchanging $f_1\leftrightarrow f_2$ and/or $f_3\leftrightarrow f_4$.}
\label{fi:hexagon-diagrams}
\end{figure}
\begin{figure}
\centerline{\footnotesize  
\input{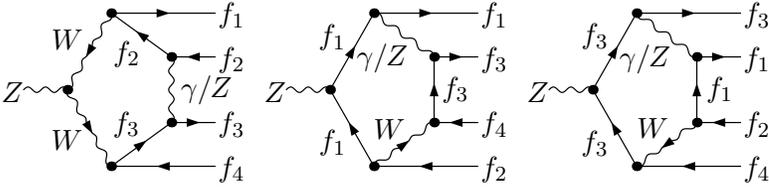}
}
\caption{Six pentagon diagrams contributing to the $Zf_1f_2f_3f_4$ vertex
  function. The remaining 18 diagrams are obtained by
  reversing the fermion flow in one or both of the fermion lines of
  the outgoing fermions and by exchanging $f_1\leftrightarrow f_2$
  and/or $f_3\leftrightarrow f_4$.}
\label{fi:ffffZ-diagrams}
\end{figure}
A set of hexagon diagrams is shown in \reffi{fi:hexagon-diagrams} and
one set of pentagon diagrams for the $Zf_1f_2f_3f_4$ vertex function in
\reffi{fi:ffffZ-diagrams}. In both cases there are three further sets
of diagrams that are obtained by reversing the fermion flow in one or
both of the fermion lines of the outgoing fermions and by exchanging
$f_1\leftrightarrow f_2$ and/or $f_3\leftrightarrow f_4$.  Those for
the $\ga f_1f_2f_3f_4$ vertex function are simply obtained by
replacing the external Z~boson by a photon in the diagrams for the
$Zf_1f_2f_3f_4$ vertex function.
\begin{figure}
\centerline{\footnotesize  
\input{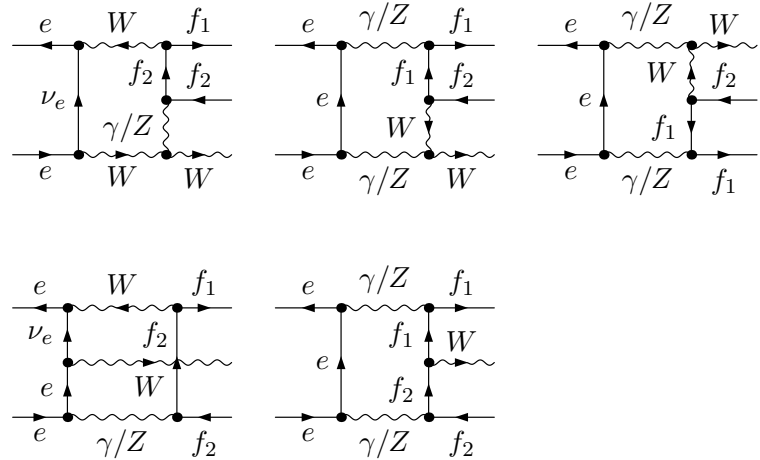}
}
\caption{A set of 16 diagrams contributing to the $eef_1f_2W$ vertex
  function. The remaining 16 diagrams contributing to this vertex
  function are obtained from those by reversing the fermion flow in
  the fermion line of the outgoing fermions and by exchanging
  $f_1\leftrightarrow f_2$.}
\label{fi:eeffW-diagrams}
\end{figure}
A set of diagrams for the $eef_1f_2W$ vertex function is listed in
\reffi{fi:eeffW-diagrams} and a further set is obtained from those 
by reversing the fermion flow in the fermion chain of the outgoing
fermions and by exchanging $f_1\leftrightarrow f_2$.
The diagrams for the $eef_3f_4W$ vertex
function can be obtained from the latter by obvious substitutions.
\begin{figure}
\centerline{\footnotesize  
\input{Vertex-Z-f1-f2-W}
}
\caption{Diagrams contributing to the $ZWf_1f_2$ vertex function}
\label{fi:ffZW-diagrams}
\end{figure}
The box diagrams of the $ZWf_1f_2$ vertex function are depicted in
\reffi{fi:ffZW-diagrams} and those for the $ZWf_3f_4$ vertex can again
be obtained from those. The diagrams for the $\ga Wf_1f_2$ and
$\ga Wf_3f_4$ boxes are obtained by replacing the external Z~boson
by a photon and omitting the diagrams with internal $H$ lines.
\begin{figure}
\centerline{\footnotesize  
\input{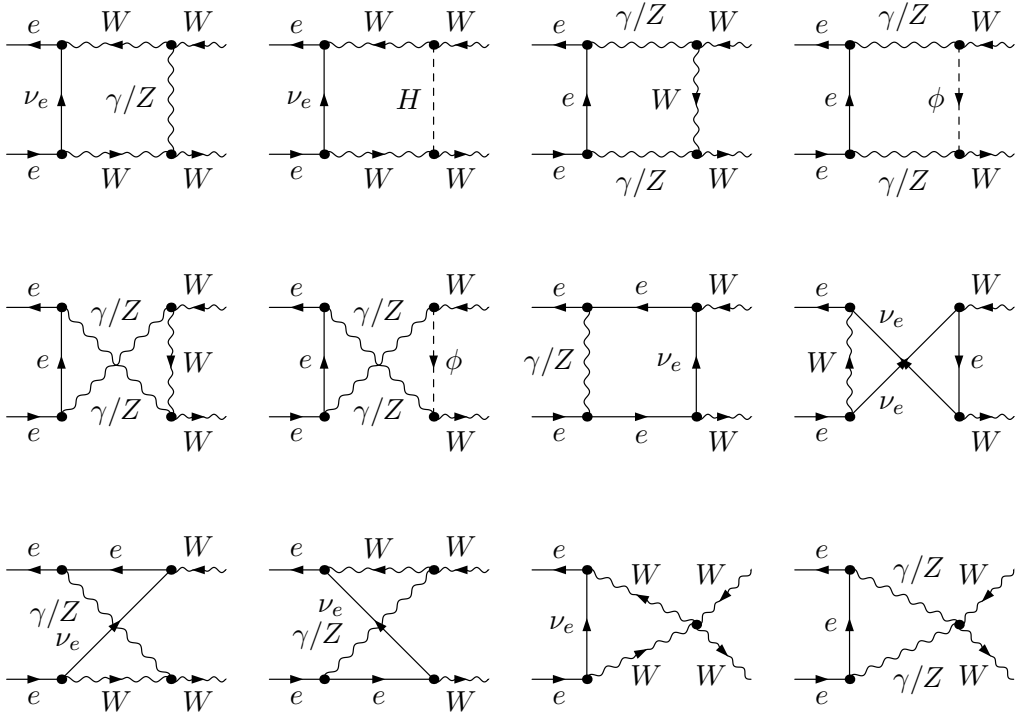}
}
\caption{Diagrams contributing to the $eeWW$ vertex function}
\label{fi:eeWW-diagrams}
\end{figure}
\begin{figure}
\centerline{\footnotesize  
\input{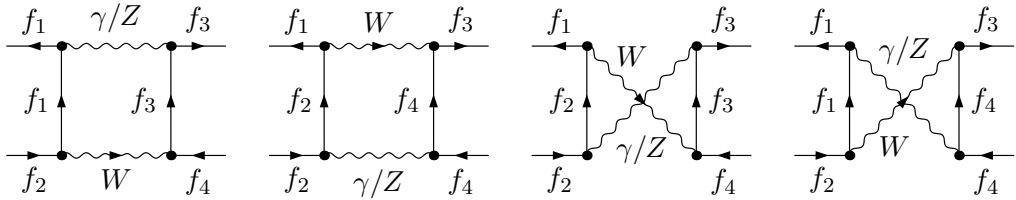}
}
\caption{Diagrams contributing to the $f_1f_2f_3f_4$ vertex function}
\label{fi:ffff-diagrams}
\end{figure}
\begin{figure}
\centerline{\footnotesize  
\input{Vertex-e-e-f1-f1}
}
\caption{Diagrams contributing to the $eef_1f_1$ vertex function}
\label{fi:eeff-diagrams}
\end{figure}
The diagrams for the $eeWW$, $f_1f_2f_3f_4$, and $eef_1f_1$ boxes are
compiled in \reffis{fi:eeWW-diagrams}, \ref{fi:ffff-diagrams}, and
\ref{fi:eeff-diagrams}. The $e\nu_ef_1f_2$ and $e\nu_ef_3f_4$ box
functions are special cases of the $f_1f_2f_3f_4$ box function.

\subsection{Calculational framework}

The amplitudes are generated with {\sc FeynArts}, using the two
independent versions 1 and 3, as described in
\citeres{Kublbeck:1990xc} and \cite{Hahn:2000kx}, respectively.  The
algebraic manipulations are performed using two independent in-house
programs implemented in {\sc Mathematica}, one of which builds upon
{\sc FormCalc}~\cite{Hahn:1998yk}.

All contributions to the matrix elements involve three spinor chains,
corresponding to the three different fermion--antifermion pairs, which
are contracted with each other and with four-momenta in many different
ways. There are ${\cal O}(10^3)$ different spinor structures to
calculate for $\cceeffff$ so that an algebraic reduction to a standard
form which involves only very few standard chains is desirable.  We
have worked out algorithms that reduce all occurring spinor chains to
${\cal O}(10)$ standard structures, the standard matrix elements
(SMEs), without introducing coefficients that lead to numerical
problems. These algorithms are described in \refse{se:spinorchains}.

It is convenient to separate the matrix elements into invariant
coefficient functions $F_n$, containing the loop integrals, and SMEs
$\Mhat_n$, containing all spinorial objects and the dependence on the
helicities of the external particles \cite{Denner:1993kt}.  After the
reduction of the spinor structures, the amplitudes take the form
\beq
\M^{\si\si_1\si_3} = \sum_n 
F^{\si\si_1\si_3}_n(\{s,s_{ij},t_{\pm i}\})
\Mhat^{\si\si_1\si_3}_n(p_+,p_-,k_1,k_2,k_3,k_4).
\eeq
The invariant functions $F_n$ are linear combinations of
one-loop integrals with coefficients that depend on scalar kinematical
variables, particle masses, and coupling factors.

The calculation of the virtual corrections has been repeated using the
background-field method \cite{Denner:1994xt}, where the individual
contributions from self-energy, vertex, box, and pentagon corrections
differ from their counterparts in the conventional formalism, apart
from those that involve only fermion--gauge-boson couplings in the
loops.  The total one-loop corrections of the conventional and of the
background-field approach were found to be in numerical agreement.

The contribution of the virtual corrections to the cross
section is given by
\beq
\de\sigma_{\virt} = \frac{1}{2s} \,
\int\rd\Phi_4 \sum_{\si=\pm\frac{1}{2}}
\frac{1}{4}(1+2P_-\si)(1-2P_+\si) \,
2\Re\left\{\M_1^{\si--}\left(\M_0^{\si--}\right)^*\right\},
\eeq
where $\M_1^{\si\si_1\si_3}$ denotes the one-loop contributions to the
helicity amplitudes.

\subsection{Calculation of loop integrals}
\label{se:loop-integrals}

For the calculation of the loop integrals we use an improved version
of the standard approach. All coefficient integrals are algebraically
reduced to a set of master integrals.  The reduction formulas are
implemented into Fortran codes, and the reduction is performed
numerically. Finally, the master integrals are either calculated from
explicit formulas or from numerical integration as explained below.

Let us first describe our standard approach that we use as long as it
yields numerically stable results. The 6-point integrals are directly
expressed in terms of six 5-point functions, as described in
\citeres{Denner:1993kt,Me65}.  The 5-point integrals are written in
terms of five 4-point functions following the method of
\citere{Denner:2002ii}.  The 3-point and 4-point tensor integrals are
algebraically reduced to the (standard) scalar 1-point, 2-point,
3-point, and 4-point functions, which are the master integrals, with
the Passarino--Veltman algorithm \cite{Passarino:1979jh}.  Finally,
the remaining scalar integrals are evaluated based on generalizations
of the methods and results of
\citeres{'tHooft:1979xw,Beenakker:1990jr,Denner:1991qq}.  UV
divergences are regulated dimensionally and IR divergences
with an infinitesimal photon mass%
\footnote{At one loop the photon-mass regularization is equivalent to
  dimensional regularization ($D=4-2\veps$) with reference mass $\mu$
  via the correspondence $\ln(m_\gamma^2)\leftrightarrow 1/\veps +
  \ln(4 \pi \mu^2) - \gamma_\rE$ as long as collinear singularities
  are regularized with fermion masses.}
$m_\gamma$.

The standard approach leads to serious numerical problems when the
Gram determinants, which appear in the denominator of the
Passarino--Veltman reduction, become small. This happens usually near
the boundary of phase space but also occurs within phase space because
of the indefinite Minkowski metric.  Since we use Passarino--Veltman
reduction only for 3-point and 4-point functions, the dangerous Gram
determinants are those involving two or three momenta. Our reduction
of 5- and 6-point functions, on the other hand, does not involve
inverse Gram determinants composed of external momenta. Instead, it
involves so-called (modified) Cayley determinants, the zeroes of which
are related to the Landau singularities of the (sub-)diagrams. We did
not encounter numerical problems with these determinants. For the
2-point tensor coefficients numerically stable explicit results exist
for arbitrary momenta \cite{Passarino:1979jh}.

For the regions where the Gram determinants become small, we have
worked out two alternative calculational methods. The precise
description of these will be given elsewhere. Here, we sketch only the
basic strategy. One method makes use of expansions of the tensor
coefficients about the limit of vanishing Gram determinants. Using
this, again all tensor coefficients can be expressed in terms of the
standard scalar 1-point, 2-point, 3-point, and 4-point functions. In
the second, alternative method we evaluate a specific tensor
coefficient, the integrand of which is logarithmic in Feynman
parametrization, by numerical integration. Then the remaining
coefficients as well as the standard scalar integral (numerator equal
to one) are algebraically derived from this coefficient. This
reduction again involves only inverse Cayley determinants, but no
inverse Gram determinants. In this approach, the set of master
integrals is not given by the standard scalar integrals anymore.  For
some specific 3-point integrals, where the Cayley determinant vanishes
exactly, analytical results have been worked out that allow for a
stable numerical evaluation.

\subsection{Checks on the calculation}

In order to prove the reliability of our results, we have performed a
number of checks, as described in more detail in
\citere{Denner:2005es}.  We have checked the structure of the (UV,
soft, and collinear) singularities, the matching between virtual and
real corrections, and the gauge independence (by repeating the
calculation in a different gauge). The most convincing check for
ourselves is the fact that we worked out the whole calculation in two
independent ways, resulting in two independent computer codes the
results of which are in good agreement. All algebraic manipulations,
including the generation of Feynman diagrams, have been done using
independent programs. In particular, one calculation uses the strategy
described in \refse{se:minimal-reduction} for the reduction of the
spinor structures, the other the strategy presented in
\refse{se:stable-reduction}. For the calculation of the loop integrals
we use the two independent in-house libraries which employ the
different calculational methods sketched in \refse{se:loop-integrals}
for the numerical stabilization.
                                                                    
\section{Algebraic reduction of spinor chains}
\label{se:spinorchains}

One-loop amplitudes for processes with six external fermions involve
$\ord(10^3)$ different spinor structures, which are products of
three spinor chains. These structures are, however, not independent,
but can be related by using the Dirac algebra, the Dirac equation,
momentum conservation, and relations that follow from the
four-dimensionality of space--time.  Of course, the latter relations
can only be used after cancelling UV divergences, which are regularized
dimensionally in our work.

Below we describe two strategies that can be exploited to express any
product of three spinor chains for a reaction involving six external
massless fermions (and no external bosons) in terms of very few
standard structures.  Note that none of the following steps leads to
Gram determinants in denominator factors, which potentially
spoil the numerical stability%
\footnote{Inverse Gram determinants, e.g., appear in the reduction
proposed in \citere{Vicini:2001pd}.}.
For generic reasons we take all six fermion momenta
$p_i$ as incoming ($\sum_{i=1}^6 p_i=0$) and use the following 
shorthand notation for a spinor chain,
\beq
\Bigl[ \DiracGamma \Bigr]^{\pm}_{ab} \, 
= \, \bar v_a(p_a) \, \DiracGamma \, \omega_{\pm} \, u_b(p_b) \,,
\eeq
where $\bar v_a(p_a)$ and $u_b(p_b)$ are the usual spinors for the
(anti-)fermions and 
\beq\label{chirality-projectors}
\omega_{\pm}=\frac{1}{2}(1\pm\gamma_5)
\eeq
the chirality projectors.  Since we deal with massless fermions, the
matrices $\DiracGamma$, which act in Dirac space, always consist of a
product of an odd number of Dirac matrices $\ga^\mu$.  The quantities
that we want to simplify have the general structure
\beq \label{SMEstructure}
\bar v_1(p_1) A \omega_\rho u_2(p_2) \;\times\;
\bar v_3(p_3) B \omega_\si  u_4(p_4) \;\times\;
\bar v_5(p_5) C \omega_\tau u_6(p_6) \;\equiv\;
\Bigl[A\Bigr]_{12}^\rho \,
\Bigl[B\Bigr]_{34}^\si  \,
\Bigl[C\Bigr]_{56}^\tau.
\eeq

We start out by deriving and summarizing some basic relations
that are used in the reduction.

\subsection{Basic relations}

\subsubsection{Identities for Dirac matrices}

The four-dimensionality of space--time leads to the Chisholm identity 
\beq\label{Chisholm-Identity}
\ga^\al \ga^\be \ga^\ga =
g^{\al\be}\ga^\ga - g^{\al\ga}\ga^\be + g^{\be\ga}\ga^\al
+\ri\epsilon^{\al\be\ga\de}\ga_\de\ga_5,
\label{eq:chisholm}
\eeq
which can be used to express products $\ga^\al \ga^\be \ga^\ga$
of Dirac matrices in terms of single Dirac matrices dressed by
$\ga_5$ factors and totally antisymmetric tensors $\epsilon^{\al\be\ga\de}$
(we adopt the convention $\eps^{0123}=+1$).

If at least two Dirac chains are involved, this identity can be applied to
shift factors of Dirac matrices from one chain to another without
introducing $\epsilon$-tensors, provided the two chains are Lorentz
contracted with each other:
\beqar\label{eq:rel}
\ga^\mu\ga^\al\ga^\be \otimes \ga_\mu &\eqref{Chisholm-Identity}&
\left( g^{\mu\al}\ga^\be - g^{\mu\be}\ga^\al + g^{\al\be}\ga^\mu
+ \ri\epsilon^{\mu\al\be\nu}\ga_\nu\ga_5 \right) \otimes \ga_\mu
\nn\\
&=& \ga^\be\otimes\ga^\al - \ga^\al\otimes\ga^\be 
+ g^{\al\be}\ga^\mu\otimes\ga_\mu
- \ga_\nu\ga_5\otimes\left( \ri\epsilon^{\nu\al\be\mu}\ga_\mu\ga_5 \right)\ga_5
\nn\\
&\eqref{Chisholm-Identity}& \ga^\be\otimes\ga^\al - \ga^\al\otimes\ga^\be 
+ g^{\al\be}\ga^\mu\otimes\ga_\mu
+ \ga^\al\ga_5\otimes\ga^\be\ga_5 
- \ga^\be\ga_5\otimes\ga^\al\ga_5 
\nn\\
&& {}
+ g^{\al\be}\ga^\mu\ga_5\otimes\ga_\mu\ga_5
- \ga_\mu\ga_5\otimes\ga^\mu\ga^\al \ga^\be\ga_5.
\eeqar
Here and in the following the symbol $\otimes$ denotes the direct product 
in Dirac space, \ie Dirac matrices are not sandwiched between 
Dirac spinors in this case.
Relation \refeq{eq:rel} can be expressed in a very compact way after
introducing the chirality projectors \refeq{chirality-projectors} and
rearranging the order of some Dirac matrices,
\beqar
\ga^\mu\ga^\al\ga^\be\omega_\pm \otimes \ga_\mu\omega_\pm &=&
\ga^\mu\omega_\pm \otimes \ga_\mu\ga^\be\ga^\al\omega_\pm,
\nn\\ 
\ga^\al\ga^\be\ga^\mu\omega_\pm \otimes \ga_\mu\omega_\pm &=&
\ga^\mu\omega_\pm \otimes \ga^\be\ga^\al\ga_\mu\omega_\pm,
\nn\\ 
\ga^\mu\ga^\al\ga^\be\omega_\pm \otimes \ga_\mu\omega_\mp &=&
\ga^\mu\omega_\pm \otimes \ga^\al\ga^\be\ga_\mu\omega_\mp,
\nn\\ 
\ga^\al\ga^\be\ga^\mu\omega_\pm \otimes \ga_\mu\omega_\mp &=&
\ga^\mu\omega_\pm \otimes \ga_\mu\ga^\al\ga^\be\omega_\mp.
\label{eq:dcrel1}
\eeqar
The second and the fourth relations are equivalent forms of the
other two.

Equations~\refeq{eq:dcrel1} lead to analogous relations for direct
products of Dirac chains with more than one
contraction.%
\footnote{Such relations can also be found in \citere{Sirlin:1981pi}.}
For two contractions we get
\beqar
\ga^\mu\ga^\al\ga^\nu\omega_\pm \otimes \ga_\mu\ga^\be\ga_\nu\omega_\pm &=&
4g^{\al\be}\ga^\mu\omega_\pm \otimes\ga_\mu\omega_\pm,
\nn\\
\ga^\mu\ga^\al\ga^\nu\omega_\pm \otimes \ga_\mu\ga^\be\ga_\nu\omega_\mp &=&
4\ga^\be\omega_\pm \otimes\ga^\al\omega_\mp.
\label{eq:dcrel2}
\eeqar
The first of these relations is obtained by multiplying
the second chain of the first relation of \refeq{eq:dcrel1} by
$\ga^\ga\ga_\be$ from the right and renaming the indices $\be\to\nu$ and
$\ga\to\be$ after some obvious algebra. The second relation is
found analogously.

Relations for direct products of Dirac chains with three contractions
directly result from \refeq{eq:dcrel2} upon contraction with
$g_{\al\be}$,
\beqar
\ga^\mu\ga^\nu\ga^\rho\omega_\pm \otimes \ga_\mu\ga_\nu\ga_\rho\omega_\pm &=&
16\ga^\mu\omega_\pm \otimes\ga_\mu\omega_\pm,
\nn\\
\ga^\mu\ga^\nu\ga^\rho\omega_\pm \otimes \ga_\mu\ga_\nu\ga_\rho\omega_\mp &=&
4\ga^\mu\omega_\pm \otimes\ga_\mu\omega_\mp.
\label{eq:dcrel3}
\eeqar
Relations for more than three contractions could be derived from the
above results recursively, but are not needed in our case.

\subsubsection{Identities for products of spinor chains}

Further interesting relations follow if we sandwich the Dirac matrices
in the above relations between spinors and perform some
simplifications.  For instance, if we contract the second relation in
\refeq{eq:dcrel2} with $p_{1\indexsep\al} p_{3\indexsep\be}$, then
from the r.h.s.\ multiply the first chain of Dirac matrices by $A$ and
the second by $B$, where here $A$ and $B$ stand for the unit matrix
${\bf 1}$ or any product of an even number of Dirac matrices, and
finally attach the external spinors, we obtain \beq
\Bigl[\ga^\mu\dsl{p}_1\ga^\nu A\Bigr]^{\pm}_{12} \,
\Bigl[\ga_\mu\dsl{p}_3\ga_\nu B\Bigr]^{\mp}_{34} = 4\Bigl[\dsl{p}_3
A\Bigr]^{\pm}_{12} \, \Bigl[\dsl{p}_1 B\Bigr]^{\mp}_{34}.  \eeq On the
l.h.s.\ we can eliminate $\dsl{p}_1$ and $\dsl{p}_3$ by the respective
Dirac equations and obtain $4(p_1p_3)\Bigl[\ga^\nu A\Bigr]^\pm_{12} \,
\Bigl[\ga_\nu B\Bigr]^\mp_{34}$.  Three similar relations can be
derived from \refeq{eq:dcrel2} with other suitable contractions.  The
resulting four relations read \beqar \Bigl[A
\dsl{p}_3\Bigr]^{\pm}_{12} \, \Bigl[\dsl{p}_2 B\Bigr]^{\pm}_{34} &=&
(p_2 p_3)\Bigl[A \ga^\mu\Bigr]^{\pm}_{12} \, \Bigl[\ga_\mu
B\Bigr]^{\pm}_{34},
\nn\\
\Bigl[\dsl{p}_4 A\Bigr]^{\pm}_{12} \,
\Bigl[B\dsl{p}_1\Bigr]^{\pm}_{34} &=& (p_1 p_4)\Bigl[\ga^\mu
A\Bigr]^{\pm}_{12} \, \Bigl[B \ga_\mu\Bigr]^{\pm}_{34},
\nn\\
\Bigl[\dsl{p}_3 A\Bigr]^{\pm}_{12} \, \Bigl[\dsl{p}_1
B\Bigr]^{\mp}_{34} &=& (p_1 p_3)\Bigl[\ga^\mu A\Bigr]^{\pm}_{12} \,
\Bigl[\ga_\mu B\Bigr]^{\mp}_{34},
\nn\\
\Bigl[A \dsl{p}_4 \Bigr]^{\pm}_{12} \, \Bigl[B \dsl{p}_2
\Bigr]^{\mp}_{34} &=& (p_2 p_4)\Bigl[A \ga^\mu\Bigr]^{\pm}_{12} \,
\Bigl[B \ga_\mu \Bigr]^{\mp}_{34}.
\label{eq:dcrel4}
\eeqar

\subsubsection{Decomposition of the metric tensor}

In four dimensions the metric tensor $g^{\mu \nu}$ can be decomposed
in terms of four independent orthonormal four-vectors $n_j^{\mu}$,
\begin{equation}
   \label{decomp-metric-tensor}
   g^{\mu\nu} 
   = \sum_{i,j=0}^3 g^{i j} \, n_i^{\mu} \, n_j^{\nu} \,,
\end{equation}
with orthonormal components $ n_i^{\mu} \, n_{j \indexsep \mu} = g_{i
  j}$, where $ g_{i j} = g^{i j} = \mathrm{diag}(1,-1,-1,-1)$, as described in
the appendices of \citeres{Denner:2003iy,Denner:2003zp}. The four
four-vectors $n^{\mu}_i$ can be constructed from three
linearly-independent momenta $p_i,p_j,p_k$.  In the case of massless
momenta, $p_i^2 = p_j^2=p_k^2=0$, they can be defined as
\begin{eqnarray}
    n^{\mu}_0(p_i, p_j, p_k)
      & =  & \frac{1}{\sqrt{2\,(p_i p_j)}} \, 
             \left( p_i^{\mu}+p_j^{\mu} \right) 
\,, \qquad \quad
    n^{\mu}_1(p_i, p_j, p_k) 
=
             \frac{1}{\sqrt{2\,(p_i p_j)}} \,
             \left( p_i^{\mu}-p_j^{\mu} \right) \nn \,, \\ \nn \\
    n^{\mu}_2(p_i, p_j, p_k) 
      & = & -\frac{1}{\sqrt{2\,(p_i p_j) \, 
                               (p_i p_k) \, (p_j p_k)}}\,
\Bigl[(p_j p_k) \, p_i^{\mu}
          +  (p_i p_k) \, p_j^{\mu} 
            -(p_i p_j) \, p_k^{\mu}\Bigr]
      \nn \,, \\ \nn \\
    n^{\mu}_3(p_i, p_j, p_k) 
      & = & -\frac{1}{\sqrt{2\,(p_i p_j) \, 
                               (p_i p_k) \, (p_j p_k)}}\,
        \epsilon^{\mu\alpha\beta\gamma} \,
      p_{i\indexsep\alpha} \, p_{j\indexsep\beta} \, p_{k\indexsep\gamma} \,.
   \label{decomp-metric-tensor-ns}
\end{eqnarray}
The  decomposition of the metric tensor \refeq{decomp-metric-tensor} 
can be used to disconnect contractions of spinor chains or other objects.
Note that the construction of the four vectors $n_i$ from the three
independent vectors $p_i$, $p_j$, $p_k$, in particular the definition
of $n_3$, avoids an inverse Gram determinant in 
\refeq{decomp-metric-tensor}; if we decomposed the metric into
four totally independent momenta, the result would get the Gram
determinant of those four momenta in the denominator.


\subsubsection[Identities involving eps-tensors]{Identities involving $\eps$-tensors}
The fact that there is no totally antisymmetric tensor of rank 5 in four
space--time dimensions leads to the Schouten identity
\begin{equation}
   \epsilon_{[\alpha\beta\gamma\delta} \, g_{\mu]\nu} = 0 \,,
   \label{Schouten-Identity}
\end{equation}
where $[\ldots]$ means antisymmetrization in $\alpha,\beta,\gamma,\delta$,
and $\mu$.

The product of totally antisymmetric tensors can be expressed
as a determinant of metric tensors,
\begin{equation}\label{eps-product}
  \epsilon^{\alpha\beta\gamma\delta}
  \epsilon^{\mu\nu\rho\sigma}
  = -\left|\begin{array}{cccc}
g^{\alpha\mu} & g^{\alpha\nu} & g^{\alpha\rho} & g^{\alpha\sigma}\\
g^{\beta\mu} & g^{\beta\nu} & g^{\beta\rho} & g^{\beta\sigma}\\
g^{\gamma\mu} & g^{\gamma\nu} & g^{\gamma\rho} & g^{\gamma\sigma}\\
g^{\delta\mu} & g^{\delta\nu} & g^{\delta\rho} & g^{\delta\sigma}\end{array}\right|
  \label{EpsSquared}
\,.
\end{equation}

\subsection{Strategy for reducing spinor structures}
\label{se:minimal-reduction}

In this section we describe one algorithm for the reduction of spinor
chains. A second one is described in \refse{se:stable-reduction}.
The algorithm described here reduces all spinor chains to a minimal set.
Although this algorithm is applicable more generally, we refer in the
description to the massless charged-current $6f$ processes 
$\Pe^+\Pe^- \bar f_1 {f}_2\bar f_3 {f}_4\to 0$, where $f_1$ and $f_3$ are
different fermions excluding electrons and electron neutrinos and
$f_2$ and $f_4$ their isospin partners.

\begin{step}
Reduction of contractions between spinor chains
\end{step}

As first step in the reduction, spinor chains contracted with other
spinor chains or $\eps$-tensors are disconnected.%
\footnote{The explicit $\eps$-tensors result from calculating the
  traces of Dirac matrices in diagrams with closed fermion loops.}
This is achieved by introducing a decomposition of the metric
tensor (\ref{decomp-metric-tensor}) between contracted Dirac matrices
or contractions between a totally antisymmetric tensor and a Dirac
matrix:
\beqar
  \gamma_{\mu} \otimes \gamma^{\mu} \,
  & = \, g^{\mu\nu} \, \gamma_{\mu} \otimes \gamma_{\nu} \,
  & \stackrel{\mbox{\scriptsize{(\ref{decomp-metric-tensor})}}}{=}
    \, \sum_{i,j=0}^3 g^{i j} \, \dsl{n}_i \otimes \dsl{n}_j \,, \nn \\
  \epsilon^{\mu\nu\rho\sigma} \, \gamma_{\mu} \,
  & = \, \epsilon^{\mu\nu\rho\sigma} \, g_{\mu \alpha} \, \gamma^{\alpha}
  & \stackrel{\mbox{\scriptsize{(\ref{decomp-metric-tensor})}}}{=}
    \, \epsilon^{\mu\nu\rho\sigma} \, 
       \sum_{i,j=0}^3 g^{i j} \, n_{i\indexsep\mu} \, \dsl{n}_j \,.
   \label{singleContraction}
   \eeqar 
The choice of momenta for a suitable decomposition of the metric
tensor depends on the positions of the contracted matrices in the
spinor chains.  Preferably the momenta $p_i,p_j,p_k$ entering
\refeq{decomp-metric-tensor-ns} and thus
\refeq{decomp-metric-tensor} are selected in such a way that the
Dirac equations $\bar v(p_i) \dsl{p}_i = 0 = \dsl{p}_i u(p_i)$ or
the mass-shell condition $\dsl{p}_i^2 = p_i^2=0$ can be used most
directly, \ie without unnecessary anticommutations of Dirac
matrices by means of the Dirac algebra, since additional terms come
with it.  Therefore, we count the number of anticommutations of the
slashed momenta $\dsl{p}_i$, introduced by the decomposition of the
metric tensor, with Dirac matrices of the chain that would be
necessary in order to apply the Dirac equation or the mass-shell
condition.  We choose the three momenta with the fewest necessary
anticommutations.%
\footnote{ More precisely: If the metric tensor between two spinor
  chains is eliminated, we count the necessary anticommutations for
  each momentum in both spinor chains, say $N_1$ and $N_2$ with $N_1
  \leq N_2$, and set $N_2$ to infinity if the momentum does not appear
  in one of the chains.  Then we choose the momenta with smallest
  $N_1-1/N_2$.  If, on the other hand, the metric tensor between a
  spinor chain and an $\eps$-tensor is replaced, we proceed
  analogously, but set the number of necessary anticommutations to
  zero for those momenta that are already contracted with the
  $\eps$-tensor and to infinity otherwise.}  

The following substeps are recursively applied until no contractions of 
spinor chains are left:
\begin{itemize}
\item Disconnect a single  Lorentz contraction between two spinor chains or 
  between a totally antisymmetric tensor and a spinor chain by 
\refeq{singleContraction}.
\item Use the Dirac algebra together with the Dirac equation or the
  mass-shell condition $\dsl{p}_i^2 = 0$ in order to shorten spinor
  chains.
\item Replace contractions between totally antisymmetric tensors and $n_3$:
\beq
  \epsilon^{\mu\alpha\beta\gamma} \, n_{3\indexsep\mu}(p_i,p_j,p_k) \,
  \stackrel{\mbox{\scriptsize{(\ref{EpsSquared})}}}{=} \,
  \frac{1}{\sqrt{2\,(p_i p_j)\,(p_i p_k)\,(p_j p_k)}} \,
    p_i^{[\alpha} \, p_j^{\beta} \, p_k^{\gamma]} \,,
  \label{N3Eps}
\eeq
where $[\ldots]$ means antisymmetrization in $\alpha,\beta$, and
$\gamma$.
\item Eliminate $\dsl{n}_3$ in spinor chains by using 
\begin{equation} \dsl{n}_3 (p_i,p_j,p_k)\,
   \eqref{Chisholm-Identity}  -\frac{\ri 
     \left[\dsl{p}_i \dsl{p}_j \dsl{p}_k-(p_i p_j) \,
     \dsl{p}_k+(p_i p_k) \, \dsl{p}_j-(p_j p_k) \, 
     \dsl{p}_i \right] \gamma_5}{\sqrt{2\,(p_i p_j)\,
                                 (p_i p_k)\,
                                 (p_j p_k)}}\,.
   \label{N3slash}
\end{equation}
\end{itemize}

After this step, the (disconnected) spinor chains we are left with are
of the form $\bigl[ \dsl{p}_j \bigr]_{ab}^{\pm}$ with $j \ne a,b$ and
$\bigl[ \dsl{p}_i\,\dsl{p}_j\,\dsl{p}_k \bigr]_{ab}^{\pm}$ with
pairwise different $i,j,k,a,b$ for processes involving six external
fermions.  Since there are many different types of multiple
contractions of spinor chains, we can only illustrate this reduction
step, and particularly the first substep, in three representative
examples:
\begin{eqnarray}
\lefteqn{    \Bigl[ \gamma^{\mu} \Bigr]_{ab}^{\sigma} \,
    \Bigl[ \gamma_{\mu} \Bigr]_{cd}^{\tau}
   \stackrel{\mbox{\scriptsize{(\ref{singleContraction})}}}{=} 
    \sum_{i,j=0}^3 g^{i j} \, 
      \Bigl[ \dsl{n}_i(p_a,p_b,p_c) \Bigr]_{ab}^{\sigma} \,
      \Bigl[ \dsl{n}_j(p_a,p_b,p_c) \Bigr]_{cd}^{\tau}
   }\qquad \nl
  & \stackrel{\mbox{\scriptsize{(\ref{N3slash})}}}{=} &
    \frac{1}{2} \, \Bigl[ \dsl{p}_c \Bigr]_{ab}^{\sigma} 
      \biggl(
        \frac{1}{(p_a p_c)}\,\Bigl[ \dsl{p}_a \Bigr]_{cd}^{\tau}
       +\frac{1}{(p_b p_c)}\,\Bigl[ \dsl{p}_b \Bigr]_{cd}^{\tau}
      \biggr) 
 \nl&&{}
    -\frac{1}{2} \, \Bigl[\dsl{p}_c  \ga_5 \Bigr]_{ab}^{\sigma} 
      \biggl(
        \frac{1}{(p_a p_c)}\,\Bigl[ \dsl{p}_a  \ga_5 \Bigr]_{cd}^{\tau}
       -\frac{1}{(p_b p_c)}\,\Bigl[ \dsl{p}_b  \ga_5 \Bigr]_{cd}^{\tau}
      \biggr) 
    \nl
  & = &
      \Bigl[ \dsl{p}_c \Bigr]_{ab}^{\sigma} \;
        \biggl(
          \frac{1}{(p_a p_c)} \, \Bigl[ \dsl{p}_a \Bigr]_{cd}^{\tau} \,
            \delta_{\sigma,-\tau}
         +\frac{1}{(p_b p_c)} \, \Bigl[ \dsl{p}_b \Bigr]_{cd}^{\tau} \,
            \delta_{\sigma,\tau} 
       \biggr) \,,\\[1ex]
%
    \lefteqn{
    \Bigl[ \gamma^{\mu} \Bigr]_{12}^+ \,
    \Bigl[ \dsl{p}_2 \, \gamma_{\mu} \, \gamma_{\nu} \Bigr]_{34}^- \,
    \Bigl[ \gamma^{\nu} \Bigr]_{56}^-
   \eqref{singleContraction} 
    \sum_{i,j=0}^3 g^{i j} \, 
      \Bigl[ \dsl{n}_i(p_2,p_1,p_3) \Bigr]_{12}^+ \,
      \Bigl[ \dsl{p}_2 \, \dsl{n}_j(p_2,p_1,p_3) \, 
            \gamma_{\nu} \Bigr]_{34}^- \,
      \Bigl[ \gamma^{\nu} \Bigr]_{56}^-
    }\qquad && \hspace{29em}\nl
  & \eqref{N3slash} &
    -\frac{1}{(p_1 p_3)} \, 
      \Bigl[ \dsl{p}_3 \Bigr]_{12}^+ \,
      \Bigl[ \dsl{p}_1 \, \dsl{p}_2 \, \gamma_{\nu} \Bigr]_{34}^-
      \Bigl[ \gamma^{\nu} \Bigr]_{56}^-
    \nl
  & \stackrel{\mbox{\scriptsize{(\ref{singleContraction})}}}{=} &
    -\frac{1}{(p_1 p_3)} \, 
    \Bigl[ \dsl{p}_3 \Bigr]_{12}^+ \,
    \sum_{i,j=0}^3 g^{i j} \, 
      \Bigl[ \dsl{p}_1 \, \dsl{p}_2 \, 
            \dsl{n}_i(p_2,p_4,p_5) \Bigr]_{34}^- \,
      \Bigl[ \dsl{n}_j(p_2,p_4,p_5) \Bigr]_{56}^-
    \nl
  & \stackrel{\mbox{\scriptsize{(\ref{N3slash})}}}{=} &
    -\frac{1}{(p_1 p_3)\,(p_4 p_5)} \, 
    \Bigl[ \dsl{p}_3 \Bigr]_{12}^+ \,
    \Bigl[ \dsl{p}_1 \, \dsl{p}_2 \, \dsl{p}_5 \Bigr]_{34}^- \,
    \Bigl[ \dsl{p}_4 \Bigr]_{56}^-
\,,\\[1ex]
%
    \lefteqn{
    \ri \, \epsilon^{\mu\alpha\gamma\delta} \,
    p_{1\indexsep\alpha} \, p_{3\indexsep\gamma} \,  p_{4\indexsep\de} \, 
    \Bigl[ \gamma_{\mu} \Bigr]_{12}^- 
   \stackrel{\mbox{\scriptsize{(\ref{singleContraction})}}}{=} 
    \ri \, \epsilon^{\mu\alpha\gamma\delta} \,
    p_{1\indexsep\alpha} \, p_{3\indexsep\gamma} \,  p_{4\indexsep\de} \, 
    \sum_{i,j=0}^3 g^{i j} \, 
      n_{i\indexsep\mu}(p_1,p_2,p_3) \, \Bigl[ \dsl{n}_j(p_1,p_2,p_3) \Bigr]_{12}^- \,
    }\qquad&&\hspace{30em}
  \nl
  &  \begin{array}{c}
      \mbox{\scriptsize{(\ref{N3Eps})}} \vspace{-3mm} \\
      = \vspace{-3mm} \\ \mbox{\scriptsize{(\ref{N3slash})}}
 \end{array}
  &
    \frac{1}{2\,(p_2 p_3)} \Bigl[ \dsl{p}_3 \Bigr]_{12}^- \, 
    \biggl( (p_1 p_2)\,(p_3 p_4)  - (p_1 p_3)(p_2p_4) + (p_1p_4) (p_2 p_3)
   \nl & & \qquad\qquad\qquad\quad\,
      -\ri \, \epsilon^{\alpha\beta\gamma\de} \,
      p_{1\indexsep\alpha} \, p_{2\indexsep\beta} \, p_{3\indexsep\gamma} \, 
 p_{4\indexsep\de} 
    \biggr)
    \nl
  & = &
    \frac{1}{2\,(p_2 p_3)} \Bigl[ \dsl{p}_3 \Bigr]_{12}^-
      A_{1\,2\,3\,4}^{\sss{+-++}} 
   \,,
\end{eqnarray}
where we introduced the abbreviation
\beq\label{abbr_A}
  A_{i\,j\,k\,l}^{a\,b\,c\,d}
  = a\, (p_i p_j)\,(p_k p_l)
   +b\, (p_i p_k)\,(p_j p_l) 
   +c\, (p_i p_l)\,(p_j p_k)
   -d\, \ri\,\epsilon_{i\,j\,k\,l} \,
  \label{DefA}
\eeq
with
$\epsilon_{i\,j\,k\,l}
 = \epsilon_{\alpha\beta\gamma\delta} \, 
        p_i^{\alpha}
        p_j^{\beta}
        p_k^{\gamma}
        p_l^{\delta}$
and upper index combinations $(\textstyle{-++\pm})$, 
$(\textstyle{+-+\pm})$, and $(\textstyle{++-\pm})$ for $(a\,b\,c\,d)$.

\begin{step}
Reduction of a spinor chain to standard form
\end{step}
Since all spinor structures are disconnected after the first step, we
can focus on single spinor chains in this step, \ie spinor chains of
the form $[B]^\si_{ab}$ where all Dirac matrices in
$B$ are contracted with momenta. These spinor chains can be
reduced to the standard form $\bigl[ \dsl{{p}} \bigr]_{ab}^{\pm}$
with a freely chosen momentum ${p} = p_n$, $p_n \ne p_a,p_b$, by
recursively applying the following substeps:
\begin{itemize}
\item For $p_m \ne p_a,p_b,{p}$, replace $\dsl{p}_m$ by
  \beq
    \dsl{p}_m \,
    \stackrel{\mbox{\scriptsize{(\ref{decomp-metric-tensor})}}}{=}
      \; p_{m\indexsep\mu} \, \sum_{i,j=0}^3 g^{i j} \, n_i^{\mu} \, \dsl{n}_j \
    \label{replaceMomentum}
  \eeq
\item Eliminate $\dsl{n}_3$ via (\ref{N3slash}).
\item Use the Dirac algebra together with the Dirac equation or the
  mass-shell condition $\dsl{p}_i^2 = 0$ in order to shorten the
  spinor chain.
\end{itemize}
Since for our case only two different types of spinor chains are left
after the first step, we can demonstrate the reduction procedure of
the two cases in full detail. The first case is reduced as:
%
\begin{eqnarray}
   \lefteqn{
    \Bigl[ \dsl{p}_m \Bigr]_{ab}^{\pm}
   \stackrel{\mbox{\scriptsize{(\ref{replaceMomentum})}}}{=} 
    p_{m\indexsep\mu} \,
    \sum_{i,j=0}^3 g^{i j} \,
                 n^{\mu}_i(p_a,p_b,p_n) \,
                 \Bigl[ \dsl{n}_j(p_a,p_b,p_n)
                 \Bigr]_{ab}^{\pm} }\qquad \nn \\
  & \stackrel{\mbox{\scriptsize{(\ref{N3slash})}}}{=} &
      \frac{  (p_a p_n)\,(p_b p_m)
           -(p_a p_b)\,(p_n p_m)
           +(p_a p_m)\,(p_n p_b)
          \pm\ri\,\epsilon_{a\,n\,b\,m}  }
       {2\,(p_a p_n)\,(p_b p_n)} \,
      \Bigl[ \dsl{p}_n \Bigr]_{ab}^{\pm} \nn \\
  &\eqref{abbr_A}&
      \frac{1}{2\,(p_a p_n)\,(p_b p_n)}\,
      A_{a\,n\,b\,m}^{\sss{+-+\mp}} \,
      \Bigl[ \dsl{p}_n \Bigr]_{ab}^{\pm}
\,.
   \label{Step2Case1}
\end{eqnarray}
The second case can be reduced to the first case as follows:
%
\begin{eqnarray}
    \lefteqn{
    \Bigl[ \dsl{p}_m \, \dsl{p}_l \, \dsl{p}_k \Bigr]_{ab}^{\pm}
   \stackrel{\mbox{\scriptsize{(\ref{replaceMomentum})}}}{=} 
   p_{l\indexsep\mu} \,
    \sum_{i,j=0}^3 g^{i j} \,
                 n^{\mu}_i(p_m,p_k,p_b) \,
                 \Bigl[ \dsl{p}_m \, \dsl{n}_j(p_m,p_k,p_b) \, \dsl{p}_k
                 \Bigr]_{ab}^{\pm}  
    }\qquad \nn \\
  & \stackrel{\mbox{\scriptsize{(\ref{N3slash})}}}{=} &
        \frac{(p_b p_m)\,(p_l p_k)
             -(p_b p_l)\,(p_m p_k) 
             +(p_b p_k)\,(p_m p_l)}
             {2\,(p_b p_m)(p_b p_k)} \,
        \Bigl[ \dsl{p}_m \, \dsl{p}_b \, \dsl{p}_k \Bigr]_{ab}^{\pm}
    \nn \\ & &  {}
      - \, \ri \, \frac{\epsilon_{b\,m\,l\,k}}{2\,(p_b p_m)(p_b p_k)} \,
      \Bigl[ \dsl{p}_m \ga_5  \dsl{p}_b \dsl{p}_k \Bigr]_{ab}^{\pm}
    \nn \\
  & = &
        \frac{ 
              (p_b p_m)\,(p_l p_k)
             -(p_b p_l)\,(p_m p_k) 
             +(p_b p_k)\,(p_m p_l)
      \mp \, \ri \, \epsilon_{b\,m\,l\,k}\,
        }{(p_b p_m)} \,
        \Bigl[ \dsl{p}_m \Bigr]_{ab}^{\pm}
    \nn \\
  &\eqref{abbr_A}&
      \frac{1}{(p_b p_m)} \, 
      A_{b\,m\,l\,k}^{\sss{+-+\pm}} \,
      \Bigl[ \dsl{p}_m \Bigr]_{ab}^{\pm} \,.
   \label{Step2Case2}
\end{eqnarray}
Thus, we end up with spinor chains in standard form and prefactors $\Sn$
containing only scalar products of external momenta $(p_i p_j)$,
$\epsilon_{i\,j\,k\,l}$, and the abbreviations \refeq{abbr_A}. Note
that \refeq{Step2Case2} could be applied recursively to longer spinor
chains.

Thus, for processes with six external fermions, all Dirac structures
occurring in one-loop amplitudes, \ie all structures of the form
\refeq{SMEstructure}, can be brought to the form
\beq  
\Sn \, \Bigl[ \dsl{p}_3 \Bigr]_{12}^{\rho} \,
           \Bigl[ \dsl{p}_1 \Bigr]_{34}^{\si} \,
           \Bigl[ \dsl{p}_1 \Bigr]_{56}^{\tau} \,,
\eeq
where we have chosen the momenta $p_3$, $p_1$, and $p_1$ in the spinor
chains by convention. Accordingly, the SMEs can be chosen as
\beq\label{eq:SME1}
\Mhat^{\rho\si\tau} =  \Bigl[ \dsl{p}_3 \Bigr]_{12}^{\rho} \,
           \Bigl[ \dsl{p}_1 \Bigr]_{34}^{\si} \,
           \Bigl[ \dsl{p}_1 \Bigr]_{56}^{\tau} \,,
      \label{Standard-Matrix-Element}
\eeq
and there is only one SME for each helicity combination of the
external fermions.  For purely W-mediated charged-current processes,
there are only two non-vanishing helicity combinations and thus
only two different SMEs.

\begin{step}
Simplification of scalar factors
\end{step}
After the reduction steps above, each spinor structure has the
form (\ref{Standard-Matrix-Element}) with prefactors $S$ containing
products of scalars products $(p_i p_j)$ in the denominator, and
possibly polynomials of scalar products, of $\epsilon_{i\,j\,k\,l}$,
and of $A_{i\,j\,k\,l}^{a\,b\,c\,\sss{\pm}}$ in the numerator.  These
prefactors can be simplified further by means of relations we describe
now.

The quantity $A_{i\,j\,k\,l}^{a\,b\,c\,\sss{\pm}}$ defined in
\refeq{DefA}
transforms under exchange of two momenta, corresponding to two of the
indices $(ijkl)$, as
\beq   \label{TransA1}
     A_{i\,j\,k\,l}^{a\,b\,c\,\sss{\pm}}
   = A_{j\,i\,k\,l}^{a\,c\,b\,\sss{\mp}}
   = A_{i\,j\,l\,k}^{a\,c\,b\,\sss{\mp}}
   = A_{k\,j\,i\,l}^{c\,b\,a\,\sss{\mp}}
   = A_{i\,l\,k\,j}^{c\,b\,a\,\sss{\mp}}
   = A_{l\,j\,k\,i}^{b\,a\,c\,\sss{\mp}}
   = A_{i\,k\,j\,l}^{b\,a\,c\,\sss{\mp}} \,,
\eeq
and is therefore invariant under exchange of two distinct pairs of 
momenta,
\begin{equation}
     A_{i\,j\,k\,l}^{a\,b\,c\,\sss{\pm}}
     = A_{j\,i\,l\,k}^{a\,b\,c\,\sss{\pm}}
     = A_{k\,l\,i\,j}^{a\,b\,c\,\sss{\pm}}
     = A_{l\,k\,j\,i}^{a\,b\,c\,\sss{\pm}} \,.
   \label{TransA2}
\end{equation}
Owing to these relations, any $A_{i'j'k'l'}^{a\,b\,c\,\sss{\pm}}$,
where $(i'j'k'l')$ is an arbitrary permutation of $(i\,j\,k\,l)$, can
be transformed into one of the six elements
\beq\label{minimal-A}
  A_{\,i\,j\,k\,l}^{\sss{-++\pm}}, \,
  A_{\,i\,j\,k\,l}^{\sss{+-+\pm}}, \, \mbox{ or } \,
  A_{\,i\,j\,k\,l}^{\sss{++-\pm}} \,.
  \label{AquivClassA}
\eeq
In the following, we use only these independent quantities
\refeq{minimal-A}, \ie for each set of indices $(i\,j\,k\,l)$ of
$A^{a\,b\,c\,\sss{\pm}}_{\,i\,j\,k\,l}$ we define a standard order.

The identity for products of totally antisymmetric tensors
(\ref{EpsSquared}) leads to relations among the
$A_{i\,j\,k\,l}^{a\,b\,c\,\sss{\pm}}$.  Relations for products
of $A_{i\,j\,k\,l}^{a\,b\,c\,\sss{\pm}}$ with the same momenta are
\beq   \label{Aproduct1}
  \begin{array}[b]{lcl}
  A_{\,i\,j\,k\,l}^{\scriptscriptstyle{-++\pm}} \,
  A_{\,i\,j\,k\,l}^{\scriptscriptstyle{-++\mp}} \,
  & = & \, 
    4\, (p_i p_k) (p_j p_l) \, (p_i p_l) (p_j p_k) \,, \\
  A_{\,i\,j\,k\,l}^{\scriptscriptstyle{+-+\pm}} \,
  A_{\,i\,j\,k\,l}^{\scriptscriptstyle{+-+\mp}} \,
  & = & \,
    4\, (p_i p_j) (p_k p_l) \, (p_i p_l) (p_j p_k) \,, \\
  A_{\,i\,j\,k\,l}^{\scriptscriptstyle{++-\pm}} \,
  A_{\,i\,j\,k\,l}^{\scriptscriptstyle{++-\mp}} \,
  & = & \, 
    4\, (p_i p_j) (p_k p_l) \, (p_i p_k) (p_j p_l) \,,
\end{array}
\eeq
and
\beq  \label{Aproduct2}
  \begin{array}[b]{lcl}
  A_{\,i\,j\,k\,l}^{\sss{+-+\pm}} \,
  A_{\,i\,j\,k\,l}^{\sss{++-\pm}} \,
  & = & \,-2\, (p_i p_j)(p_k p_l) \, 
    A_{\,i\,j\,k\,l}^{\sss{-++\mp}} \, \\
  A_{\,i\,j\,k\,l}^{\sss{-++\pm}} \,
  A_{\,i\,j\,k\,l}^{\sss{++-\pm}} \, 
  & = & \,-2\, (p_i p_k)(p_j p_l) \,
    A_{\,i\,j\,k\,l}^{\sss{+-+\mp}} \,, \\
  A_{\,i\,j\,k\,l}^{\sss{-++\pm}} \,
  A_{\,i\,j\,k\,l}^{\sss{+-+\pm}} \,
  & = & \,-2\, (p_i p_l)(p_j p_k) \,
    A_{\,i\,j\,k\,l}^{\sss{++-\mp}} \,.
  \end{array}
\eeq
Note that in these and the following formulas double Latin indices are
not summed.  In \refeqs{Aproduct1} and \refeqf{Aproduct2} the second
and third lines are obtained from the first by the substitutions
($j\leftrightarrow k$) and ($i\leftrightarrow k$), respectively, and
subsequent transformation to the six elements \refeq{minimal-A}.

Relations for products of
$A_{i\,j\,k\,l}^{a\,b\,c\,\sss{\pm}}$ which differ in one momentum read
\beq
  \begin{array}[b]{lclcl}
   A_{\,i\,j\,k\,l}^{\sss{-++\pm}}
   & = & \, \frac{1}{2\, (p_i p_m) (p_j p_m)} \,
     A_{\,i\,j\,k\,m}^{\sss{-++\pm}} \,
     A_{\,i\,j\,l\,m}^{\sss{-++\mp}} \,
   & = & \, \frac{1}{2\, (p_k p_m) (p_l p_m)} \,
     A_{\,i\,k\,l\,m}^{\sss{++-\pm}} \,
     A_{\,j\,k\,l\,m}^{\sss{++-\mp}} \,, 
  \end{array}
  \label{Aproduct3}
\eeq
and
\beq
  \begin{array}[b]{lclclcl}
  A_{\,i\,j\,k\,l}^{\sss{+-+\pm}}
  A_{\,i\,j\,k\,m}^{\sss{++-\pm}}
  & = & -\frac{(p_i p_j)}{(p_i p_m)}
        A_{\,i\,j\,k\,m}^{\sss{-++\mp}}
        A_{\,i\,k\,l\,m}^{\sss{-++\pm}}
  & = & \frac{(p_i p_j)}{(p_l p_m)}
        A_{\,i\,k\,l\,m}^{\sss{++-\mp}}
        A_{\,j\,k\,l\,m}^{\sss{+-+\mp}}
  & = & -\frac{(p_i p_j)}{(p_j p_l)}
        A_{\,i\,j\,k\,l}^{\sss{-++\mp}}
        A_{\,j\,k\,l\,m}^{\sss{-++\pm}} 
  \,.
  \end{array}
  \label{Aproduct4}
\eeq
Two further relations can be obtained from each of the equations in
\refeq{Aproduct3} and \refeq{Aproduct4} via the substitutions
($j\leftrightarrow k$) and ($i\leftrightarrow k$). Three further sets
of relations can finally be constructed by substituting
($l\leftrightarrow m$) in all these relations derived from
\refeq{Aproduct4}.

Step 3 consists of two parts:
\begin{itemize}
\item First, we try to eliminate {\em sums} containing
  $A_{i\,j\,k\,l}^{a\,b\,c\,\sss{\pm}}$, $(p_i p_j)\,(p_k p_l)$,
  $\epsilon_{i\,j\,k\,l}$, and products thereof. To this end, we use
  the Schouten identity (\ref{Schouten-Identity}), the relations
  \refeq{Aproduct1}--\refeq{Aproduct4}, and linear relations like
$A_{i\,j\,k\,l}^{\sss{+-+\pm}} = 2\,(p_i p_j)\,(p_k
p_l)-A_{i\,j\,k\,l}^{\sss{++-\mp}}$ or $(p_i p_j)\,(p_k p_l) \pm
\ri\,\epsilon_{i\,j\,k\,l} = A_{i\,j\,k\,l}^{\sss ++-\mp} -(p_i
p_k)\,(p_j p_l) + (p_i p_l)\,(p_j p_k)$ that follow from the
definition of the $A_{i\,j\,k\,l}^{a\,b\,c\,\sss{\pm}}$ \refeq{abbr_A}.
All these relations are applied recursively to parts of $\Sn$ in
(\ref{Standard-Matrix-Element}) as long as they lead to
simplifications.
  
When starting from structures of the form \refeq{SMEstructure}, we
succeeded in this way to eliminate all sums, and the resulting $\Sn$
involve only products of $A_{i\,j\,k\,l}^{a\,b\,c\,\sss{\pm}}$ and
scalar products in the numerator, and products of scalar products in
the denominator.
  
\item Second, the {\em products} of
  $A_{i\,j\,k\,l}^{a\,b\,c\,\sss{\pm}}$ are simplified further and
  brought into a standard form by using the relations
  \refeq{Aproduct1}--\refeq{Aproduct4}.

\end{itemize}

Typical results are
\beqar\label{SME-results}
  \Bigl[ \gamma^{\mu} \gamma^{\nu} \gamma^{\rho} \Bigr]_{12}^{\sigma}
  \Bigl[ \gamma_{\mu} \gamma_{\nu} \gamma_{\rho} \Bigr]_{34}^{\tau} \,
  & = &\,
   \frac{4}{(p_1 p_3)} \,
  \Bigl[ \dsl{p}_3 \Bigr]_{12}^{\sigma}
  \Bigl[ \dsl{p}_1 \Bigr]_{34}^{\tau}
  \Biggl(
    \frac{2\, A_{1\,2\,3\,4}^{\sss{-++}\sigma}}{(p_1 p_4)\,(p_2 p_3)} 
    \,\delta_{\sigma,\tau}
    +\delta_{\sigma,-\tau}
  \Biggr) \,,
\vspace{1.5mm} \nl
  \Bigl[ \dsl{p}_3 \dsl{p}_5 \gamma^{\mu}  \Bigr]_{12}^-
  \Bigl[ \gamma_{\mu} \Bigr]_{56}^- \,
  & = & \, 0 \,,
\vspace{1.5mm} \nl
  \Bigl[ \dsl{p}_3 \gamma^{\mu} \gamma^{\nu}  \Bigr]_{12}^+ \,
  \Bigl[ \gamma_{\mu} \gamma^{\rho} \gamma^{\kappa}  \Bigr]_{34}^-
  \Bigl[ \gamma_{\nu} \gamma_{\rho} \gamma_{\kappa}  \Bigr]_{56}^- \,
  & = & \, 0 \,,
\vspace{1.5mm} \nl
  \Bigl[ \dsl{p}_3 \gamma^{\mu} \gamma^{\nu}  \Bigr]_{12}^{\sigma}
  \Bigl[ \gamma_{\nu} \gamma^{\rho} \gamma^{\kappa}  \Bigr]_{34}^-
  \Bigl[ \gamma_{\mu} \gamma_{\rho} \gamma_{\kappa}  \Bigr]_{56}^- \,
  & = & \, \frac{-4\,A_{\,1\,3\,4\,5}^{\sss{-+++}} \,
                     A_{\,1\,3\,4\,6}^{\sss{+-++}}}
               {(p_1 p_3)\,(p_1 p_4)\,(p_1 p_5)\,(p_1 p_6)\,(p_3 p_4)}
         \Bigl[ \dsl{p}_3 \Bigr]_{12}^{\sigma}
         \Bigl[ \dsl{p}_1 \Bigr]_{34}^-
         \Bigl[ \dsl{p}_1 \Bigr]_{56}^-
\nl&&\times
  \Bigl( \delta_{\sigma,\sss{-}} +2\,\delta_{\sigma,\sss{+}} \Bigr)\,,
\vspace{1.5mm} \nl
  \Bigl[ \dsl{p}_3 \gamma^{\mu} \gamma^{\nu} \gamma^{\rho} \gamma^{\kappa}
         \Bigr]_{12}^- \,
  \Bigl[ \gamma_{\nu} \gamma_{\rho} \gamma_{\kappa}  \Bigr]_{34}^- \,
  \Bigl[ \gamma_{\mu}   \Bigr]_{56}^- \,
  & = & \, 
\frac{8\,A_{\,1\,2\,3\,4}^{\sss{-++-}}\,A_{\,1\,3\,5\,6}^{\sss{-++-}}}
               {(p_1 p_3)\,(p_1 p_4)\,(p_1 p_5)\,(p_1 p_6)\,(p_2 p_3)} \,
         \Bigl[ \dsl{p}_3 \Bigr]_{12}^-
         \Bigl[ \dsl{p}_1 \Bigr]_{34}^-
         \Bigl[ \dsl{p}_1 \Bigr]_{56}^- \,,
\vspace{1.5mm} \nl
-\ri \epsilon^{\mu \nu \rho \sigma}\, p_{1,\sigma} \,
  \Bigl[ \gamma_{\mu} \Bigr]_{12}^+ \,
  \Bigl[ \gamma_{\nu} \Bigr]_{34}^- \,
  \Bigl[ \gamma_{\rho} \Bigr]_{56}^- \,
  & = & \, \frac{A_{\,1\,2\,3\,5}^{\sss{+-+-}}}
               {2\,(p_1 p_3)\,(p_1 p_5)\,(p_2 p_3)} \,
         \Bigl[ \dsl{p}_3 \Bigr]_{12}^+
         \Bigl[ \dsl{p}_1 \Bigr]_{34}^-
         \Bigl[ \dsl{p}_1 \Bigr]_{56}^- \,.
\eeqar

The entire reduction algorithm, described in the three steps above,
reduces the spinor chains of the considered processes to the SMEs
\refeq{Standard-Matrix-Element} and about 35 different
$A_{i\,j\,k\,l}^{a\,b\,c\,\sss{\pm}}$.  When inserting the results into
the amplitudes, further simplifications between contributions of
different spinor structures can be performed owing to the simple
structure of the SMEs. This speeds up the calculations considerably.

The spinor structures could alternatively be evaluated directly in the
Weyl--van der Waerden spinor formalism, as described in some detail in
the appendix. Although most of the relations given in this section can
be easily derived using the spinor formalism, many simplifications
that are based on four-momenta are harder to perform in that approach.

\subsection{Alternative strategy for reducing Dirac structures}
\label{se:stable-reduction}

In this section we describe an alternative strategy for the reduction
of Dirac structures. In this approach we take care that no quantities
are introduced in the denominator that can lead to numerical problems,
like scalar products or Gram determinants.  As a consequence, the
Dirac structures are not reduced to the minimal set, but to a small
set of standard structures.  Moreover, the coefficients in front of
these standard structures only involve Lorentz products, but no
contractions with $\eps$-tensors, facilitating further simplifications
in amplitudes.%
\footnote{In this approach all explicit $\eps$-tensors are eliminated
  with the Chisholm identity \refeq{Chisholm-Identity} at the
  beginning of the reduction.}

\setcounter{step}{0}
\begin{step}
Reduction of multiple contractions between spinor chains
\end{step}
Multiple Lorentz contractions between two spinor chains
can be reduced to single contractions using \refeq{eq:dcrel2} and
\refeq{eq:dcrel3}. A typical example is given by
\beq
\Bigl[A \ga^\mu\dsl{p}_i\ga^\nu\Bigr]^\pm_{12} \,
\Bigl[B \ga_\mu\dsl{p}_j\ga_\nu\Bigr]^\pm_{34}
\eqref{eq:dcrel2}
4(p_i p_j) 
\Bigl[A \ga^\mu\Bigr]^\pm_{12} \, \Bigl[B \ga_\mu\Bigr]^\pm_{34}.
\eeq
This procedure leaves only products of spinor
chains with (i) no contractions, (ii) exactly one
contraction between two spinor chains, or (iii) the case in which
one spinor chain is singly contracted with 
each of the two other chains.

Case (iii) can be reduced to (i) and (ii) as follows.  Each spinor
chain of type (iii) can be brought to the form $\bigl[A
\ga^\mu\dsl{p}_i\ga^\nu\bigr]_{12}^\rho \, \bigl[B
\ga_\mu\bigr]_{34}^\si \, \bigl[C \ga_\nu\bigr]_{56}^\tau$ 
or to one of the analogous forms with external fermions interchanged.
The transformation to this standard form produces only terms of types
(i) and (ii). The factor $\dsl{p}_i$ in the first chain is always
present, because the number of Dirac matrices in the chain is at least
three.  We can assume that $p_i$ is the momentum of a spinor belonging
to the other chains, \ie $i=3,4,5,$ or 6.  Otherwise we could
eliminate it with the Dirac equation after anticommuting it to the
left or right in the spinor chain $\bigl[\dots\bigr]_{12}$.  The trick
to get rid of the two Lorentz contractions is to apply
\refeq{eq:dcrel1} in order to shift the factor $\dsl{p}_i$ to the
spinor chain where it can be eliminated with its Dirac equation. This
means, if $i=3$ or 4, $\dsl{p}_i$ is shifted into
$\bigl[\dots\bigr]_{34}$; if $i=5$ or 6, it is shifted into
$\bigl[\dots\bigr]_{56}$. We give a simple example:
\beqar
\lefteqn{\Bigl[A \ga^\mu\dsl{p}_4\ga^\nu\Bigr]^\pm_{12} \,
\Bigl[B \ga_\mu\Bigr]^\mp_{34} \, \Bigl[C \ga_\nu\Bigr]_{56}^\tau
\eqref{eq:dcrel1} \Bigl[A \ga^\mu\Bigr]^\pm_{12} \,
\Bigl[B \dsl{p}_4\ga^\nu\ga_\mu \Bigr]^\mp_{34} \, 
\Bigl[C \ga_\nu\Bigr]_{56}^\tau
}\qquad
\nn\\
&=&
\Bigl[A \ga^\mu\Bigr]^\pm_{12} \,
\Bigl[B (2p_4^\nu\ga_\mu-2p_{4,\mu}\ga^\nu)\Bigr]^\mp_{34} \, 
\Bigl[C \ga_\nu\Bigr]_{56}^\tau
\nn\\
&=&
2\Bigl[A \ga^\mu\Bigr]^\pm_{12} \,
\Bigl[B \ga_\mu\Bigr]^\mp_{34} \, 
\Bigl[C \dsl{p}_4\Bigr]_{56}^\tau
-2\Bigl[A \dsl{p}_4\Bigr]^\pm_{12} \,
\Bigl[B \ga^\nu\Bigr]^\mp_{34} \, 
\Bigl[C \ga_\nu\Bigr]_{56}^\tau.
\eeqar

\begin{step}
Reduction of single contractions to standard form
\end{step}
\begin{sloppypar}
  After the first step, the only contractions between spinor chains
  are of the form $\bigl[A \ga^\mu\bigr]_{12}^\rho \, \bigl[B
  \ga_\mu\bigr]_{34}^\si \, \bigl[C\bigr]_{56}^\tau$ (or with fermions
  interchanged), where the matrices $A$, $B$, $C$ are products of
  slashed momenta. The matrices $A$ and $B$ can be systematically
  eliminated as follows.
\end{sloppypar}

We reduce the number of slashed momenta in $A$ until $A={\bf 1}$.  If
$A={\bf 1}$, there is nothing to do.  Otherwise $A$ is of the form
$A'\dsl{p}_i\dsl{p}_j$ with $A'$ containing two slashed momenta less
than $A$.  If either $i$ or $j$ is 3 or 4, we shift
$\dsl{p}_i\dsl{p}_j$ to the spinor chain $\bigl[\dots]_{34}$ with the
help of \refeq{eq:dcrel1} and eliminate $\dsl{p}_3$ or $\dsl{p}_4$
with its Dirac equation.  This leaves the cases
$\dsl{p}_i\dsl{p}_j=\dsl{p}_5\dsl{p}_6$ or $\dsl{p}_6\dsl{p}_5$ (since
$\dsl{p}_i\dsl{p}_i=0$), which can be reduced by momentum
conservation, e.g.,
$\dsl{p}_5\dsl{p}_6=\dsl{p}_5(-\dsl{p}_1-\dsl{p}_2-\dsl{p}_3-\dsl{p}_4)$.
The terms $\dsl{p}_3$ and $\dsl{p}_4$ are again shifted to
$\bigl[\dots]_{34}$, while $\dsl{p}_1$ and $\dsl{p}_2$ can be
eliminated with their Dirac equations in $\bigl[\dots]_{12}$ directly.
The whole procedure reduces the number of slashed momenta in $\bigl[A
\ga^\mu\bigr]_{12}^\rho$ by two and can be repeated until $A={\bf 1}$.

The same procedure can be applied to $\bigl[B \ga_\mu\bigr]_{34}^\si$
until $B={\bf 1}$. Finally, all products of spinor chains containing a
Lorentz contraction between two chains are of the form
$\bigl[\ga^\mu\bigr]^\rho_{12} \, \bigl[\ga_\mu\bigr]^\si_{34} \,
\bigl[C\bigr]_{56}^\tau$ (or with fermions interchanged).  Again, we
give an illustrative example:
\beqar
\lefteqn{\Bigl[\dsl{p}_4\dsl{p}_5\ga^\mu\Bigr]^\pm_{12} \,
\Bigl[\dsl{p}_5\dsl{p}_6\ga_\mu\Bigr]^\mp_{34} \, \Bigl[C\Bigr]_{56}^\tau}
\qquad\nn\\
&\eqref{eq:dcrel1}&
\Bigl[\ga^\mu\Bigr]^\pm_{12} \,
\Bigl[\dsl{p}_5\dsl{p}_6\ga_\mu\dsl{p}_4\dsl{p}_5\Bigr]^\mp_{34} \, 
\Bigl[C\Bigr]_{56}^\tau
\nn\\
&=& 2(p_4 p_5) \,
\Bigl[\ga^\mu\Bigr]^\pm_{12} \,
\Bigl[\dsl{p}_5\dsl{p}_6\ga_\mu\Bigr]^\mp_{34} \, 
\Bigl[C\Bigr]_{56}^\tau
\nn\\
&=& -2(p_4 p_5) \,
\Bigl[\ga^\mu\Bigr]^\pm_{12} \,
\Bigl[(\dsl{p}_1+\dsl{p}_2+\dsl{p}_3+\dsl{p}_4)\dsl{p}_6
\ga_\mu\Bigr]^\mp_{34} \, 
\Bigl[C\Bigr]_{56}^\tau
\nn\\
&\eqref{eq:dcrel1}& -2(p_4 p_5) \, \Bigl\{
\Bigl[\ga^\mu(\dsl{p}_1+\dsl{p}_2)\dsl{p}_6\Bigr]^\pm_{12} \,
\Bigl[\ga_\mu\Bigr]^\mp_{34} 
+\Bigl[\ga^\mu\Bigr]^\pm_{12} \,
\Bigl[(\dsl{p}_3+\dsl{p}_4)\dsl{p}_6\ga_\mu\Bigr]^\mp_{34} \, 
\Bigr\} \Bigl[C\Bigr]_{56}^\tau
\nn\\
&=& -4(p_4 p_5) \, \Bigl\{
\Bigl[\dsl{p}_6\Bigr]^\pm_{12} \, \Bigl[\dsl{p}_1\Bigr]^\mp_{34} 
+(p_2 p_6) \Bigl[\ga^\mu\Bigr]^\pm_{12} \, \Bigl[\ga_\mu\Bigr]^\mp_{34} 
\nn\\
&& \qquad\qquad {}
+(p_4 p_6) \Bigl[\ga^\mu\Bigr]^\pm_{12} \,
\Bigl[\ga_\mu\Bigr]^\mp_{34} \, 
-\Bigl[\dsl{p}_4\Bigr]^\pm_{12} \, \Bigl[\dsl{p}_6\Bigr]^\mp_{34} \, 
\Bigr\} \Bigl[C\Bigr]_{56}^\tau.
\eeqar

\begin{step}
Reduction of products of slashed momenta
\end{step}
\begin{sloppypar}
  First we write all products of slashed momenta in a standard form.
  After the preceding steps, we are left with spinor chains of the
  form $\bigl[A\bigr]_{12}^\rho \, \bigl[\ga^\mu\bigr]^\si_{34} \,
  \bigl[\ga_\mu\bigr]_{56}^\tau$, $\bigl[\ga^\mu\bigr]^\rho_{12} \,
  \bigl[B\bigr]_{34}^\si \, \bigl[\ga_\mu\bigr]_{56}^\tau$,
  $\bigl[\ga^\mu\bigr]^\rho_{12} \, \bigl[\ga_\mu\bigr]^\si_{34} \,
  \bigl[C\bigr]_{56}^\tau$, or $\bigl[A\bigr]_{12}^\rho \,
  \bigl[B\bigr]_{34}^\si \, \bigl[C\bigr]_{56}^\tau$.  None of the
  $A$, $B$, $C$ contains open Dirac indices, \ie they are products
  of slashed momenta and chirality projectors.  With the help of Dirac
  algebra and of momentum conservation the spinor chains can be
  brought into the form of a linear combination of terms
  $\bigl[\dsl{p}_i\bigr]^\rho_{ab}$ and
  $\bigl[\dsl{p}_i\dsl{p}_j\dsl{p}_k\bigr]^\rho_{ab}$.  For
  $\bigl[A\bigr]_{12}^\rho$, e.g., we substitute
  $p_6=-p_1-p_2-p_3-p_4-p_5$, eliminate all $\dsl{p}_1$ and
  $\dsl{p}_2$ with their Dirac equations by anticommuting them to the
  left or right and reorder the remaining $\dsl{p}_i$ terms according
  to increasing $i$.  Owing to the relation $\dsl{p}\dsl{p}=p^2$, this
  leaves only $\bigl[\dsl{p}_i\bigr]^\rho_{12}$ with $i=3,4,5$ and
  $\bigl[\dsl{p}_3\dsl{p}_4\dsl{p}_5\bigr]^\rho_{12}$.  In the same
  way we reduce the other uncontracted spinor chains to
  $\bigl[\dsl{p}_j\bigr]^\rho_{34}$,
  $\bigl[\dsl{p}_5\dsl{p}_6\dsl{p}_1\bigr]^\rho_{34}$,
  $\bigl[\dsl{p}_k\bigr]^\rho_{56}$, and
  $\bigl[\dsl{p}_1\dsl{p}_2\dsl{p}_3\bigr]^\rho_{56}$ with $j=5,6,1$
  and $k=1,2,3$.
\end{sloppypar}

\begin{sloppypar}
We now eliminate all chains of the form
$\bigl[\dsl{p}_i\dsl{p}_j\dsl{p}_k\bigr]^\rho_{ab} \,
\bigl[\dsl{p}_l\dsl{p}_m\dsl{p}_n\bigr]^\si_{cd}$
by making use of the Chisholm identity \refeq{eq:chisholm}
which can be written as
\beq
\dsl{p}_i\dsl{p}_j\dsl{p}_k = E_{ijk} + G_{ijk}
\eeq
with the shorthands
\beqar
E_{ijk} &=& \ri\epsilon^{\al\be\ga\de} p_{i,\al} p_{j,\be} p_{k,\ga}
\ga_\de\ga_5,
\nn\\
G_{ijk} &=& (p_i p_j)\dsl{p}_k-(p_i p_k)\dsl{p}_j+(p_j p_k)\dsl{p}_i.
\eeqar
Thus, we can write
\beqar
\Bigl[\dsl{p}_i\dsl{p}_j\dsl{p}_k\Bigr]^\rho_{ab} \,
\Bigl[\dsl{p}_l\dsl{p}_m\dsl{p}_n\Bigr]^\si_{cd}
&=&
\Bigl[E_{ijk}\Bigr]^\rho_{ab} \, \Bigl[E_{lmn}\Bigr]^\si_{cd} 
+ \Bigl[\dsl{p}_i\dsl{p}_j\dsl{p}_k\Bigr]^\rho_{ab} \,
  \Bigl[G_{lmn}\Bigr]^\si_{cd}
\nn\\
&& {}
+ \Bigl[G_{ijk}\Bigr]^\rho_{ab} \,
  \Bigl[\dsl{p}_l\dsl{p}_m\dsl{p}_n\Bigr]^\si_{cd}
- \Bigl[G_{ijk}\Bigr]^\rho_{ab} \, \Bigl[G_{lmn}\Bigr]^\si_{cd}.
\hspace{2em}
\label{eq:dcrel5}
\eeqar
In the first term on the r.h.s.\ we can express the product of
$\eps$-tensors,
\beq
\Bigl[E_{ijk}\Bigr]^\rho_{ab} \, \Bigl[E_{lmn}\Bigr]^\si_{cd} =
-\epsilon^{\al\be\ga\de} \epsilon^{\al'\be'\ga'\de'} 
p_{i,\al} p_{j,\be} p_{k,\ga} p_{l,\al'} p_{m,\be'} p_{n,\ga'}
\Bigl[\ga_\de\ga_5\Bigr]^\rho_{ab} \,
\Bigl[\ga_{\de'}\ga_5\Bigr]^\si_{cd},
\eeq
in terms of ordinary Lorentz products because of \refeq{eps-product}.
This means, after exploiting \refeq{eq:dcrel5} products of slashed
momenta only occur in terms of the forms: (i)
$\bigl[\dsl{p}_3\dsl{p}_4\dsl{p}_5\bigr]^\rho_{12} \,
\bigl[\ga^\mu\bigr]^\si_{34} \, \bigl[\ga_\mu\bigr]^\tau_{56}$, (ii)
$\bigl[\dsl{p}_3\dsl{p}_4\dsl{p}_5\bigr]^\rho_{12} \,
\bigl[\dsl{p}_j\bigr]^\si_{34} \, \bigl[\dsl{p}_k\bigr]^\tau_{56}$,
and similar terms with products of slashed momenta in the other spinor
chains.
\end{sloppypar}

Terms of type (ii) can always be simplified with \refeq{eq:dcrel4}.
If one of the indices $j,k$ is 1 or 2, then
$\dsl{p}_3\dsl{p}_4\dsl{p}_5$ can always be reordered such that one of
the relations in \refeq{eq:dcrel4} applies (if necessary momentum
conservation has to be used).  We illustrate this by a simple example:
\beqar
\lefteqn{\Bigl[\dsl{p}_3\dsl{p}_4\dsl{p}_5\Bigr]^+_{12} \, 
\Bigl[\dsl{p}_2\Bigr]^-_{34} \, 
\Bigl[\dsl{p}_k\Bigr]^\tau_{56}
=
2(p_4 p_5) \Bigl[\dsl{p}_3\Bigr]^+_{12} \, 
\Bigl[\dsl{p}_2\Bigr]^-_{34} \, 
\Bigl[\dsl{p}_k\Bigr]^\tau_{56}
-\Bigl[\dsl{p}_3\dsl{p}_5\dsl{p}_4\Bigr]^+_{12} \, 
\Bigl[\dsl{p}_2\Bigr]^-_{34} \, 
\Bigl[\dsl{p}_k\Bigr]^\tau_{56}
}\qquad\nn\\
&\eqref{eq:dcrel4}&
2(p_4 p_5) \Bigl[\dsl{p}_3\Bigr]^+_{12} \, 
\Bigl[\dsl{p}_2\Bigr]^-_{34} \, 
\Bigl[\dsl{p}_k\Bigr]^\tau_{56}
-(p_2 p_4) \Bigl[\dsl{p}_3\dsl{p}_5\ga^\mu\Bigr]^+_{12} \, 
\Bigl[\ga_\mu\Bigr]^-_{34} \, 
\Bigl[\dsl{p}_k\Bigr]^\tau_{56}
\nn\\
&\eqref{eq:dcrel1}&
2(p_4 p_5) \Bigl[\dsl{p}_3\Bigr]^+_{12} \, 
\Bigl[\dsl{p}_2\Bigr]^-_{34} \, 
\Bigl[\dsl{p}_k\Bigr]^\tau_{56}
-(p_2 p_4) \Bigl[\ga^\mu\Bigr]^+_{12} \, 
\Bigl[\ga_\mu\dsl{p}_3\dsl{p}_5\Bigr]^-_{34} \, 
\Bigl[\dsl{p}_k\Bigr]^\tau_{56}
\nn\\
&=&
2(p_4 p_5) \Bigl[\dsl{p}_3\Bigr]^+_{12} \, 
\Bigl[\dsl{p}_2\Bigr]^-_{34} \, 
\Bigl[\dsl{p}_k\Bigr]^\tau_{56}
-2(p_2 p_4) \Bigl[\dsl{p}_3\Bigr]^+_{12} \, 
\Bigl[\dsl{p}_5\Bigr]^-_{34} \, 
\Bigl[\dsl{p}_k\Bigr]^\tau_{56}.
\eeqar
If neither $j$ or $k$ has the value 1 or 2 in terms of type (ii), then
one of the relations in \refeq{eq:dcrel4} applies to the product
$\bigl[\dsl{p}_j\bigr]^\si_{34} \, \bigl[\dsl{p}_k\bigr]^\tau_{56}$
(if necessary momentum conservation has to be used).  In this case
terms of type (ii) transform into terms of type (i) and terms without
products of slashed momenta.

Finally, we have to reduce terms of type (i), which turns out to be a
tedious task. We illustrate the basic steps for the following example,
\beqar
\lefteqn{\Bigl[\dsl{p}_3\dsl{p}_4\dsl{p}_5\Bigr]^-_{12} \, 
\Bigl[\ga^\mu\Bigr]^-_{34} \, 
\Bigl[\ga_\mu\Bigr]^-_{56}
=
\frac{1}{2}\Bigl[\dsl{p}_3\dsl{p}_4\ga^\nu\Bigr]^-_{12} \, 
\Bigl[\ga^\mu\Bigr]^-_{34} \, 
\Bigl[\ga_\nu\dsl{p}_5\ga_\mu\Bigr]^-_{56}
}\qquad\nn\\
&=&
-\frac{1}{2}\Bigl[\dsl{p}_3\dsl{p}_4\ga^\nu\Bigr]^-_{12} \, 
\Bigl[\ga^\mu\Bigr]^-_{34} \, 
\Bigl[\ga_\nu\dsl{p}_6\ga_\mu\Bigr]^-_{56}
\nn\\
&& {}
-\frac{1}{2}\Bigl[\dsl{p}_3\dsl{p}_4\ga^\nu\Bigr]^-_{12} \,
\Bigl[\ga^\mu\Bigr]^-_{34} \,
\Bigl[\ga_\nu(\dsl{p}_1+\dsl{p}_2+\dsl{p}_3+\dsl{p}_4)
\ga_\mu\Bigr]^-_{56}
\nn\\
&\eqref{eq:dcrel1}&
-\Bigl[\dsl{p}_3\dsl{p}_4\ga^\nu\Bigr]^-_{12} \, 
\Bigl[\dsl{p}_6\Bigr]^-_{34} \, 
\Bigl[\ga_\nu\Bigr]^-_{56}
\nn\\
&& {}
-\frac{1}{2}\Bigl[\dsl{p}_3\dsl{p}_4\ga^\nu
\ga_\mu(\dsl{p}_1+\dsl{p}_2+\dsl{p}_3+\dsl{p}_4) \Bigr]^-_{12} \,
\Bigl[\ga^\mu\Bigr]^-_{34} \, \Bigl[\ga_\nu\Bigr]^-_{56}.
\hspace{2em}
\label{eq:dcrelaux}
\eeqar
At first sight, we have made the expression more complicated by using
the Dirac equation and some rearrangements for $\dsl{p}_5$, which was
then eliminated by momentum conservation in the second equality.  The
last expression in \refeq{eq:dcrelaux} can be simplified by
anticommuting the terms in $(\dsl{p}_1+\dsl{p}_2+\dsl{p}_3+\dsl{p}_4)$
in such a way that the Dirac equations for $\dsl{p}_1$ and $\dsl{p}_2$
apply and that the terms $\dsl{p}_3$ and $\dsl{p}_4$ drop out via
$\dsl{p}_i\dsl{p}_i=0$. The various terms that survive this procedure
can all be further reduced as described in the previous steps. After
some tedious algebra, we get
\beq
\Bigl[\dsl{p}_3\dsl{p}_4\dsl{p}_5\Bigr]^-_{12} \, 
\Bigl[\ga^\mu\Bigr]^-_{34} \, 
\Bigl[\ga_\mu\Bigr]^-_{56}
= -2(p_3 p_5) \Bigl[\ga^\mu\Bigr]^-_{12} \,
\Bigl[\ga_\mu\Bigr]^-_{34} \, \Bigl[\dsl{p}_4\Bigr]^-_{56}
+2(p_4 p_5) \Bigl[\dsl{p}_3\Bigr]^-_{12} \,
\Bigl[\ga^\mu\Bigr]^-_{34} \, 
\Bigl[\ga_\mu\Bigr]^-_{56}.
\eeq
Note, however, that the explicit form of this result is not unique,
\ie it might look different if other algebraic manipulations had been
made.

\begin{step}
Reducing products of the form
$\bigl[\dsl{p}_i\bigr]^\rho_{12} \, 
 \bigl[\dsl{p}_j\bigr]^\si_{34} \, 
 \bigl[\dsl{p}_k\bigr]^\tau_{56}$
\end{step}
Obviously, momentum conservation can be used such that the indices
$i,j,k$ take only the following values, $i=3,4,5$, $j=5,6,1$, and
$k=1,2,3$.  Equations~\refeq{eq:dcrel4} provide $2\times3\times3=18$
relations per chirality which can be used to reduce the number of
these 27 $(ijk)$ values to 9.

There are, however, still some relations among the structures
$\bigl[\dsl{p}_i\bigr]^\rho_{12} \,
 \bigl[\dsl{p}_j\bigr]^\si_{34} \,
 \bigl[\dsl{p}_k\bigr]^\tau_{56}$
and $\bigl[\dsl{p}_i\bigr]^\rho_{12} \,
\bigl[\ga^\mu\bigr]^\si_{34} \, \bigl[\ga_\mu\bigr]^\tau_{56}$, etc.,
which can be exploited to eliminate some $(ijk)$ values.
Such relations can be deduced as in the following example,
\beqar
0 &=& \Bigl[\dsl{p}_1\dsl{p}_6\ga^\nu\dsl{p}_4\ga^\mu\Bigr]^-_{12} \,
 \Bigl[\ga_\mu\Bigr]^-_{34} \,
 \Bigl[\ga_\nu\Bigr]^-_{56}
\nn\\
&\eqref{eq:dcrel1}&
\Bigl[\ga^\nu\dsl{p}_4\ga^\mu\Bigr]^-_{12} \,
\Bigl[\ga_\mu\Bigr]^-_{34} \,
\Bigl[\dsl{p}_6\dsl{p}_1\ga_\nu\Bigr]^-_{56}
\nn\\
&\eqref{eq:dcrel1}&
\Bigl[\ga^\mu\Bigr]^-_{12} \,
\Bigl[\dsl{p}_4\ga^\nu\ga_\mu\Bigr]^-_{34} \,
\Bigl[\dsl{p}_6\dsl{p}_1\ga_\nu\Bigr]^-_{56}
\nn\\
&=&
\Bigl[\ga^\mu\Bigr]^-_{12} \,
\Bigl[2p_4^\nu\ga_\mu-2p_{4,\mu}\ga^\nu\Bigr]^-_{34} \,
\Bigl[2(p_1 p_6)\ga_\nu-2p_{6,\nu}\dsl{p}_1\Bigr]^-_{56}
\nn\\
&=&
4(p_1 p_6) \Bigl[\ga^\mu\Bigr]^-_{12} \,
\Bigl[\ga_\mu\Bigr]^-_{34} \, \Bigl[\dsl{p}_4\Bigr]^-_{56}
-4(p_4 p_6) \Bigl[\ga^\mu\Bigr]^-_{12} \,
\Bigl[\ga_\mu\Bigr]^-_{34} \, \Bigl[\dsl{p}_1\Bigr]^-_{56}
\nn\\
&& {}
-4(p_1 p_6) \Bigl[\dsl{p}_4\Bigr]^-_{12} \,
\Bigl[\ga^\nu\Bigr]^-_{34} \, \Bigl[\ga_\nu\Bigr]^-_{56}
+4\Bigl[\dsl{p}_4\Bigr]^-_{12} \,
\Bigl[\dsl{p}_6\Bigr]^-_{34} \, \Bigl[\dsl{p}_1\Bigr]^-_{56}.
\eeqar
This expresses the last term in terms of 
$\bigl[\dsl{p}_i\bigr]^-_{12} \,
\bigl[\ga^\mu\bigr]^-_{34} \, \bigl[\ga_\mu\bigr]^-_{56}$, etc.
In total we find 7 such relations per chirality, so that the number of 
structures 
$\bigl[\dsl{p}_i\bigr]^\rho_{12} \,
 \bigl[\dsl{p}_j\bigr]^\si_{34} \,
 \bigl[\dsl{p}_k\bigr]^\tau_{56}$
can be reduced to 2 combinations $(ijk)$ per chirality configuration
$(\rho\si\tau)$.

\begin{step}
Reducing products of the form
$\bigl[\dsl{p}_i\bigr]^\rho_{12} \, 
 \bigl[\ga^\mu\bigr]^\si_{34} \, 
 \bigl[\ga_\mu\bigr]^\tau_{56}$, etc.
\end{step}
\begin{sloppypar}
  As in the previous step, we use momentum conservation to constrain
  the indices $i,j,k$ in the structures
  $\bigl[\dsl{p}_i\bigr]^\rho_{12} \, \bigl[\ga^\mu\bigr]^\si_{34} \,
  \bigl[\ga_\mu\bigr]^\tau_{56}$, $\bigl[\ga^\mu\bigr]^\rho_{12} \,
  \bigl[\dsl{p}_j\bigr]^\pm_{34} \, \bigl[\ga_\mu\bigr]^\tau_{56}$,
  and $\bigl[\ga^\mu\bigr]^\rho_{12} \, \bigl[\ga_\mu\bigr]^\si_{34}
  \, \bigl[\dsl{p}_k\bigr]^\tau_{56}$ to the values $i=3,4,5$,
  $j=5,6,1$, and $k=1,2,3$.  One out of these 9 resulting structures
  (per chirality) is, however, redundant and can be easily eliminated
  as follows,
\beqar
0 &=&
\Bigl[\ga^\mu(\dsl{p}_1+\dsl{p}_2+\dsl{p}_3+\dsl{p}_4+\dsl{p}_5+\dsl{p}_6)
\ga^\nu\Bigr]^\pm_{12} \,
 \Bigl[\ga_\mu\Bigr]^\pm_{34} \, \Bigl[\ga_\nu\Bigr]^\pm_{56}
\nn\\
&\eqref{eq:dcrel1}&
\Bigl[\ga^\mu(\dsl{p}_1+\dsl{p}_2) \ga^\nu\Bigr]^\pm_{12} \,
 \Bigl[\ga_\mu\Bigr]^\pm_{34} \, \Bigl[\ga_\nu\Bigr]^\pm_{56}
+\Bigl[\ga^\mu\Bigr]^\pm_{12} \,
 \Bigl[\ga_\mu\ga^\nu(\dsl{p}_3+\dsl{p}_4)\Bigr]^\pm_{34} \,
 \Bigl[\ga_\nu\Bigr]^\pm_{56}
\nn\\
&& {}
+\Bigl[\ga^\nu\Bigr]^\pm_{12} \,
 \Bigl[\ga_\mu\Bigr]^\pm_{34} \,
 \Bigl[(\dsl{p}_5+\dsl{p}_6)\ga^\mu\ga_\nu\Bigr]^\pm_{56}
\nn\\
&=&
2\Bigl[\ga^\mu\Bigr]^\pm_{12} \,
 \Bigl[(\dsl{p}_1+\dsl{p}_6)\Bigr]^\pm_{34} \,
 \Bigl[\ga_\mu\Bigr]^\pm_{56}
+2\Bigl[\ga^\mu\Bigr]^\pm_{12} \,
 \Bigl[\ga_\mu\Bigr]^\pm_{34} \,
 \Bigl[(\dsl{p}_2+\dsl{p}_3)\Bigr]^\pm_{56}
\nn\\
&& {}
-2\Bigl[(\dsl{p}_3+\dsl{p}_6)\Bigr]^\pm_{12} \,
 \Bigl[\ga^\mu\Bigr]^\pm_{34} \, \Bigl[\ga_\mu\Bigr]^\pm_{56}
\nn\\
&=&
2\Bigl[\ga^\mu\Bigr]^\pm_{12} \,
 \Bigl[(\dsl{p}_1+\dsl{p}_6)\Bigr]^\pm_{34} \,
 \Bigl[\ga_\mu\Bigr]^\pm_{56}
+2\Bigl[\ga^\mu\Bigr]^\pm_{12} \,
 \Bigl[\ga_\mu\Bigr]^\pm_{34} \,
 \Bigl[(\dsl{p}_2+\dsl{p}_3)\Bigr]^\pm_{56}
\nn\\
&& {}
+2\Bigl[(\dsl{p}_4+\dsl{p}_5)\Bigr]^\pm_{12} \,
 \Bigl[\ga^\mu\Bigr]^\pm_{34} \, \Bigl[\ga_\mu\Bigr]^\pm_{56}.
\eeqar
Relations for the other chirality configurations can be obtained
in the same way.
\end{sloppypar}

We note that there are further relations among the spinor structures
considered in this step. The remaining relations, however, involve
coefficients with Lorentz products, so that their use to further
eliminate some structures would lead to Lorentz products in the
denominator, which we want to avoid. Nevertheless we give two such
relations for illustration:
\beqar
\lefteqn{(p_1 p_5) \Bigl[\dsl{p}_3\Bigr]^\pm_{12} \,
  \Bigl[\ga^\mu\Bigr]^\pm_{34} \, \Bigl[\ga_\mu\Bigr]^\pm_{56}
+(p_3 p_5) \Bigl[\ga^\mu\Bigr]^\pm_{12} \, 
  \Bigl[\dsl{p}_1\Bigr]^\pm_{34} \, \Bigl[\ga_\mu\Bigr]^\pm_{56}}
\qquad\nn\\
&=&
(p_1 p_3) \Bigl[\ga^\mu\Bigr]^\pm_{12} \,
 \Bigl[\dsl{p}_5\Bigr]^\pm_{34} \, \Bigl[\ga_\mu\Bigr]^\pm_{56} 
+(p_1 p_5) \Bigl[\ga^\mu\Bigr]^\pm_{12} \, 
 \Bigl[\ga_\mu\Bigr]^\pm_{34} \, \Bigl[\dsl{p}_3\Bigr]^\pm_{56} 
\nn\\
&=&
(p_3 p_5) \Bigl[\ga^\mu\Bigr]^\pm_{12} \,
 \Bigl[\ga_\mu\Bigr]^\pm_{34} \, \Bigl[\dsl{p}_1\Bigr]^\pm_{56}
+(p_1 p_3) \Bigl[\dsl{p}_5\Bigr]^\pm_{12} \,
 \Bigl[\ga^\mu\Bigr]^\pm_{34} \, \Bigl[\ga_\mu\Bigr]^\pm_{56}.
\eeqar
\\[.5em]
{\bf Summary:} {\it Final set of spinor structures}
\\[.5em]
After applying the steps described above, all Dirac structures
occurring in one-loop amplitudes for processes with six external
fermions can be expressed as linear combinations of the following SMEs
with coefficients that do not involve any denominators,
\beqar
\Mhat^{\rho\si\tau}_{\{1,2\}} &=& \Bigl[\ga^\mu\Bigr]^\rho_{12} \,
 \Bigl[\ga_\mu\Bigr]^\si_{34} \, 
 \Bigl[\dsl{p}_{\{1,2\}}\Bigr]^\tau_{56},
\nn\\
\Mhat^{\rho\si\tau}_{\{3,4,5\}} &=& 
 \Bigl[\dsl{p}_{\{3,4,5\}}\Bigr]^\rho_{12} \,
 \Bigl[\ga^\mu\Bigr]^\si_{34} \, \Bigl[\ga_\mu\Bigr]^\tau_{56},
\nn\\
\Mhat^{\rho\si\tau}_{\{6,7,8\}} &=& \Bigl[\ga^\mu\Bigr]^\rho_{12} \,
 \Bigl[\dsl{p}_{\{5,6,1\}}\Bigr]^\si_{34} \, 
 \Bigl[\ga_\mu\Bigr]^\tau_{56},
\nn\\
\Mhat^{\rho\si\tau}_{9} &=& \Bigl[\dsl{p}_4\Bigr]^\rho_{12} \,
 \Bigl[\dsl{p}_6\Bigr]^\si_{34} \, 
 \Bigl[\dsl{p}_2\Bigr]^\tau_{56},
\nn\\
\Mhat^{\rho\si\tau}_{10} &=& \Bigl[\dsl{p}_3\Bigr]^\rho_{12} \,
 \Bigl[\dsl{p}_5\Bigr]^\si_{34} \, 
 \Bigl[\dsl{p}_1\Bigr]^\tau_{56}.
\eeqar

\section{The complex-mass scheme at one loop}
\label{se:complex-masses}

\newcommand{\cmps}{\mu^2_P}
\newcommand{\cmhs}{\mu^2_\PH}
\newcommand{\cmt}{\mu_\Pt}
\newcommand{\cmws}{\mu^2_\PW}
\newcommand{\cmzs}{\mu^2_\PZ}
\newcommand{\cmw}{\mu_\PW}
\newcommand{\cmz}{\mu_\PZ}
\newcommand{\csw}{\mathswitch {s_\rw}}
\newcommand{\ccw}{\mathswitch {c_\rw}}
\newcommand{\cZ}{\mathcal{Z}}

The description of resonances in perturbation theory requires at least
a partial Dyson summation of self-energy insertions. This leads to a
mixing of perturbative orders and, if done carelessly, can easily
jeopardise gauge invariance \cite{Berends:1969nt,Argyres:1995ym}.
Therefore, the proper introduction of finite-width effects is a
non-trivial problem.  While several solutions have been described for
lowest-order predictions
\cite{Denner:1999gp,Argyres:1995ym,Aeppli:1993cb,Beenakker:1996kn,Stuart:1991xk,Aeppli:1993rs,Baur:1995aa,Passarino:1999zh,Beenakker:1999hi,Beneke:2003xh},
no viable, universally valid scheme exists so far for a consistent
evaluation of radiative corrections in the presence of resonances.  A
pole expansion
\cite{Aeppli:1993cb,Stuart:1991xk,Aeppli:1993rs,Beneke:2003xh}
provides a gauge-invariant answer and is applicable to radiative
corrections, but restricts the validity of the result to the resonance
region only and is not reliable in threshold regions.  In our
calculation we want to cover both the threshold region, where the pole
approximation is not applicable, and the continuum above threshold,
where threshold expansions are not valid. Moreover, the calculation
should be valid both for resonant and non-resonant regions in phase
space.  In other words, we are after a unified description that is
applicable in the complete phase space and does not require any
matching between different treatments for different regions.

Such a description is provided by the ``complex-mass
scheme'' (CMS), which was introduced in \citere{Denner:1999gp} for
lowest-order calculations.  In this approach
the W- and Z-boson masses are consistently considered as complex
quantities, defined as the locations of the poles in the complex $k^2$
plane of the corresponding propagators with momentum $k$.  Gauge
invariance is preserved if the complex masses are introduced
everywhere in the Feynman rules, in particular in the definition of
the weak mixing angle, 
\beq\label{eq:compl-mixing} 
\cos^2\theta_\PW \equiv \ccw^2 = 1-\csw^2 =
\frac{\cmws}{\cmzs},
\eeq
which is derived from the ratio of the complex mass squares
of the gauge bosons,%
\footnote{While it is generally accepted that the mass and width of
  unstable particles are related to the pole of the propagator in the
  complex plane, this does not define the mass and the width
  separately.  This arbitrariness is discussed in detail in
  \citere{Bohm:2004zi}, where also a definition of the mass and width
  is proposed such that the width is given by the inverse lifetime.
  We have chosen the popular definition \refeq{eq:complex-masses},
  but the complex renormalization scheme is applicable to other
  definitions as well.}
\beqar\label{eq:complex-masses} 
\cmws&=& \MW^2 - \ri\MW\GW, \qquad \cmzs= \MZ^2 - \ri\MZ\GZ.  
\eeqar 
The (algebraic) relations, such as Ward identities, that follow from
gauge invariance remain valid, because the gauge-boson masses are
modified only by an analytic continuation. As a consequence unitarity
cancellations are respected, and the amplitudes have a decent
high-energy behaviour.

While necessary in the resonant propagators, the consistent
introduction of complex gauge-boson masses introduces spurious terms
in other places, as \eg in the weak mixing angle
\refeq{eq:compl-mixing}.  When using the CMS at tree level, which
amounts to replacing the real gauge-boson masses by the complex masses
\refeq{eq:complex-masses} and the weak mixing angle by
\refeq{eq:compl-mixing} in tree-level amplitudes, the spurious terms
are of order $\ord(\GW/\MW)=\Oa$ relative to the lowest-order term
(both in resonant and non-resonant regions).

Here we propose a generalization of the CMS to higher
orders.  The complex masses are introduced directly at the level of
the Lagrangian by splitting the bare masses into complex renormalized
masses and complex counterterms. This scheme has the following properties:
\begin{itemize}
\item From the Lagrangian we obtain Feynman rules with complex masses
  and counter\-terms with which we can perform perturbative calculations
  as usual.  Since we do not change the theory at all, but only
  rearrange its perturbative expansion, no double counting of terms
  occurs.
\item For each unstable particle mass, we add and subtract the same
  imaginary part in the Lagrangian.  One of these terms provides the
  imaginary part for the mass parameter and becomes part of the free
  propagator, while the other becomes part of a counterterm vertex.
  The first term is, thus, resummed but the second is not.
  Independently of the imaginary part that is added and subtracted,
  this procedure does not spoil the algebraic relations that govern
  gauge invariance, and unitarity cancellations are exactly respected.
  In practice, this means that we can insert values for the
  gauge-boson widths that are not directly related to the one-loop
  order to which the corrections for the process are calculated. We
  could even go beyond one loop in the calculation of the widths or
  take an empirical value.
\item Performing an $\Oa$ calculation in the CMS yields $\Oa$ accuracy
  everywhere in phase space provided the width that enters in the
  resonant propagators via the complex mass is calculated including at
  least $\Oa$ corrections. This is evident away from the resonances,
  where one could expand in terms of the width, thus recovering the
  usual perturbative expansion. In the resonance region, where the
  resonant contributions dominate, both the prefactors of the resonant
  propagators and the resonant propagators themselves are taken into
  account in $\Oa$ and our results differ by $\Oaa$ terms from a
  leading pole approximation where this is applicable. Thus, any
  spurious terms are of order $\Oaa$.
\end{itemize}

Introducing complex masses and couplings seems to violate unitarity.
Obviously, the Cutkosky cutting equations \cite{Cutkosky:1960sp} are
no longer valid, and unitarity cannot simply be proven order by order
anymore.  However, since we do not modify the bare Lagrangian, the
unitarity-violating terms are of higher order, \ie of $\Oaa$ in an
$\Oa$ calculation.  Moreover, this unitarity violation cannot be
enhanced, because all Ward identities are exactly preserved. In this
respect one should also mention that unstable particles should be
excluded as external states and only the $S$-matrix connecting stable
particle states needs to be unitary, as has already been pointed out
by Veltman in the sixties \cite{Veltman:1963th}.  Of course, before
the described CMS can be viewed as a rigorous procedure to define a
renormalized quantum field theory it has to be clarified whether one
can directly prove unitarity order by order in this formalism. In
particular, it is an interesting question whether one can construct
modified cutting equations in the CMS.

\paragraph{Complex renormalization -- 't Hooft--Feynman gauge}

The consistent introduction of complex masses in loop calculations
necessitates the formulation of an appropriate renormalization
prescription. To this end, we generalize the on-shell renormalization
scheme formulated in \citeres{Denner:1994xt,Denner:1993kt,Aoki:1980ix}
at the one-loop level in a straight-forward way.  A generalization to
higher orders should be possible, but this is beyond the scope of this
paper.

Following the conventions of \citere{Denner:1993kt}, the renormalized
transverse ($\rT$) gauge-boson self-energies read
\beqar\label{eq:ren-se}
\hat\Si^{W}_{\rT}(k^2) &=& \Si^{W}_{\rT}(k^2) - \de\cmws 
+(k^2-\cmws)\de \cZ_{W}, \nl
\hat\Si^{ZZ}_{\rT}(k^2) &=& \Si^{ZZ}_{\rT}(k^2) - \de\cmzs 
+(k^2-\cmzs)\de \cZ_{ZZ}, \nl
\hat\Si^{AA}_{\rT}(k^2) &=& \Si^{AA}_{\rT}(k^2)
+ k^2 \de \cZ_{AA}, \nl
\hat\Si^{AZ}_{\rT}(k^2) &=& \Si^{AZ}_{\rT}(k^2) 
+ k^2 \frac{1}{2}\de \cZ_{AZ} +(k^2-\cmzs) \frac{1}{2}\de \cZ_{ZA},
\eeqar
where $A$ denotes the photon field, and the hat indicates renormalized
self-energies. Compared to \citere{Denner:1993kt}, the renormalized
on-shell masses and mass counterterms are replaced by the renormalized
complex masses $\cmw$ and $\cmz$ everywhere, \ie also within the
self-energies. We denote the field renormalization constants in the
CMS by calligraphic letters.  The complex renormalized masses and mass
counterterms result from a splitting of the real bare masses squared,%
\footnote{Similar ideas were proposed in \citere{Stuart:1990vk}.}
\beq
M_{\PW,0}^2=\cmws+\de\cmws, \qquad M_{\PZ,0}^2=\cmzs+\de\cmzs,
\eeq
where here and in the following bare quantities are indicated by a
subscript 0.  Similarly, splitting the bare fields in complex field
renormalization constants and renormalized fields
\beq
\begin{array}[b]{cll}
W_{0}^{\pm}  & = & 
(1+\frac{1}{2}\delta \cZ_{W}) W^{\pm} , \\[1em]
\left(\barr{l} Z_{0} \\ A_{0} \earr \right)  & = &
\left(\barr{cc} 1 + \frac{1}{2}\delta \cZ_{ZZ} & \frac{1}{2}\delta \cZ_{ZA}
\\ [1ex]
               \frac{1}{2}\delta \cZ_{AZ}  & 1 + \frac{1}{2}\delta \cZ_{AA}
\end{array} \right)
\left(\barr{l} Z \\[1ex] A \earr \right)  ,
\earr
\eeq
implies that the bare and renormalized fields have different phases.
Thus, for instance, the renormalized Z-boson field becomes complex,
while the corresponding bare field is real. As a consequence, the
renormalized Lagrangian, \ie the Lagrangian in terms of renormalized
fields without counterterms, is not hermitian, but the total
Lagrangian (which is equal to the bare Lagrangian) of course is.

In order to fix the counterterms, we generalize the renormalization
conditions of the complete on-shell scheme
\cite{Denner:1993kt,Aoki:1980ix} and require
\beqar \label{eq:ren-cond-CMS}
\hat\Si^{W}_{\rT}(\cmws) &=& 0, \qquad 
\hat\Si^{ZZ}_{\rT}(\cmzs) = 0, \nl
\hat\Si^{AZ}_{\rT}(0) &=& 0, \qquad
\hat\Si^{AZ}_{\rT}(\cmzs) = 0, \nl
\hat\Si^{\prime W}_{\rT}(\cmws) &=& 0, \qquad
\hat\Si^{\prime ZZ}_{\rT}(\cmzs) = 0, \qquad
\hat\Si^{\prime AA}_{\rT}(0) = 0,
\eeqar
where the prime denotes differentiation with respect to the argument.
The conditions \refeq{eq:ren-cond-CMS}, in particular the first two,
fix the mass counterterms in such a way that the renormalized mass is
equal to the location of the propagator pole in the complex plane.
This is a gauge-invariant quantity, as pointed out and shown in
\citeres{Stuart:1991xk,Sirlin:1991fd}. The last five renormalization
conditions in \refeq{eq:ren-cond-CMS} fix the field renormalization
constants.  Note that the field renormalization constants of the
gauge-boson fields exactly drop out in all $S$-matrix elements that do
not involve external gauge bosons, but allow to render all vertex
functions finite.  This generally holds for all field renormalization
constants of unstable particles as long as one does not consider
$S$-matrix elements for external unstable particles. Unlike in
\citeres{Denner:1993kt,Aoki:1980ix}, we did not take real parts in the
renormalization conditions \refeq{eq:ren-cond-CMS}, and thus not only
the mass renormalization constants but also the field renormalization
constants become in general complex. This ansatz is supported by the
fact that the imaginary part of one-loop scattering amplitudes
involving unstable external particles becomes gauge dependent if the
imaginary parts of the counterterms are not included
\cite{Denner:1997kq}.  For the definition of the renormalized mass and
width this scheme is exactly the one described in Appendix~D of
\citere{Beenakker:1996kn}.  We stress the fact that the
renormalization constant $\de \cZ_{W}$ applies to both the $W^+$ and
$W^-$ field, \ie the imaginary part of $\de \cZ_{W}$ is fixed by the
renormalization condition and does not change sign when going from the
$W^+$ to the $W^-$ field.

The renormalization conditions \refeq{eq:ren-cond-CMS} have the
solutions
\beqar\label{exact-complex-ren-const}
\de\cmws &=& \Si^{W}_{\rT}(\cmws), \qquad 
\de\cmzs = \Si^{ZZ}_{\rT}(\cmzs), \nl
\de \cZ_{ZA} &=& \frac{2}{\cmzs}\Si^{AZ}_{\rT}(0), \qquad
\de \cZ_{AZ} = -\frac{2}{\cmzs}\Si^{AZ}_{\rT}(\cmzs), \nl
\de \cZ_{W} &=& - \Si^{\prime W}_{\rT}(\cmws), \qquad
\de \cZ_{ZZ} = -\Si^{\prime ZZ}_{\rT}(\cmzs), \qquad
\de \cZ_{AA} = -\Si^{\prime AA}_{\rT}(0),
\eeqar
which require to calculate the self-energies for complex squared
momenta. 

Owing to its definition \refeq {eq:compl-mixing}, the renormalization
of the complex weak mixing angle is determined by
\beq\label{eq:ren-mixing-angle}
\frac{\de\csw}{\csw} = -\frac{\ccw^2}{\csw^2}\frac{\de\ccw}{\ccw}
=-\frac{\ccw^2}{2\csw^2}
\left(\frac{\de\cmws}{\cmws}-\frac{\de\cmzs}{\cmzs}\right).
\eeq

The electric charge is fixed in the on-shell scheme by requiring that
there are no higher-order corrections to the $ee\ga$ vertex in the
Thomson limit.  In the CMS this condition reads
\beq\label{eq:ren-charge}
\frac{\de e}{e} = \frac{1}{2}\Si^{\prime AA}(0) -
  \frac{\sw}{\cw}\frac{\Si^{AZ}_{\rT}(0)}{\cmzs}.
\eeq
Because of the presence of the complex masses and couplings in the
loop integrals and explicitly in \refeq{eq:ren-charge},
the charge renormalization constant and thus the renormalized charge
become complex. Since the imaginary part of the bare charge vanishes,
the imaginary part of the charge renormalization constant is directly
fixed by the imaginary part of self-energies. In a one-loop
calculation, the imaginary part of the renormalized charge drops out
in the corrections to the absolute square of the matrix element,
because the charge factorizes from the lowest-order matrix element.
Starting from the two-loop level, the imaginary part has to be taken
into account.

For a correct description of the resonances at the $\Oa$ level, we
need the width including $\Oa$ corrections. The width of the W and Z
bosons is implicitly defined via the first two equations in
\refeq{exact-complex-ren-const}. Using
$\de\cmws=M_{\PW,0}^2-\cmws$ and taking the imaginary part results in 
\beq
\MW\GW =\Im\{\Si^{W}_{\rT}(\MW^2-\ri\MW\GW )\},
\eeq
which can be iteratively solved for $\GW$.
In $\Oaa$, \ie including first-order corrections to the width, the
result is equivalent to the one obtained in the usual on-shell scheme
(see below).

In our calculation we consider external fermions only in the massless
limit, in which these fermions are stable.  Therefore, the
corresponding on-shell self-energies do not involve any absorptive
parts. Nevertheless they become complex via the complex renormalized
weak mixing angle and the complex gauge-boson masses, and thus the
field renormalization constants $\de\cZ_{f,\si}$ of the fermion fields
$f^\si$, defined by
\beq
\barr{cll}
f_{0}^{\si} & = & 
(1+\frac{1}{2}\delta \cZ_{f,\si})  f^{\si} , \qquad \sigma=\mathrm{R,L},
\earr
\eeq
become complex. 
As in the case of $\de\cZ_{W}$, also these complex field
renormalization constants apply both for fermions and antifermions,
\ie fields and antifields are not connected by complex conjugation
anymore.  Explicitly the $\de\cZ_{f,\si}$ are given by
\beq
\de\cZ_{f,\sigma} = -\Sigma^{f,\sigma}(m_f^2)
-m_f^2\left[  \Sigma^{\prime f,\mathrm{R}}(m_f^2)
             +\Sigma^{\prime f,\mathrm{L}}(m_f^2)
            +2\Sigma^{\prime f,\mathrm{S}}(m_f^2) \right],
\eeq
where $\sigma=\mathrm{R,L}$ refers to the right- and left-handed
components of the fermion self-energy $\Sigma^{f}(p)$ following the
conventions of \citere{Denner:1993kt}. Note that again no real part
was taken in this relation in contrast to the usual on-shell
renormalization. In contrast to $\de\cZ_{W}$, there are soft IR
divergences in the field renormalization constants $\de\cZ_{f}$ that
are not regularized by finite widths but by the usual IR regulators
such as $m_\gamma$.

In the massless limit, the fermion-mass renormalization constant $\de
m_f$ tends to zero, and the quark-mixing matrix, if assumed to be
different from the unit matrix, need not be renormalized.

Since the top~quark and the Higgs~boson do not appear in the
lowest-order matrix elements for the processes under consideration,
the corresponding mass and field counterterms are not needed in our
calculation. Nevertheless, for completeness we define the
corresponding renormalization constants here.

For the Higgs boson the whole renormalization proceeds along the same lines
as for the gauge bosons above.
The complex Higgs mass  squared
\beq
\cmhs= \MH^2 - \ri\MH\GH = M_{\PH,0}^2 - \de\cmhs
\eeq
is defined as the location of the zero in $k^2$ in the renormalized
Higgs-boson self-energy
\beq
\hat\Si^{H}(k^2) = \Si^{H}(k^2) - \de\cmhs 
+(k^2-\cmhs)\de \cZ_{H}. 
\eeq
Fixing the Higgs field renormalization as above, we obtain the
renormalization constants 
\beqar \label{eq:complex-ren-const-Higgs-mass}
\de\cmhs &=& \Si^{H}(\cmhs), \qquad
\de \cZ_{H} = - \Si^{\prime H}(\cmhs).
\eeqar

For the top quark the renormalization procedure works analogously, \ie
we introduce the complex mass via
\beq
\cmt^2= \Mt^2 - \ri\Mt\Gt, 
\qquad
m_{\Pt,0}=\cmt+\de\cmt.
\eeq
The renormalized top-quark self-energy reads 
\beqar
\hat\Sigma^{t}(p) &=& 
  \left[\Sigma^{t,\mathrm{R}}(p^2)+\de \cZ_{t,\mathrm{R}}\right]\dsl{p}\omega_+
+ \left[\Sigma^{t,\mathrm{L}}(p^2)+\de \cZ_{t,\mathrm{L}}\right]\dsl{p}\omega_-
\nn\\
&& {}
+ \cmt\left[ \Sigma^{t,\mathrm{S}}(p^2) 
- \frac{1}{2}(\de \cZ_{t,\mathrm{R}} + \de \cZ_{t,\mathrm{L}})
-\frac{\de\cmt}{\cmt} \right]
\eeqar
in the conventions of \citere{Denner:1993kt}.
Generalizing the on-shell renormalization conditions to complex
renormalization as 
\beqar \label{eq:complex-ren-const-Top-mass}
\de\cmt &=& \frac{\cmt}{2}
\left[  \Sigma^{t,\mathrm{R}}(\cmt^2)
       +\Sigma^{t,\mathrm{L}}(\cmt^2)
      +2\Sigma^{t,\mathrm{S}}(\cmt^2) \right],
\nn\\
\de \cZ_{t,\sigma} &=& -\Sigma^{t,\sigma}(\cmt^2)
-\cmt^2\left[  \Sigma^{\prime t,\mathrm{R}}(\cmt^2)
             +\Sigma^{\prime t,\mathrm{L}}(\cmt^2)
            +2\Sigma^{\prime t,\mathrm{S}}(\cmt^2) \right], \qquad
\sigma=\mathrm{R,L},
\hspace{2em}
\eeqar
fixes $\cmt^2$ as the location of the complex pole in $p^2$ in the top
propagator.  

Finally, we complete the renormalization in the scalar sector.
The field renormalization for the would-be Goldstone bosons
can simply be set equal to the Higgs field renormalization constant
$\de \cZ_H$, which is sufficient to cancel all UV divergences in
vertex functions. The tadpole counterterm $\de t$ is again introduced
to cancel explicitly occurring tadpole graphs, \ie we set $\de t=-t$,
where $\Gamma^H=\ri t$ is the one-point vertex function for the
Higgs boson at one loop. The gauge-fixing term need not be renormalized.

In summary, in the CMS the usual renormalization conditions of the
on-shell scheme can be used, but without taking any real parts. All
parameters, in particular also the renormalization points, become
complex. 

\paragraph{A simplified version of the complex renormalization}

As shown above, renormalization in the complex-mass scheme requires
to calculate the self-energies for complex squared
momenta. This demands an analytic continuation of the 2-point
functions entering the self-energies in the momentum variable to the
unphysical Riemann sheet. This complication can be avoided by
expanding the self-energies appearing in the renormalization constants
about real arguments such that one-loop accuracy is retained.

We schematically illustrate the procedure for a scalar resonance $P$
with pole mass $M_P$ and pole width $\Gamma_P$, i.e.\
the location of the complex pole in the propagator is
$\mu_P^2=M_P^2-\ri M_P\Gamma_P$.
The one-loop self-energy correction in the complex-mass
scheme is proportional to 
\beq\label{eq:selffactor}
f(k^2)=\frac{\Sigma(k^2)-\de\mu_P^2}{k^2-\mu_P^2}
+\de\cZ_P
=\frac{\Sigma(k^2)-\Sigma(\mu_P^2)}{k^2-\mu_P^2}
-\Sigma'(\mu_P^2),
\eeq
where we used the on-shell counterterms
\begin{eqnarray}
\de\mu^2_P  &=& \Si(\cmps) ,\qquad
\de \cZ_{P} = - \Si^{\prime}(\cmps).
\end{eqnarray}
Note that the pole at $k^2=\mu_P^2$ cancels exactly in
\refeq{eq:selffactor}, and $f(k^2)$ is well-behaved in the vicinity of
the resonance ($k^2\approx M_P^2$), where $k^2-\mu_P^2\approx \ri
M_P\Gamma_P$ is of one-loop order, as the width $\Gamma_P$.  This is
crucial for one-loop precision, and when approximating $f(k^2)$ we
have to make sure that the approximation is one-loop exact in the
vicinity of the resonance.

If the self-energy can be expanded as
\begin{eqnarray}\label{eq:expansion}
\Sigma(\mu_P^2)&=&\Sigma(M_P^2)+(\mu_P^2-M_P^2)
\Sigma'(M_P^2)\;+\; {\cal O}\left((\mu_P^2-M_P^2)^2\right)\nl
&=&\Sigma(M_P^2)-\ri M_P\Gamma_P
\Sigma'(M_P^2)\;+\; {\cal O}\left((M_P\Gamma_P)^2\right),
\end{eqnarray}
we can approximate the mass and wave-function renormalization
counterterms as
\begin{eqnarray}
\de\mu^2_P  &=& 
\Si(M_P^2) +  (\cmps-M_P^2) \Si^{\prime}(M_P^2) + \Oaaa,
\nn\\
\de \cZ_{P} &=& - \Si^{\prime}(M_P^2) + \Oaa,
\end{eqnarray}
and the resulting approximation
\beqar\label{eq:selffactorappr}
f(k^2)&=&
\frac{\Sigma(k^2)-\Sigma(M_P^2)-(\cmps-M_P^2) \Si^{\prime}(M_P^2)}{k^2-\mu_P^2}
-\Sigma'(M_P^2) + \Oaa\nl
&=&
\frac{\Sigma(k^2)-\Sigma(M_P^2)-(k^2-M_P^2) \Si^{\prime}(M_P^2)}{k^2-\mu_P^2}
 + \Oaa
\eeqar
is correct at one-loop accuracy.  The $\Oaa$ and $\Oaaa$
contributions in \refeqs{eq:expansion} and \refeqf{eq:selffactorappr}
result from products of terms $\Si=\Oa$ and $(\mu_P^2-M_P^2)=\Oa$ and
are UV finite by construction at the one-loop level.

While the expansion \refeq{eq:expansion} holds true for neutral and
colourless fields, it breaks down for charged or coloured fields in
the presence of photon or gluon exchange due to the contributions with
a branch cut at $k^2=\mu_P^2$.
We explicitly see this upon considering
\beq
\Sigma(k^2) = a\left(\frac{\mu_P^2}{k^2}-1\right)\ln\left(1-\frac{k^2}{\mu_P^2}\right) 
\;+\; \mbox{regular terms near $k^2\sim M_P^2$}
\eeq
with a constant $a$, which is the typical functional form
for a self-energy diagram with $P$ emitting and reabsorbing a
photon or gluon.

In this case, the difference between the exact self-energy and the
expansion is given by
\beq\label{eq:ccexpansion}
[\Sigma(M_P^2)-\ri M_P\Gamma_P \Sigma'(M_P^2)]
-\Sigma(\mu_P^2)
= \ri a \frac{\Gamma_P}{M_P}
\;+\; a\,{\cal O}(\Gamma_P^2 \ln\Gamma_P)
\eeq
in the limit $\Gamma_P\ll M_P$, i.e.\ it is of two-loop order and thus
of one order lower than required. Substituting \refeq{eq:ccexpansion}
into \refeq{eq:selffactorappr} shows that $f(k^2)$ does not have
one-loop accuracy anymore.  However, the failure of the expansion can
be easily corrected by adding the missing term back to the expanded
counterterm:
\begin{eqnarray}\label{eq:expansioncorr}
\Sigma(\mu_P^2)&=&\Sigma(M_P^2)+(\mu_P^2-M_P^2)
\left[\Sigma'(M_P^2)+ \frac{a}{M_P^2}\right]\;+\; {\cal O}\left((\mu_P^2-M_P^2)^2\right)\nn\\
&=&\Sigma(M_P^2)-\ri M_P\Gamma_P
\Sigma'(M_P^2) -\ri a\frac{\Gamma_P}{M_P}
\;+\; {\cal O}
\left((M_P\Gamma_P)^2\right).
\end{eqnarray}
In this way the counterterms can be consistently expressed in terms of
self-energies at real momentum arguments.

The described procedure can be applied to the complex-mass scheme in
the Standard Model as follows. The gauge-boson self-energies 
at the complex pole positions can be approximated as
\beqar
\Si^{W}_{\rT}(\cmws) &=& \Si^{W}_{\rT}(\MW^2) +  (\cmws-\MW^2)
\Si^{\prime W}_{\rT}(\MW^2)  + c^{W}_{\rT} + \Oaaa,\nl
\Si^{ZZ}_{\rT}(\cmzs) &=& \Si^{ZZ}_{\rT}(\MZ^2) +  (\cmzs-\MZ^2)
\Si^{\prime ZZ}_{\rT}(\MZ^2) + \Oaaa,\nl
\frac{1}{\cmzs}\Si^{AZ}_{\rT}(\cmzs) &=& 
\frac{1}{\cmzs}\Si^{AZ}_{\rT}(0)
+\frac{1}{\MZ^2}\Si^{AZ}_{\rT}(\MZ^2)  
-\frac{1}{\MZ^2}\Si^{AZ}_{\rT}(0) + \Oaa,
\eeqar
as done similarly in Appendix~D of \citere{Beenakker:1996kn}.
The constant
\beq
c^{W}_{\rT} \;=\; \frac{\ri\alpha}{\pi} \MW\GW
 \;=\; \frac{\alpha}{\pi}(\MW^2-\cmws).
\eeq
compensates for the failure of the expansion of the photon-exchange
diagram in the W-boson self-energy as described above

By neglecting the  $\Oaa$  and $\Oaaa$ terms, we can 
replace \refeq{exact-complex-ren-const} by
\beqar\label{complex-ren-const-mass}
\de\cmws &=& \Si^{W}_{\rT}(\MW^2)+  (\cmws-\MW^2)
\Si^{\prime W}_{\rT}(\MW^2) + c^{W}_{\rT} , \nl 
\de\cmzs &=& \Si^{ZZ}_{\rT}(\MZ^2)+  (\cmzs-\MZ^2)
\Si^{\prime ZZ}_{\rT}(\MZ^2), 
\eeqar
and 
\beqar\label{complex-ren-const-field}
\de \cZ_{ZA} &=& \frac{2}{\cmzs}\Si^{AZ}_{\rT}(0), \qquad
\de \cZ_{AZ} = -\frac{2}{\MZ^2}\Si^{AZ}_{\rT}(\MZ^2)
+\left(\frac{\cmzs}{\MZ^2}-1\right) \de \cZ_{ZA}
, \nl
\de \cZ_{W} &=& - \Si^{\prime W}_{\rT}(\MW^2), \qquad
\de \cZ_{ZZ} = -\Si^{\prime ZZ}_{\rT}(\MZ^2).
\eeqar
While the missing $\Oaa$
terms in $\de \cZ_{AZ}$ do not influence our results, since the
gauge-boson field renormalization constants drop out as there is no
external gauge boson in the process under consideration, the missing
(finite) $\Oaaa$ terms in the mass counterterms are beyond the
accuracy of a one-loop calculation. The counterterms
\refeq{complex-ren-const-mass} involve only functions that appear also
in the usual on-shell renormalization scheme
\cite{Denner:1993kt,Aoki:1980ix}, but consistently take into account
the imaginary parts.

When inserting the counterterms \refeq{complex-ren-const-mass} and
\refeq{complex-ren-const-field} into
\refeq{eq:ren-se}, we can
rewrite the renormalized self-energies in the CMS as
\beqar\label{eq:ren-se-cms}
\hat\Si^{W}_{\rT}(k^2) &=& \Si^{W}_{\rT}(k^2) - \de\MW^2
+(k^2-\MW^2)\de Z_{W} - c^{W}_{\rT}, \nl
\hat\Si^{ZZ}_{\rT}(k^2) &=& \Si^{ZZ}_{\rT}(k^2) - \de\MZ^2 
+(k^2-\MZ^2)\de Z_{ZZ}, \nl
\hat\Si^{AA}_{\rT}(k^2) &=& \Si^{AA}_{\rT}(k^2)
+ k^2 \de Z_{AA}, \nl
\hat\Si^{AZ}_{\rT}(k^2) &=& \Si^{AZ}_{\rT}(k^2) 
+ k^2 \frac{1}{2}\de Z_{AZ} +(k^2-\MZ^2) \frac{1}{2}\de Z_{ZA}
\eeqar
with 
\beqar\label{complex-ren-const-ons}
\de\MW^2 &=& \Si^{W}_{\rT}(\MW^2), \qquad 
\de\MZ^2 = \Si^{ZZ}_{\rT}(\MZ^2), \nl
\de Z_{ZA} &=& \frac{2}{\MZ^2}\Si^{AZ}_{\rT}(0), \qquad
\de Z_{AZ} = -\frac{2}{\MZ^2}\Si^{AZ}_{\rT}(\MZ^2), \nl
\de Z_{W} &=& - \Si^{\prime W}_{\rT}(\MW^2), \qquad
\de Z_{ZZ} = -\Si^{\prime ZZ}_{\rT}(\MZ^2), \qquad
\de Z_{AA} = -\Si^{\prime AA}_{\rT}(0).
\eeqar
Apart from the terms $ c^{W}_{\rT}$, equations \refeq{eq:ren-se-cms} with \refeq{complex-ren-const-ons}
have exactly the form of the renormalized self-energies in the usual
on-shell scheme, but without taking the real part of the counterterms.
While in the on-shell scheme the self-energies are calculated
in terms of the real renormalized masses $\MZ^2$ and $\MW^2$, in
\refeq{eq:ren-se-cms} and \refeq{complex-ren-const-ons} the
self-energies are to be calculated in terms of the complex internal masses
$\cmzs$ and $\cmws$, although with real squared momenta.  Note that
this difference between usual on-shell and complex renormalization
also changes the form of the IR divergence appearing in the W-field
renormalization constant $\de Z_{W}$.  In the former scheme, it
appears as logarithm $\ln m_\gamma$ of an infinitesimally small photon
mass (or as the related $1/(4-D)$ pole in dimensional regularization);
in the latter, the W~width regularizes the singularity via $\ln\GW$.

The renormalization of  the complex weak mixing angle is given by
\refeq{eq:ren-mixing-angle} with mass counterterms from
\refeq{complex-ren-const-mass}. The renormalization of the 
electric charge stays the same as in 
\refeq{eq:ren-charge}.

The width is now defined via
\refeq{complex-ren-const-mass}. Using $\de\cmws=M_{\PW,0}^2-\cmws$ and
taking the imaginary part of \refeq{complex-ren-const-mass} yields
\beq
\MW\GW =\Im\{\Si^{W}_{\rT}(\MW^2)\} 
- \MW\GW \Re\{\Si^{\prime W}_{\rT}(\MW^2) \} + \Oaaa,
\eeq
which can be iteratively solved for $\GW$. Note that the self-energies
depend on $\GW$ via the complex W-boson mass. In $\Oaa$, \ie including
first-order corrections to the width, the result is equivalent to the
one obtained in the usual on-shell scheme. To this order the imaginary
part of the self-energy is required in two-loop accuracy, but it can
be more easily obtained by calculating the one-loop corrections to the
decay processes $\PW\to \bar f f'$ \cite{Beenakker:1996kn}. In our
numerical calculation, we calculate the width from the decay processes
including $\Oa$ corrections \cite{Bardin:1986fi}.

For the (neutral and colourless) Higgs boson, the approximate
renormalization works as for the generic scalar $P$ discussed above.
The renormalization constants can be approximated as
\beqar \label{eq:exact-complex-ren-const-Higgs}
\de\cmhs &=& 
\Si^{H}(\MH^2) +  (\cmhs-\MH^2) \Si^{\prime H}(\MH^2) + \Oaaa,
\nn\\
\de \cZ_{H} &=&  - \Si^{\prime H}(\MH^2) + \Oaa,
\eeqar
so that the renormalized Higgs-boson self-energy up to finite $\Oaa$ terms
can be written as
\beq
\hat\Si^{H}(k^2) = \Si^{H}(k^2) - \de\MH^2 +(k^2-\MH^2) \de Z_{H}.
\eeq
with
\beqar
\de\MH^2 &=& \Si^{H}(\MH^2),
\qquad
\de Z_{H} = - \Si^{\prime H}(\MH^2).
\eeqar
Note, however, that 
for large Higgs-boson masses ($\MH\gsim400\GeV$) the Higgs-boson width grows
drastically, so that the expansion of the mass counterterm
for $\GH/\MH\to0$ will not be justified anymore.

Expanding the self-energies appearing in the renormalization constants
\refeq{eq:complex-ren-const-Top-mass} for the top quark about $\Mt^2$
and neglecting (UV-finite) terms of order $\Oaaa$ in the mass
counterterm and of order $\Oaa$ in the field renormalization constant,
the renormalized top-quark self-energy can be expressed as
\beqar
\hat\Sigma^{t}(p) &=& 
  \left\{\Sigma^{t,\mathrm{R}}(p^2)+\de Z_{t,\mathrm{R}}\right\}\dsl{p}\omega_+
+ \left\{\Sigma^{t,\mathrm{L}}(p^2)+\de Z_{t,\mathrm{L}}\right\}\dsl{p}\omega_-
\nn\\
&& {}
+ \cmt\biggl\{ \Sigma^{t,\mathrm{S}}(p^2) 
- \frac{1}{2}(\de Z_{t,\mathrm{R}} + \de Z_{t,\mathrm{L}})
-\frac{\de\Mt}{\Mt} 
\nn\\
&& \qquad\quad {}
+\ri\Mt\Gt \left[  \frac{1}{2}\Sigma^{\prime t,\mathrm{R}}(\Mt^2)
             +\frac{1}{2}\Sigma^{\prime t,\mathrm{L}}(\Mt^2)
            +\Sigma^{\prime t,\mathrm{S}}(\Mt^2) \right]
-c^t \biggr\}
\eeqar
with
\beqar
\de\Mt &=& \frac{\Mt}{2}
\left[  \Sigma^{t,\mathrm{R}}(\Mt^2)
       +\Sigma^{t,\mathrm{L}}(\Mt^2)
      +2\Sigma^{t,\mathrm{S}}(\Mt^2) \right],
\nn\\
\de Z_{t,\sigma} &=& -\Sigma^{t,\sigma}(\Mt^2)
-\Mt^2\left[  \Sigma^{\prime t,\mathrm{R}}(\Mt^2)
             +\Sigma^{\prime t,\mathrm{L}}(\Mt^2)
            +2\Sigma^{\prime t,\mathrm{S}}(\Mt^2) \right], \qquad
\sigma=\mathrm{R,L}.
\hspace{2em}
\eeqar
Note that for QCD corrections the neglected terms are of order
$\ord{(\alphas^3)}$ or $\ord{(\alphas^2)}$, respectively.
The constant $c^t$ again originates from the
non-analytic terms from photon and gluon exchange, as explained above.
Taking electroweak and QCD corrections into account, it reads
$$
c^t = \frac{\ri\alpha Q_\Pt^2}{\pi} \frac{\Gt}{\Mt}
+ \frac{\ri\alpha_{\mathrm{s}}C_{\mathrm{F}}}{\pi} \frac{\Gt}{\Mt}
$$
with the top-quark charge $Q_\Pt=2/3$ and $C_{\mathrm{F}}=4/3$.

In summary, the calculation of self-energies with complex momentum
arguments can be avoided by carefully expanding these about real
values.  In case of charged or coloured particles, extra constants
must be added to the expanded self-energies in order not to spoil the
one-loop accuracy of the results. All the mass arguments of the
self-energies are complex and no real parts must be taken.

\paragraph{Complex renormalization -- background-field gauge}

Using the background-field method, the gauge-boson field
renormalization constants can be determined in terms of the parameter
renormalization in such a way that Ward identities possess the same form
before and after renormalization \cite{Denner:1994xt}. 
Real parameters have to be substituted by the corresponding complex
parameters everywhere when the complex renormalization is employed.
The complex parameter renormalization is fixed as above in
\refeq{complex-ren-const-mass}, \refeq{eq:ren-mixing-angle},
\refeq{eq:ren-charge}, \refeq{eq:complex-ren-const-Higgs-mass}, and
\refeq{eq:complex-ren-const-Top-mass}.  Note that
$\Si^{AZ}_{\rT}(0)$ vanishes in the background-field gauge as a
consequence of the background-field gauge invariance of the effective
action, which in particular simplifies the charge renormalization
constant \refeq{eq:ren-charge}.

Since the gauge-boson field renormalization constants drop out in the
$S$-matrix elements without external gauge-boson fields, we can
alternatively also use the definitions in
\refeq{complex-ren-const-field} in the calculation.

\paragraph{Loop integrals with complex masses}

The consistent use of complex gauge-boson masses requires to use these
also in the loop integrals.  Thus, we need one-loop integrals with
complex internal masses.  The IR-singular integrals can be found in
\citere{Beenakker:1990jr}.  Concerning the non-IR singular cases, we
have analytically continued the results of \citere{'tHooft:1979xw} for
the 2-point and 3-point functions\footnote{Note that the result of
  \citere{'tHooft:1979xw} for the scalar two-point function is not
  valid in general for complex masses. In this case an extra $\eta$
  function has to be added. The same comment applies to the results
  for the 2-point tensor integrals in
  \citere{Passarino:1979jh}.}%
, and the relevant results of
\citere{Denner:1991qq} for the 4-point functions. We have checked all
these results by independent direct calculation of the
Feynman-parameter integrals. These results will be published
elsewhere.%
\footnote{Meanwhile general results for scalar 4-point functions with
  complex mass parameters have been published in \citere{Denner:2010tr}.}

\section{Numerical results}
\label{se:numres}

\subsection{Input parameters and setup}
\label{se:setup}

The numerical results are based on the same set of input parameters as
in \citere{Denner:2005es}:
\beq\arraycolsep 2pt
\begin{array}[b]{lcllcllcl}
\GF & = & 1.16637 \times 10^{-5} \GeV^{-2}, \quad&
\alpha(0) &=& 1/137.03599911, &
\alpha_{\mathrm{s}} &=& 0.1187,\\
\MW & = & 80.425\GeV, &
\MZ & = & 91.1876\GeV, &
\GZ & = & 2.4952\GeV, \\
\MH & = & 115\GeV, \\
\Me & = & 0.51099892\MeV, &
m_\mu &=& 105.658369\MeV,\quad &
m_\tau &=& 1.77699\GeV, \\
\Mu & = & 66\MeV, &
\Mc & = & 1.2\GeV, &
\Mt & = & 178\;\GeV, 
\\
\Md & = & 66\MeV, &
\Ms & = & 150\MeV, &
\Mb & = & 4.3\GeV, 
\end{array}
\label{eq:SMpar}
\eeq
which essentially follows \citere{Eidelman:2004wy}.  For the top-quark
mass $\Mt$ we have taken the more recent value of
\citere{Azzi:2004rc}.  The masses of the light quarks are adjusted to
reproduce the hadronic contribution to the photonic vacuum
polarization of \citere{Jegerlehner:2001ca}. Since we parametrize the
lowest-order cross section with the Fermi constant $\GF$ ($\GF$
scheme), \ie we derive the electromagnetic coupling $\alpha$ according
to $ \alpha_{\GF} = \sqrt{2}\GF\MW^2(1-\MW^2/\MZ^2)/\pi$, the results are
practically independent of the masses of the light quarks.  Moreover,
this procedure absorbs the corrections proportional to $\Mt^2/\MW^2$
in the fermion--W-boson couplings and the running of $\al(Q^2)$ from
$Q^2=0$ to the electroweak scale. In the relative radiative
corrections, we use, however, $\alpha(0)$ as coupling parameter, which
is the correct effective coupling for real photon emission.

QCD corrections are treated in the ``naive'' approach of multiplying cross
sections and partial decay rates by factors
$(1+\alpha_{\mathrm{s}}/\pi)$ per hadronically decaying W~boson.  The
W-boson width $\GW$ is calculated from the above input including
electroweak ${\cal O}(\alpha)$ and QCD corrections, yielding
\beq
\GW = 2.09269848\ldots\GeV.
\eeq
This procedure ensures that the effective branching ratios for the
leptonic, semileptonic, and hadronic W~decays, which result from the
integration over the decay fermions, add up to 1.
The value for the Z~decay width $\GZ$, which is needed because of 
the Z~resonance in the ISR convolution below the W-pair threshold,
is taken from experiment \cite{Eidelman:2004wy}. All other particles,
including the top quark and the Higgs boson, are taken as stable.

The setup differs from the one of \citere{Denner:2005es} only in the
event selection. In contrast to \citere{Denner:2005es}, where no
phase-space cuts were applied at all, we now impose selection cuts
in conjunction with a photon recombination procedure. In detail,
we adopt the same procedure as in \citere{Denner:2000bj}:
\begin{enumerate}
\item All bremsstrahlung photons within a cone of 5 degrees around the
  beams are treated as invisible, \ie their momenta are disregarded
  when calculating angles, energies, and invariant masses.
\item Next, the invariant masses $M_{f\gamma}$ of the photon with each
  of the charged final-state fermions are calculated. If the smallest
  $M_{f\gamma}$ is smaller than $M_\recomb= 25\GeV$ or if the energy
  of the photon is smaller than $1\GeV$, the photon is combined with
  the charged final-state fermion that leads to the smallest
  $M_{f\gamma}$, \ie the momenta of the photon and the fermion are
  added and associated with the momentum of the fermion, and the
  photon is discarded%
\footnote{Except for the $1\GeV$ cut, the described cut and
  recombination procedure coincides with the one used in the first two
  papers of \citere{Denner:2000kn}.}.  
\item Finally, all events are discarded in which one of the charged
  final-state fermions is within a cone of 10 degrees around the
  beams (after a possible recombination with a photon).  
  No other cuts are applied.
\end{enumerate}

The presented results have been obtained with $10^8$ events, using the
subtraction method.  All but the lowest-order
predictions include naive QCD corrections and
improvements by ISR beyond ${\cal O}(\alpha)$, as described in
\citere{Denner:2000bj}.

\subsection{Results for differential cross sections}

In this section we consider results for various distributions at
$\sqrt{s}=200\GeV$ and $\sqrt{s}=500\GeV$ in the setup described
above.  For reference, in \refta{ta:cstot} we provide the
corresponding integrated cross sections for the final states
$\nu_\tau\tau^+\mu^-\bar\nu_\mu$, $\Pu\bar\Pd\mu^-\bar\nu_\mu$, and
$\Pu\bar\Pd\Ps\bar\Pc$ in various approximations for different CM
energies $\sqrt{s}$.  The numbers in parentheses represent the
uncertainties from Monte Carlo integration in the last digits of the
predictions. 
\begin{table}
\arraycolsep 10pt
\centerline{$\begin{array}{rcccc}
\phantom{\sqrt{s}/\mathrm{GeV}}
\\
\hline
\sqrt{s}/\mathrm{GeV} & \born(\FW) & \born(\CMS) &
 \DPA & \eefourf 
\\
\hline
\rlap{$\Pep\Pem\to\nu_\tau\tau^+\mu^-\bar\nu_\mu$}\hfill\\
\hline
200  & 211.52(3)  & 211.40(3)  & 191.98(3) & 192.18(3)
\\[-.3em]
     &              & [-0.06\%] & [-9.24(1)\%]& [-9.09(1)\%]
\\
500  & ~62.17(1)  & ~62.14(1)   &   ~65.48(2) &  ~65.24(2)
\\[-.3em]
     &              & [-0.05\%] &  [+5.32(1)\%]& [+4.99(1)\%]
\\
\hline
\rlap{$\Pep\Pem\to\Pu\bar\Pd\mu^-\bar\nu_\mu$}\hfill\\
\hline
200  & 628.72(9)  & 628.37(9)  &  591.55(9) & 592.20(10)
\\[-.3em]
     &              & [-0.06\%] &  [-5.91(1)\%]& [-5.76(1)\%]
\\
500  &  180.83(4)  & 180.73(4)   &   197.87(5) &  197.06(5)
\\[-.3em]
     &              & [-0.06\%] & [+9.42(1)\%]& [+9.03(2)\%]
\\
\hline
\rlap{$\Pep\Pem\to\Pu\bar\Pd\Ps\bar\Pc$}\hfill\\
\hline
200  & 1868.9(3)  & 1867.8(3)  &  1822.6(3) & 1824.9(3)
\\[-.3em]
     &              & [-0.06\%] &  [-2.48(1)\%]& [-2.30(1)\%]
\\
500  &  ~526.1(1)  & ~525.8(1)   &  ~597.1(2) &  ~594.5(2)
\\[-.3em]
     &              & [-0.06\%] & [+13.50(2)\%]~& [+13.07(2)\%]~
\end{array}$}
\caption{Integrated cross sections in fb for 
$\Pep\Pem\to\nu_\tau\tau^+\mu^-\bar\nu_\mu$,
$\Pu\bar\Pd\mu^-\bar\nu_\mu$, and $\Pu\bar\Pd\Ps\bar\Pc$
in Born approximation (in the fixed-width and complex-mass schemes),
in DPA, and using the full ${\cal O}(\alpha)$ correction (\eefourf); all
but the Born cross sections include higher-order ISR 
and (if relevant) naive QCD corrections.}
\label{ta:cstot}
\end{table}%
Columns two and three in each table contain the two versions of the
lowest-order cross section for the full $\Pep\Pem\to4f$ processes
corresponding to the different treatments of finite-width effects as
provided by the ``fixed-width scheme'' (FW) and the complex-mass
scheme (CMS). In the FW scheme the finite constant width, and thus the
complex mass, is only inserted into the propagators.  The relative
difference $\si_{\born}(\CMS)/\si_{\born}(\FW)-1$ of the schemes in
lowest order is given by the numbers in square brackets in the third
columns.  We have not given an error on this difference, because the
two Born predictions are strongly correlated.  The last-but-one
columns show the DPA of {\sc RacoonWW} which also includes some
effects beyond DPA, as described in \citere{Denner:2000bj}; the
numbers in square brackets are defined as $\de_{\DPA} =
\si_{\DPA}/\si_{\born}(\FW)-1$.  We normalize $\si_{\DPA}$ to
$\si_{\born}(\FW)$, because the lowest-order part of the DPA is per
default evaluated in the FW scheme in {\sc RacoonWW}.  Finally, the
last columns (\eefourf) contain the full one-loop corrections to
$\Pep\Pem\to4f$; the numbers in square brackets are defined as
$\de_{\eefourf} = \si_{\eefourf}/\si_{\born}(\CMS)-1$.  Here we
normalize to $\si_{\born}(\CMS)$, because the full $\Pep\Pem\to4f$
calculation is consistently performed in the CMS. For
$\sqrt{s}=500\GeV$ the difference between the predictions based on the
full $\Oa$ corrections and on the DPA slightly increases by $\sim
0.1\%$ with respect to the results without cuts
\cite{Denner:2005es}. This tendency can be attributed to the fact that
at high energies the cross section is dominated by forward-scattered
nearly on-shell W-bosons. Such events are discarded by the cuts thus
reducing the contribution of on-shell W-boson pairs and worsening the
quality of the DPA. 

For the differential distributions, we focus on the semileptonic
process $\Pep\Pem\to\Pu\bar\Pd\mu^-\bar\nu_\mu$ in the following. The
respective results for the final states
$\nu_\tau\tau^+\mu^-\bar\nu_\mu$ and $\Pu\bar\Pd\Ps\bar\Pc$ look
similar, up to an offset resulting from the QCD corrections; in
particular, the difference between the DPA and the full ${\cal
  O}(\alpha)$ calculation are almost identical.  We always display the
lowest-order prediction (Born) and the result of the full one-loop
calculation (ee4f) in the upper row of each figure. The relative
corrections (in per cent) in the DPA approach (DPA), $\delta_{\DPA} =
\rd\sigma_{\DPA} / \rd \sigma_{\Born}(\FW)-1$, and in the full
one-loop calculation (ee4f), $\delta_{\eefourf} = \rd\sigma_{\eefourf} /
\rd \sigma_{\Born}(\CMS)-1$, are shown in the lower rows of the
figures. These additionally include an inset depicting the relative
difference between the full one-loop and the DPA calculation with
respect to the DPA calculation, $\Delta = \rd\sigma_{\mathrm{ee4f}} /
\rd \sigma_{\mathrm{DPA}}-1$.  We define all angles in the laboratory
system, which is the CM system of the initial state.  The momenta of
the $\PWp$ and $\PWm$ bosons are defined as
\beq
k_+=k_1+k_2, \qquad k_-=k_3+k_4,
\eeq
respectively, after a possible photon recombination. From these momenta the
invariant masses of the virtual W~bosons and their angles are
calculated.

\begin{figure}
{\unitlength 1cm
\begin{picture}(16,16)
\put(-4.9,- 9.7){\includegraphics{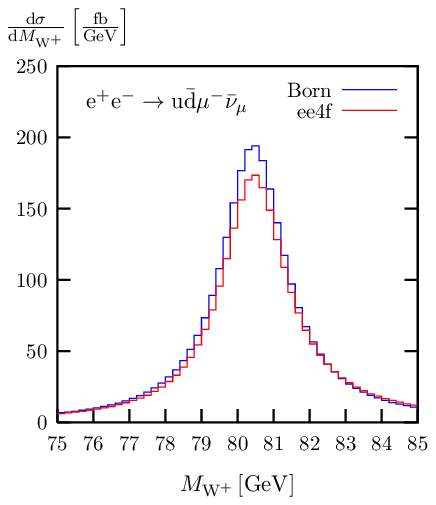}}
\put( 3.3,- 9.7){\includegraphics{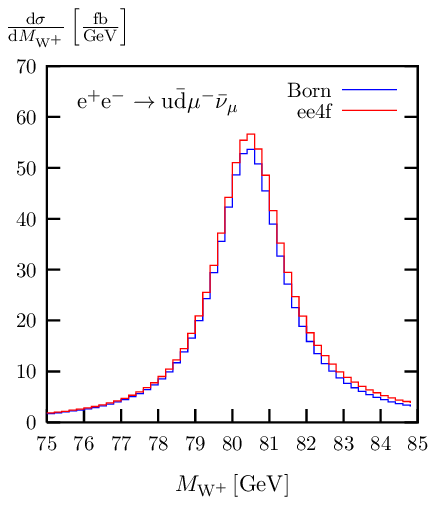}}
\put(-4.9,-17.7){\includegraphics{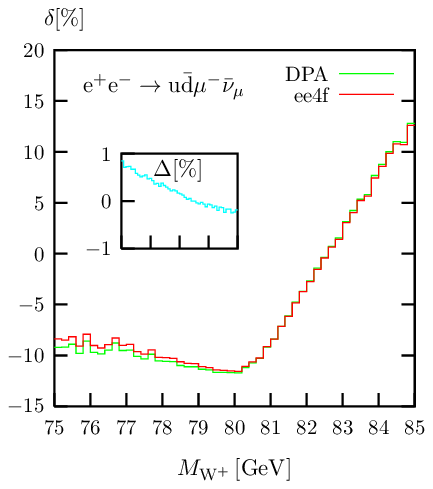}}
\put( 3.3,-17.7){\includegraphics{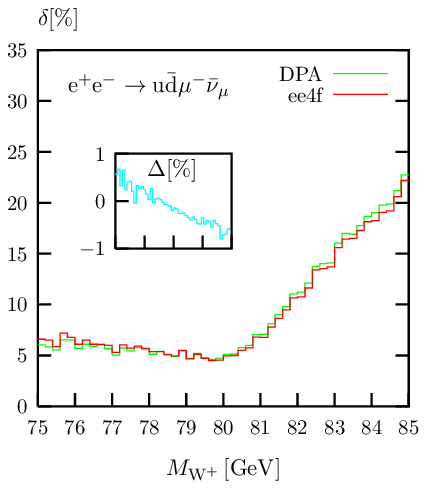}}
\put(3,15.5){$\sqrt{s}=200\GeV$}
\put(11.2,15.5){$\sqrt{s}=500\GeV$}
\end{picture}}%
\caption{Distribution in the invariant mass of the $\protect\PWp$
  boson (upper row) and the corresponding corrections (lower row) at
  $\sqrt{s}=200\GeV$ (l.h.s.) and  $\sqrt{s}=500\GeV$ (r.h.s.)  for
  $\Pep\Pem\to \Pu \Pdbar \mu^- \bar\nu_{\mu}$. The inset plot shows
  the difference between the full $\Oa$ corrections and those in DPA.}
\label{fi:mwp}
\end{figure}

\begin{figure}
{\unitlength 1cm
\begin{picture}(16,16)
\put(-4.9,- 9.7){\includegraphics{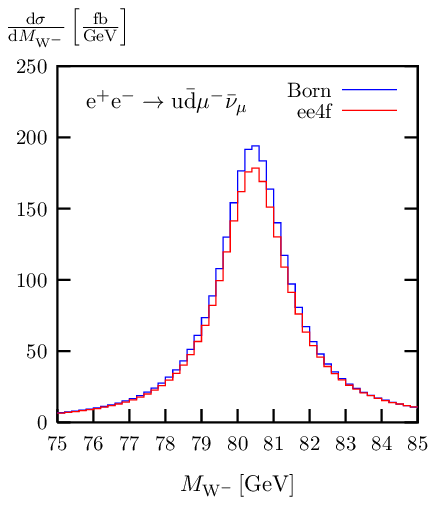}}
\put( 3.3,- 9.7){\includegraphics{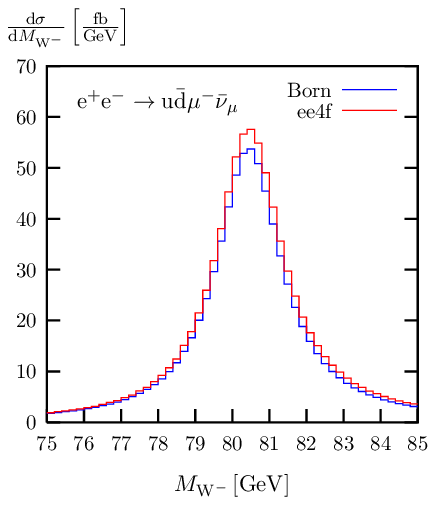}}
\put(-4.9,-17.7){\includegraphics{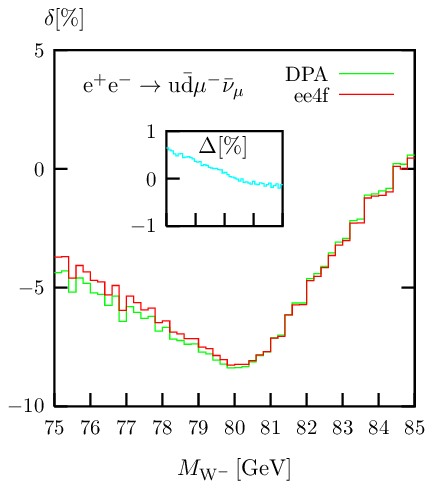}}
\put( 3.3,-17.7){\includegraphics{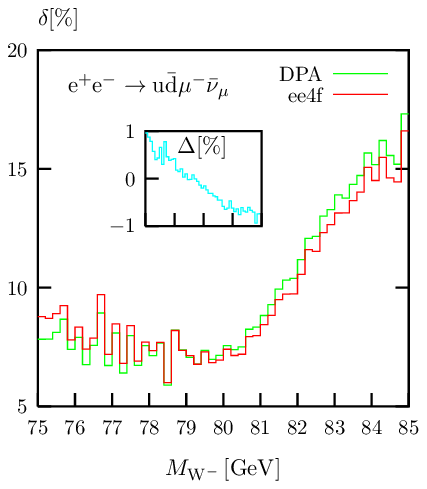}}
\put(3,15.5){$\sqrt{s}=200\GeV$}
\put(11.2,15.5){$\sqrt{s}=500\GeV$}
\end{picture}}%
\caption{Distribution in the invariant mass of the $\protect\PWm$
  boson (upper row) and the corresponding corrections (lower row) at
  $\sqrt{s}=200\GeV$ (l.h.s.) and 
  $\sqrt{s}=500\GeV$ (r.h.s.)  for
  $\Pep\Pem\to \Pu \Pdbar \mu^- \bar\nu_{\mu}$. The inset plot shows
  the difference between the full $\Oa$ corrections and those in DPA.}
\label{fi:mwm}
\end{figure}

\begin{figure}
{\unitlength 1cm
\begin{picture}(16,16)
\put(-4.9,- 9.7){\includegraphics{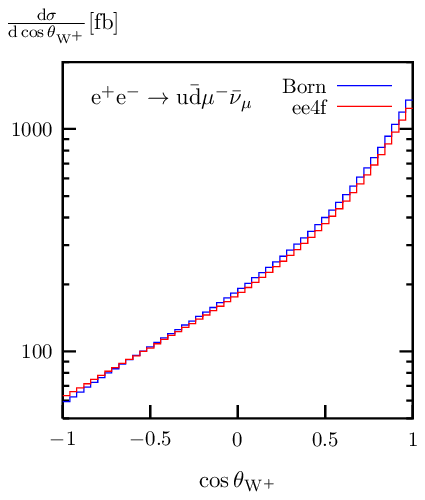}}
\put( 3.3,- 9.7){\includegraphics{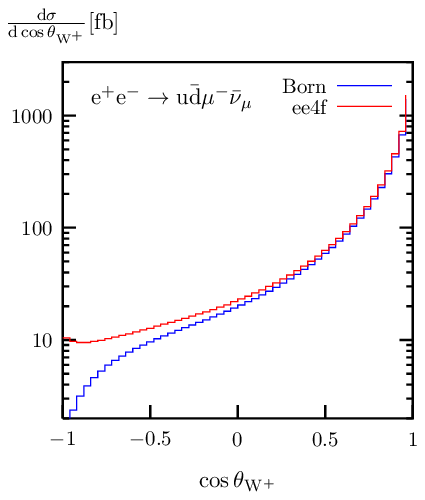}}
\put(-4.9,-17.7){\includegraphics{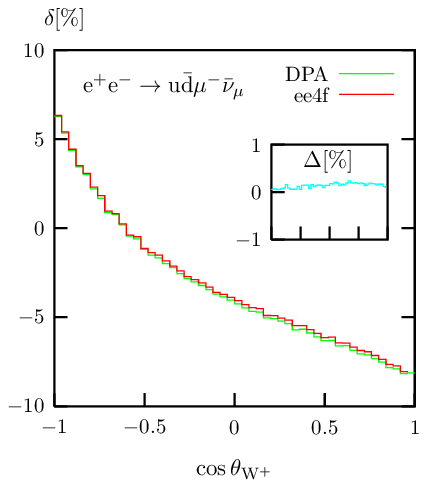}}
\put( 3.3,-17.7){\includegraphics{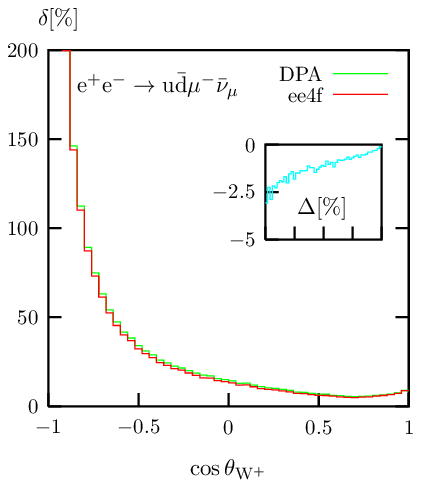}}
\put(3,15.5){$\sqrt{s}=200\GeV$}
\put(11.2,15.5){$\sqrt{s}=500\GeV$}
\end{picture}}%
\caption{Distribution in the cosine of the  $\protect\PWp$ production angle
  with respect to the $\protect\Pep$ beam (upper row) and the
  corresponding corrections (lower row) at $\sqrt{s}=200\GeV$ (l.h.s.)
  and $\sqrt{s}=500\GeV$ (r.h.s.)  for $\Pep\Pem\to \Pu \Pdbar \mu^-
  \bar\nu_{\mu}$. The inset plot shows the difference between the full
  $\Oa$ corrections and those in DPA.}
\label{fi:cthwp}
\end{figure}

\begin{figure} 
{\unitlength 1cm
\begin{picture}(16,16)
\put(-4.9,- 9.7){\includegraphics{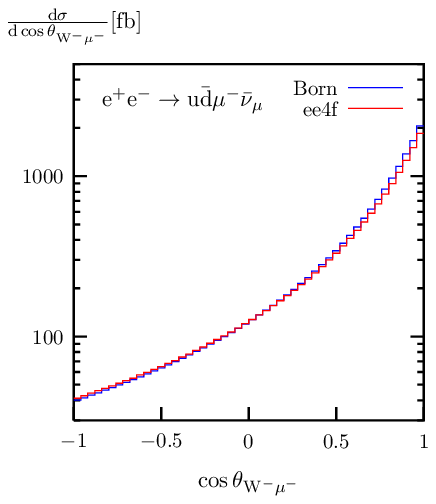}}
\put( 3.3,- 9.7){\includegraphics{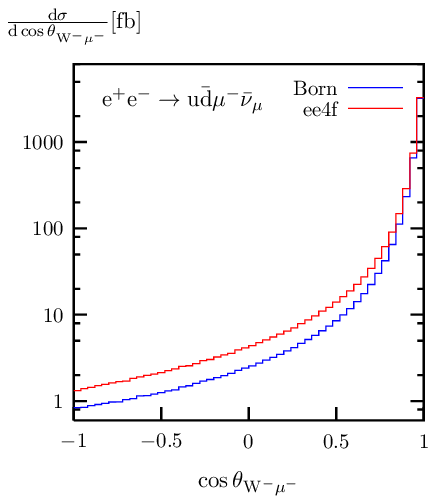}}
\put(-4.9,-17.7){\includegraphics{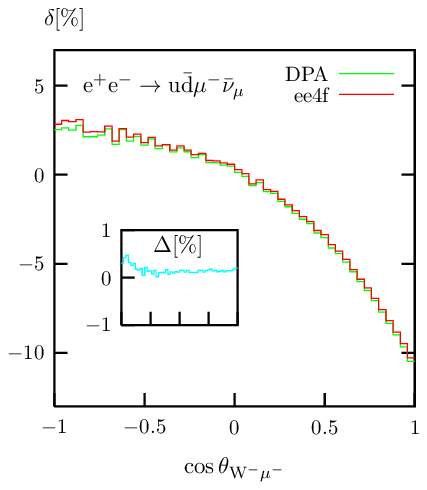}}
\put( 3.3,-17.7){\includegraphics{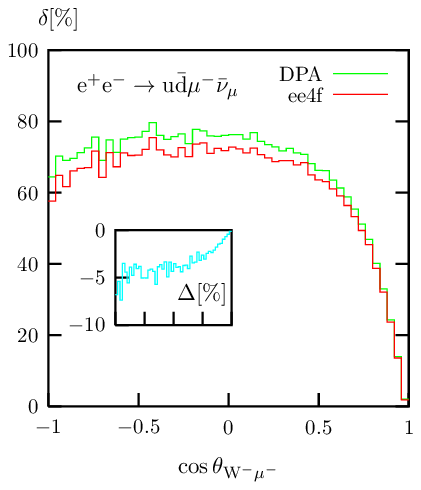}}
\put(3,15.5){$\sqrt{s}=200\GeV$}
\put(11.2,15.5){$\sqrt{s}=500\GeV$}
\end{picture}}%
\caption{Distribution in the cosine of the  $\protect\mu^-$ decay angle
  with respect to the $\protect\PWm$ direction (upper row) and the
  corresponding corrections (lower row) at $\sqrt{s}=200\GeV$ (l.h.s.)
  and $\sqrt{s}=500\GeV$ (r.h.s.)  for $\Pep\Pem\to \Pu \Pdbar \mu^-
  \bar\nu_{\mu}$. The inset plot shows the difference between the full
  $\Oa$ corrections and those in DPA.}
\label{fi:cthwmmu}
\end{figure}

\begin{figure}
{\unitlength 1cm
\begin{picture}(16,16)
\put(-4.9,- 9.7){\includegraphics{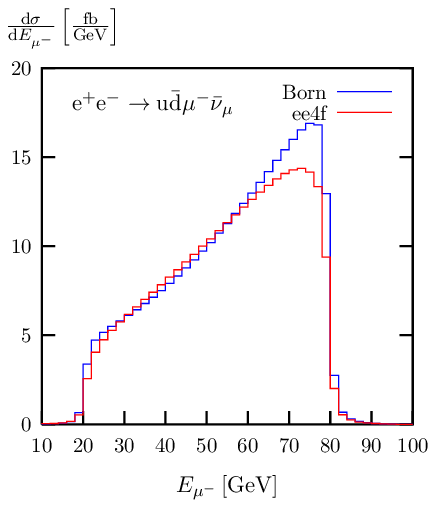}}
\put( 3.3,- 9.7){\includegraphics{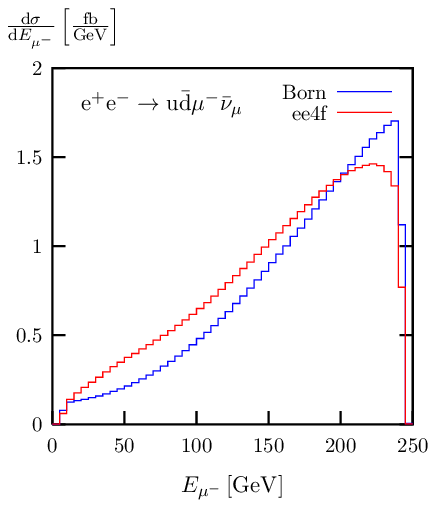}}
\put(-4.9,-17.7){\includegraphics{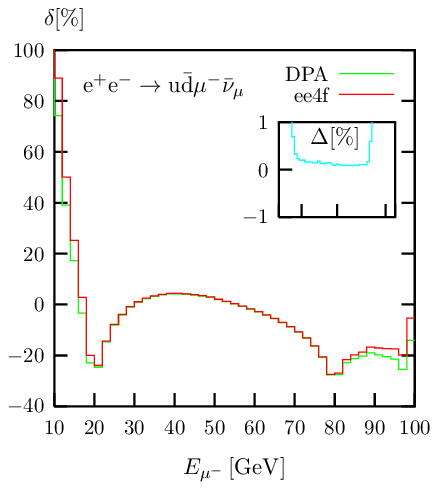}}
\put( 3.3,-17.7){\includegraphics{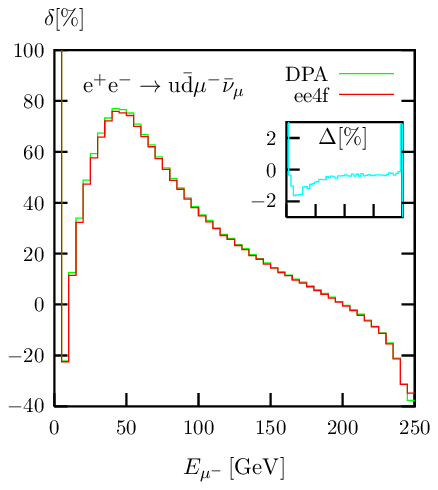}}
\put(3,15.5){$\sqrt{s}=200\GeV$}
\put(11.2,15.5){$\sqrt{s}=500\GeV$}
\end{picture}}%
\caption{Distribution in the energy of the $\protect\mu^-$ (upper row) and the
  corresponding corrections (lower row) at $\sqrt{s}=200\GeV$ (l.h.s.)
  and $\sqrt{s}=500\GeV$ (r.h.s.)  for $\Pep\Pem\to \Pu \Pdbar \mu^-
  \bar\nu_{\mu}$. The inset plot shows the difference between the full
  $\Oa$ corrections and those in DPA}
\label{fi:Em}
\end{figure}

The invariant-mass distributions for the $\PWp$ and $\PWm$ bosons are
shown in \reffis{fi:mwp} and \ref{fi:mwm}. From the plots for the
relative corrections, it can be seen that the full one-loop
corrections are smaller than the DPA corrections for invariant masses
bigger than $\MW$ and vice versa for invariant masses smaller than
$\MW$. If neglected, this effect will give rise to a small shift in the
direct reconstruction of the W-boson mass.
The distribution in the cosine of the $\PWp$ production angle
$\theta_{\PWp}$ is shown in \reffi{fi:cthwp}.
While at LEP energies there is hardly any distortion of the shape
induced by corrections beyond DPA, at $500\GeV$ the difference of the
corrections in DPA and the complete $\Oa$ corrections rises from
$0\%$ to about $-2.5\%$ with increasing scattering angle. Note that
such a distortion of the shape of the angular distribution can be a
signal for anomalous triple gauge-boson couplings.
The angular dependence of $\De$ is even more pronounced in the
distribution in the decay angle $\theta_{\PWm\mu^-}$ presented in
\reffi{fi:cthwmmu}.  Note, however, that as a general feature the
cross section is smallest where the corrections beyond DPA are
largest.
The distribution in the energy $E_{\mu^-}$ of the muon can be found
in \reffi{fi:Em}. Again, for $\sqrt{s}=200\GeV$ the corrections beyond
DPA hardly depend on the muon energy in the interval $20\GeV\lsim
E_{\mu}\lsim80\GeV$, where two resonant W~bosons are kinematically
possible. Outside this interval, the DPA runs out of control, and the
difference to the full calculation becomes big; but there the cross
section is very small.

\begin{figure}
{\unitlength 1cm
\begin{picture}(16,16)
\put(-4.9,- 9.7){\includegraphics{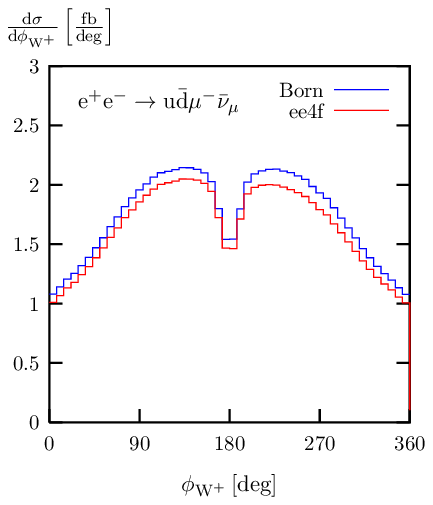}}
\put( 3.3,- 9.7){\includegraphics{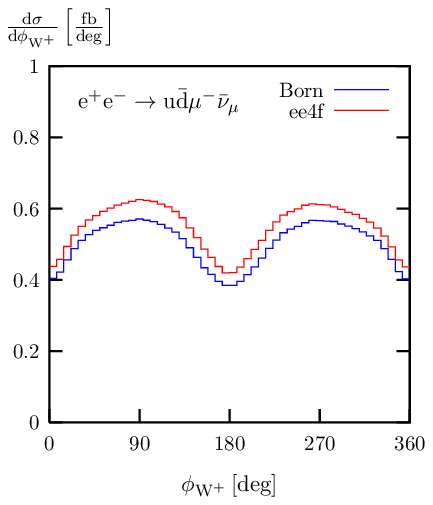}}
\put(-4.9,-17.7){\includegraphics{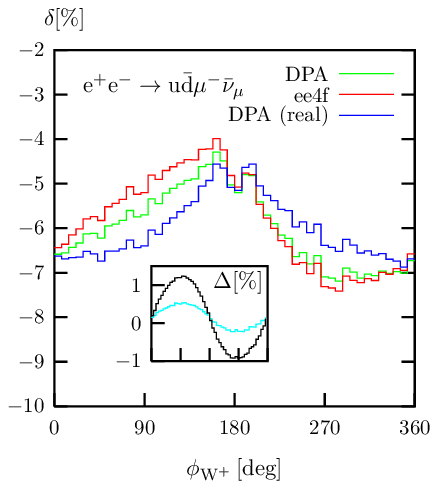}}
\put( 3.3,-17.7){\includegraphics{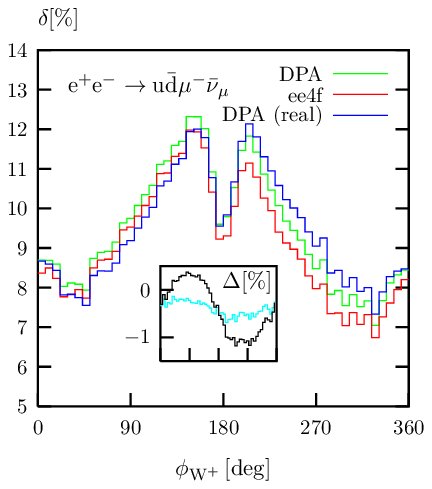}}
\put(3,15.5){$\sqrt{s}=200\GeV$}
\put(11.2,15.5){$\sqrt{s}=500\GeV$}
\end{picture}}%
\caption{Distribution in the azimuthal decay angle of the $\protect\PWp$
  (upper row) and the corresponding corrections (lower row) at
  $\sqrt{s}=200\GeV$ (l.h.s.)  and $\sqrt{s}=500\GeV$ (r.h.s.)  for
  $\Pep\Pem\to \Pu \Pdbar \mu^- \bar\nu_{\mu}$. The inset plot shows
  the difference of the full $\Oa$ corrections to those in DPA and the
  larger difference to those 
  in DPA (real), which is calculated without imaginary parts.}
\label{fi:phiwp}
\end{figure}

Finally, we consider the distributions in azimuthal angles that were
also discussed in \citere{Denner:2000bj} including corrections in DPA.
In \reffi{fi:phiwp} we show the distributions 
in the azimuthal decay
angle $\phi_{\PWp}$ of the $\PWp$ boson, \ie the angle between the
decay plane of the $\PWp$ and the plane of \PW-pair production,
\newcommand{\bp}{{\bf p}}
\newcommand{\bk}{{\bf k}}
\beqar
\cos\phi_{\PWp} &=& \frac{(\bk_+\times \bp_+)(\bk_+\times \bk_1)}
{|\bk_+\times \bp_+||\bk_+\times \bk_1|},\nl
\sgn(\sin\phi_{\PWp}) &=& \sgn\left\{\bk_+\cdot
\left[(\bk_+\times \bp_+)\times(\bk_+\times \bk_1)\right]\right\}.
\eeqar
Here, in particular at $200\GeV$ the difference between DPA and the
full $\Oa$ corrections is approximately proportional to
$\sin\phi_{\PWp}$ plus some constant offset. In the DPA, it can be
deduced from Appendix~A of \citere{Beenakker:1998gr}, that the
contributions of the imaginary parts of the one-loop coefficient
functions always involve a factor $\sin\phi_{\PWp}$ or a factor
$\sin\phi_{\PWm}$ together with symmetric functions in these
angles. Contributions of real parts, on the other hand, are symmetric
in these angles. In order to illustrate this feature, we have included
an extra curve in the plots for the relative corrections to this
distribution labelled ``DPA (real)''. In this curve, we have switched
off all imaginary parts in the DPA calculation, apart from the finite
width in the resonant propagators. As expected, these results are
symmetric in the angle $\phi_{\PWp}$ about $180^\circ$. The
contribution proportional to $\sin\phi_{\PWp}$ in the difference of
the full $\Oa$ calculation with respect to the DPA (real) is even
larger than the corresponding difference with respect to the complete
DPA. These properties of the DPA suggest, that also the sinusoidal
dependence on $\phi_{\PWp}$ in the latter difference results from 
imaginary parts. A possible source could be the missing imaginary
parts of the counterterms in DPA.  

\begin{figure}
{\unitlength 1cm
\begin{picture}(16,16)
\put(-4.9,- 9.7){\includegraphics{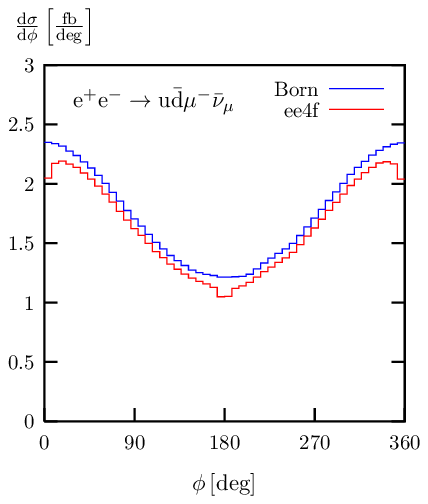}}
\put( 3.3,- 9.7){\includegraphics{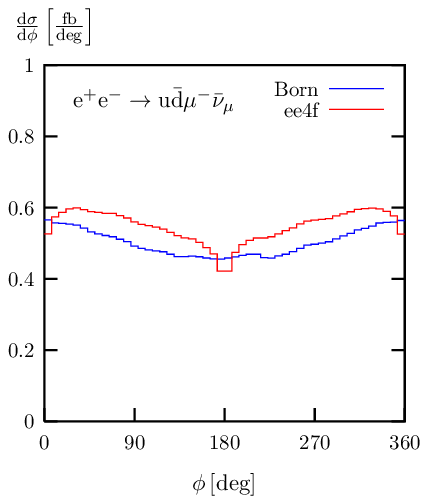}}
\put(-4.9,-17.7){\includegraphics{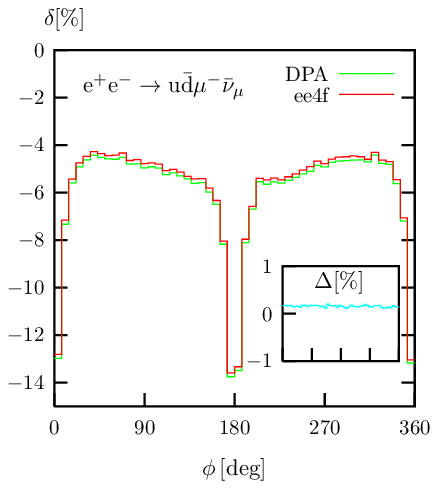}}
\put( 3.3,-17.7){\includegraphics{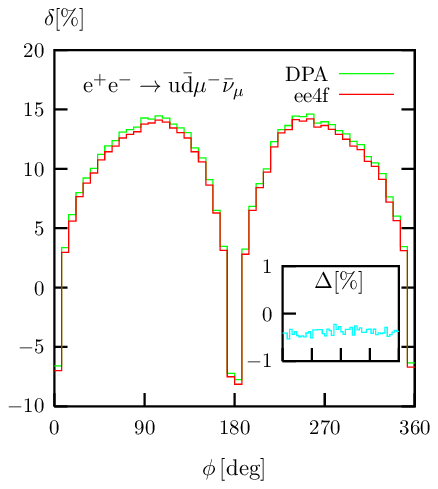}}
\put(3,15.5){$\sqrt{s}=200\GeV$}
\put(11.2,15.5){$\sqrt{s}=500\GeV$}
\end{picture}}%
\caption{Distribution in the azimuthal angle $\phi$  (upper row) and the 
  corresponding corrections (lower row) at $\sqrt{s}=200\GeV$ (l.h.s.)
  and $\sqrt{s}=500\GeV$ (r.h.s.)  for $\Pep\Pem\to \Pu \Pdbar \mu^-
  \bar\nu_{\mu}$. The inset plot shows the difference between the full
  $\Oa$ corrections and those in DPA.}
\label{fi:phi}
\end{figure}

The distributions in the angle $\phi$ between the two planes spanned
by the momenta of the two fermion pairs in which the \PW~bosons decay, \ie
(note that $\bk_+=-\bk_-$ for non-photonic events)
\beqar
\cos\phi &=& \frac{(\bk_+\times \bk_1)(-\bk_-\times \bk_3)}
{|\bk_+\times \bk_1||{-\bk_-}\times \bk_3|},\nl
\sgn(\sin\phi) &=& \sgn\left\{\bk_+\cdot
\left[(\bk_+\times \bk_1)\times(-\bk_-\times \bk_3)\right]\right\},
\eeqar
are presented in \reffi{fi:phi}.
The large corrections for angles $\phi$ near $0^\circ$ or $180^\circ$,
\ie if the two decay planes coincide, result from the suppression of
hard photonic corrections \cite{Denner:2000bj}. The corrections beyond
DPA do hardly depend on $\phi$.

\section{Conclusions}
\label{se:concl}

We have presented technical and conceptual details as well as further
numerical results of a calculation of the complete electroweak ${\cal
  O}(\alpha)$ corrections to the charged-current four-fermion
production processes $\Pep\Pem\to\nu_\tau\tau^+\mu^-\bar\nu_\mu$,
$\Pu\bar\Pd\mu^-\bar\nu_\mu$, and $\Pu\bar\Pd\Ps\bar\Pc$.

In particular, we have described methods how the ${\cal O}(10^3)$
occurring different spinor structures can be algebraically reduced to
a few simple standard structures. The presented algorithms, which
shorten the analytical results considerably and thereby render the
resulting computer code relatively short, should be useful for other
future calculations.

Moreover, a concept for consistently performing one-loop calculations
with complex masses for unstable particles is presented.  Technically
the complex masses are introduced via an appropriate complex
renormalization prescription at the level of the Lagrangian, so that
the usual machinery of perturbation theory (Feynman rules etc.)  can
be used to organize the calculation. Since the theory is not changed
at all, there is no danger of double counting terms.  The complex mass
parameters, which also enter the weak mixing angle and coupling
constants, can be viewed as an analytical continuation of the real
masses. Consequently, all algebraic relations that follow from gauge
invariance (Slavnov--Taylor identities, Ward identities, cancellation
of gauge-parameter dependences) are valid in spite of the complex
masses which incorporate finite-width effects.  The price to pay for
this achievement is that all one-loop integrals have to be performed
with complex mass parameters.

Finally, we have completed our discussion of numerical results, which
was started in \citere{Denner:2005es} by considering total cross
sections, by showing the effects of the complete $\Oa$ corrections to
various differential cross sections of physical interest.  In
particular, we have considered differential cross sections sensitive
to imaginary parts.  Our results have also been compared to
predictions based on the double-pole approximation, revealing that the
latter approximation is not sufficient to fully exploit the potential
of a future linear collider in an analysis of W-boson pairs at high
energies.  Specifically, at (and above) a CM energy of $500\GeV$ the
corrections beyond DPA induce a non-negligible distortion in angular
distributions that could be misinterpreted as signal for anomalous
triple gauge-boson couplings if not taken into account.  The remaining
theoretical uncertainties for the total cross section have been
discussed in \citere{Denner:2005es}. For differential distributions
the uncertainties from QCD effects are even more relevant.

In this paper we have presented methods that were successfully used
for the first complete $\Oa$ calculation of electroweak corrections to a
process with six external particles and involving unstable particles in
intermediate states. Processes of this type will become more and more
important in the future. Our methods are not specifically adapted to
the considered process and therefore should be helpful in precision
calculations for similar processes at future colliders.

\par
\vskip 1cm

\section*{Acknowledgements}

This work was supported in part by the Swiss National Science
Foundation, by the Swiss Bundesamt f\"ur Bildung und Wissenschaft, and
by the European Union under contract HPRN-CT-2000-00149.  A.D. thanks
the Lehrstuhl f\"ur Theoretische Physik II of the University of
W\"urzburg, where part of this work has been done, for its kind
hospitality.

\appendix
\section*{Appendix}
\renewcommand{\theequation}{A.\arabic{equation}}

\section*{Interpretation of the reduction algorithm of 
\refse{se:minimal-reduction}
in the Weyl--van der Waerden spinor formalism}
\label{se:WvdWreducction}

In this appendix we inspect the reduction algorithm of
\refse{se:minimal-reduction} within the Weyl--van der Waerden (WvdW) spinor 
formalism, which is particularly simple for massless particles.
We consistently employ the conventions of
\citere{Dittmaier:1998nn} for the spinor method and highlight
only those features that are crucial for our manipulations.

In this formalism, Dirac spinors and momenta of massless fermions are 
described by Weyl spinors $p_A$, $p_{\dot A}=(p_A)^*$ which are related to
the four-momentum $p_\mu$ according to 
\beq
p_\mu \si^\mu_{\dot A B} = p_{\dot A} p_{B},
\eeq
where $\si^\mu_{\dot A B}$ denote the unit matrix ($\mu=0$) and the 
Pauli matrices ($\mu=1,2,3$). The matrices $\si^\mu$ also appear as
the non-vanishing blocks in the Dirac matrices $\gamma^\mu$.
Thus, Dirac spinor chains translate into chains
involving Weyl spinors and $\si^\mu$ matrices. Coupled chains can be
separated by using the relation
\beq
\si^\mu_{\dot A B}\si_{\mu,\dot C D} = 2 \eps_{\dot A \dot C}\eps_{BD},
\label{eq:sisi}
\eeq
where $\eps_{AB}$ is the totally antisymmetric tensor in two
dimensions and $\eps_{\dot A \dot B}$ is its complex conjugate.
In this way, all Weyl spinors get contracted in so-called spinor products
\beq
\langle p_i p_j \rangle = \eps^{AB} p_{i,A} p_{j,B}
\eeq
or their complex conjugates. Using \refeq{eq:sisi}, Lorentz products can
also be translated into spinor products. We illustrate the above
considerations by the simplest examples:
\beqar
(p_ip_j) &=& \frac{1}{2}\langle p_ip_j\rangle\langle p_ip_j\rangle^*
,\nl
\Bigl[ \dsl{p}_k \Bigr]_{ij}^- &=& \langle p_kp_i\rangle\langle p_kp_j\rangle^*
,\nl
\Bigl[ \dsl{p}_k \Bigr]_{ij}^+ &=& 
\langle p_kp_i\rangle^*\langle p_kp_j\rangle.
\eeqar
Dirac chains of the form \refeq{SMEstructure} involving no additional
$\eps^{\mu\nu\rho\si}$ can thus be directly reduced to a product of
(at least four) spinor products.  The totally antisymmetric tensor
$\eps^{\mu\nu\rho\si}$ is translated into WvdW objects according to
\beq\label{eps-spinor}
\text \eps^{\mu\nu\rho\kappa}
\;=\; \disp\frac{\ri}{4} \left(
\si^{\mu}_{\dot A B}\si^{\nu,B \dot C}
\si^{\kappa}_{\dot C D}\si^{\rho,D \dot A}
-\si^{\mu}_{\dot A B}\si^{\rho,B \dot C}
\si^{\kappa}_{\dot C D}\si^{\nu,D \dot A}
\right),
\eeq
so that contractions with $\eps^{\mu\nu\rho\si}$ can also be expressed
in terms of WvdW spinor products.
The two-dimensionality of the WvdW spinor space is rather restrictive
and leads to the identity
\beq\label{eq:fierz}
\langle p_i p_j\rangle\langle p_k p_l\rangle
+\langle p_i p_k\rangle\langle p_l p_j\rangle
+\langle p_i p_l\rangle\langle p_j p_k\rangle = 0,
\eeq
which frequently admits simplifications in complicated expressions.
In the cases relevant for the considered processes, these identities
allow to combine the sums that result from the application of
\refeq{eps-spinor}, so that also all spinorial expressions involving
contractions with $\eps^{\mu\nu\rho\si}$ can be reduced to a product
of spinor products.

The identities \refeq{eps-spinor} and \refeq{eq:fierz} also allow to write
the combinations $A_{i\,j\,k\,l}^{a\,b\,c\,d}$ of \refeq{DefA} as
simple factors of spinor products:
\beqar\label{relations-A-spinor}
A_{i\,j\,k\,l}^{\sss{++-+}}&=& 
\frac{1}{2}\langle p_ip_j\rangle\langle p_kp_l\rangle
\langle p_ip_k\rangle^*\langle p_jp_l\rangle^* ,\nl
A_{i\,j\,k\,l}^{\sss{+-++}}&=& 
\frac{1}{2}\langle p_ip_l\rangle\langle p_kp_j\rangle
\langle p_ip_j\rangle^*\langle p_kp_l\rangle^* ,\nl
A_{i\,j\,k\,l}^{\sss{-+++}}&=& 
\frac{1}{2}\langle p_ip_k\rangle\langle p_jp_l\rangle
\langle p_ip_l\rangle^*\langle p_jp_k\rangle^* ,\nl
A_{i\,j\,k\,l}^{a\,b\,c\,\sss{-}}&=& (A_{i\,j\,k\,l}^{a\,b\,c\,\sss{+}})^*.
\eeqar

When expressing the $A_{i\,j\,k\,l}^{a\,b\,c\,d}$ in terms of spinor
products, the relations \refeq{Aproduct1}--\refeq{Aproduct4} become
trivial.  The relations \refeq{relations-A-spinor} and the fact that
in the WvdW formalism all spinor chains can be expressed in terms of
products of spinor products explains why in the formalism of 
\refse{se:minimal-reduction} all spinor chains could be reduced to a
simple product of $A_{i\,j\,k\,l}^{a\,b\,c\,d}$ and scalar products.  The
results of \refeq{SME-results} read in the WvdW formalism
\beqar
  \Bigl[ \gamma^{\mu} \gamma^{\nu} \gamma^{\rho} \Bigr]_{12}^{\sigma}
  \Bigl[ \gamma_{\mu} \gamma_{\nu} \gamma_{\rho} \Bigr]_{34}^{\tau} \,
  & = &  32 \langle p_2 p_4\rangle\langle p_1 p_3\rangle^*
           \de_{\si+}\de_{\tau+}
         +32 \langle p_2 p_4\rangle^*\langle p_1 p_3\rangle
           \de_{\si-}\de_{\tau-}
  \nl&&{}
           +8 \langle p_2 p_3\rangle\langle p_1 p_4\rangle^*
           \de_{\si+}\de_{\tau-}
         +8 \langle p_2 p_3\rangle^*\langle p_1 p_4\rangle
           \de_{\si-}\de_{\tau+}  ,\nl
  \Bigl[ \dsl{p}_3 \gamma^{\mu} \gamma^{\nu}  \Bigr]_{12}^{\sigma}
  \Bigl[ \gamma_{\nu} \gamma^{\rho} \gamma^{\kappa}  \Bigr]_{34}^-
  \Bigl[ \gamma_{\mu} \gamma_{\rho} \gamma_{\kappa}  \Bigr]_{56}^- \,
  & = &   32 \langle p_3 p_5\rangle\langle p_4 p_6\rangle^*
\nl&&{}\times
          ( \de_{\si+} \langle p_2 p_3\rangle\langle p_1 p_3\rangle^*
      +\de_{\si-} \langle p_1 p_3\rangle\langle p_2 p_3\rangle^*) \,,
\vspace{1.5mm} \nl
  \Bigl[ \dsl{p}_3 \gamma^{\mu} \gamma^{\nu} \gamma^{\rho} \gamma^{\kappa}
         \Bigr]_{12}^- \,
  \Bigl[ \gamma_{\nu} \gamma_{\rho} \gamma_{\kappa}  \Bigr]_{34}^- \,
  \Bigl[ \gamma_{\mu}   \Bigr]_{56}^- \,
  & = &  
64\langle p_1 p_3\rangle\langle p_3 p_5\rangle 
\langle p_2 p_4\rangle^*\langle p_3 p_6\rangle^*\,,
\vspace{1.5mm} \nl
-\ri  \epsilon^{\mu \nu \rho \sigma}\, p_{1,\sigma} \,
  \Bigl[ \gamma_{\mu} \Bigr]_{12}^+ \,
  \Bigl[ \gamma_{\nu} \Bigr]_{34}^- \,
  \Bigl[ \gamma_{\rho} \Bigr]_{56}^- \,
  & = & \, -2\langle p_1 p_2\rangle\langle p_3 p_5\rangle 
\langle p_1 p_4\rangle^*\langle p_1 p_6\rangle^*
\,.
\vspace{1.5mm}
\eeqar

Finally, the relevant SMEs \refeq{eq:SME1} read in the WvdW formalism
\beqar
\Mhat^{---} &=& \langle p_1p_3\rangle \langle p_2p_3\rangle^*
 \langle p_1p_3\rangle \langle p_1p_4\rangle^*
 \langle p_1p_5\rangle \langle p_1p_6\rangle^*,\nl
\Mhat^{+--} &=& \langle p_1p_3\rangle^* \langle p_2p_3\rangle
 \langle p_1p_3\rangle \langle p_1p_4\rangle^*
 \langle p_1p_5\rangle \langle p_1p_6\rangle^*.
\eeqar

Of course, it would have been possible to translate each spinor chain
into WvdW spinor products directly after performing the loop
integration, \ie skipping the steps described in
\refse{se:minimal-reduction}. However, it would have been hard to
perform all the manipulations with WvdW objects made there in terms of
Lorentz products. For instance, it is practically impossible to make
full use of momentum conservation in very involved expressions in
terms of WvdW spinor products.  It turned out in many examples given
in the literature that no algorithmic recipe has been found yet
yielding the most compact expressions in complicated
amplitudes. Moreover, the generalization to massive fermions is more
complicated in the WvdW formalism than in the method described in
\refse{se:minimal-reduction}.  In fact a variant of this method has
already been used in \citere{Denner:2003zp} for the process
$\Pep\Pem\to\Pt\bar\Pt\PH$.


\begin{thebibliography}{99}
\frenchspacing
\newcommand{\ap}[3]{{\sl Ann.~Phys.} {\bf #1} (19#2) #3}
\newcommand{\zp}[3]{{\sl Z.~Phys.} {\bf #1} (19#2) #3}
\newcommand{\np}[3]{{\sl Nucl.~Phys.} {\bf #1} (19#2) #3}
\newcommand{\pl}[3]{{\sl Phys.~Lett.} {\bf #1} (19#2) #3}
\newcommand{\pr}[3]{{\sl Phys.~Rev.} {\bf #1} (19#2) #3}
\newcommand{\prl}[3]{{\sl Phys.~Rev.~Lett.} {\bf #1} (19#2) #3}
\newcommand{\fp}[3]{{\sl Fortschr.~Phys.} {\bf #1} (19#2) #3}
\newcommand{\jp}[3]{{\sl J.~Phys.} {\bf #1} (19#2) #3}
\newcommand{\cpc}[3]{{\sl Comput.~Phys.~Commun.} {\bf #1} (19#2) #3}
\newcommand{\ijmp}[3]{{\sl Int.~J.~Mod.~Phys.} {\bf #1} (19#2) #3}
\newcommand{\nim}[3]{{\sl Nucl.~Instr.~Meth.} {\bf #1} (19#2) #3}
\newcommand{\nc}[3]{{\sl Nuovo Cimento} {\bf #1} (19#2) #3}
\newcommand{\sjnp}[3]{\vj{Sov. J. Nucl. Phys.}{#1}{#2}{#3}}
\newcommand{\ej}[3]{{\bf #1 }\ifnum#2<100 (19#2) \else (#2) \fi #3}
\newcommand{\vj}[4]{{\sl #1} {\bf #2} (19#3) #4}

\bibitem{lep2}
The LEP Collaborations ALEPH, DELPHI, L3, OPAL, the LEP EWWG,
and the SLD Heavy Flavor and Electroweak Groups,
hep-ex/0412015.

\bibitem{Aguilar-Saavedra:2001rg}
J.~A.~Aguilar-Saavedra {\it et al.},
TESLA Technical Design Report Part III: Physics at an $\mathrm{e^+e^-}$
Linear Collider,
hep-ph/0106315.

\bibitem{Abe:2001wn}
T.~Abe {\it et al.}  [American Linear Collider Working Group Collaboration],
in {\it Proc. of the APS/DPF/DPB Summer Study on the Future of
  Particle Physics (Snowmass 2001) } ed. R.~Davidson and C.~Quigg,
SLAC-R-570, {\it Resource book for Snowmass 2001},
[hep-ex/0106055, hep-ex/0106056, hep-ex/0106057, hep-ex/0106058].

\bibitem{Abe:2001gc}
K.~Abe {\it et al.}  [ACFA Linear Collider Working Group Collaboration],
ACFA Linear Collider Working Group report,
[hep-ph/0109166].

\bibitem{talkKM}
K.~M\"onig and A.~Tonazzo,
talk given by K.~M\"onig at the {\it 2nd ECFA/DESY Study on Physics and
Detectors for a Linear Electron--Positron Collider},
Padova, Italy, 2000.

\bibitem{Alles:1976qv}
  W.~Alles, C.~Boyer and A.~J.~Buras,
  Nucl.\ Phys.\ B {\bf 119} (1977) 125;\\
%
  K.~J.~F.~Gaemers and G.~J.~Gounaris,
  Z.\ Phys.\ C {\bf 1} (1979) 259.

\bibitem{Lemoine:1979pm}
  M.~Lemoine and M.~J.~G.~Veltman,
  Nucl.\ Phys.\ B {\bf 164} (1980) 445;\\
%
  R.~Philippe,
  Phys.\ Rev.\ D {\bf 26} (1982) 1588;\\
%
%
  J.~Fleischer, F.~Jegerlehner and M.~Zra\l{}ek,
  Z.\ Phys.\ C {\bf 42} (1989) 409.

\bibitem{Bohm:1987ck}
M.~B\"ohm {\it et al.},
  Nucl.\ Phys.\ B {\bf 304} (1988) 463.

\bibitem{Beenakker:1990sf}
  W.~Beenakker, K.~Ko\l odziej and T.~Sack,
  Phys.\ Lett.\ B {\bf 258} (1991) 469;\\
%
  W.~Beenakker, F.~A.~Berends and T.~Sack,
  Nucl.\ Phys.\ B {\bf 367} (1991) 287;\\
%
  H.~Tanaka, T.~Kaneko and Y.~Shimizu,
  Comput.\ Phys.\ Commun.\  {\bf 64} (1991) 149;\\
%
  K.~Ko\l odziej and M.~Zra\l{}ek,
  Phys.\ Rev.\ D {\bf 43} (1991) 3619;\\
%
  J.~Fleischer, K.~Ko\l odziej and F.~Jegerlehner,
  Phys.\ Rev.\ D {\bf 47} (1993) 830.

\bibitem{Dittmaier:1991np}
  S.~Dittmaier, M.~B\"ohm and A.~Denner,
  Nucl.\ Phys.\ B {\bf 376} (1992) 29
  [Erratum-ibid.\ B {\bf 391} (1993) 483];\\
%
  M.~Kuroda, I.~Kuss and D.~Schildknecht,
  Phys.\ Lett.\ B {\bf 409} (1997) 405
  [hep-ph/9705294].

\bibitem{Beenakker:1993tt}
  W.~Beenakker {\it et al.},
  Nucl.\ Phys.\ B {\bf 410} (1993) 245;
%
  Phys.\ Lett.\ B {\bf 317} (1993) 622;\\
%
  M.~Kuroda and D.~Schildknecht,
  Nucl.\ Phys.\ B {\bf 531} (1998) 24
  [hep-ph/9807250].

\bibitem{Bardin:1986fi}
  D.~Y.~Bardin, S.~Riemann and T.~Riemann,
  Z.\ Phys.\ C {\bf 32} (1986) 121;\\
%
  F.~Jegerlehner,
  Z.\ Phys.\ C {\bf 32} (1986) 425
  [Erratum-ibid.\ C {\bf 38} (1988) 519];\\
%
  A.~Denner and T.~Sack,
  Z.\ Phys.\ C {\bf 46} (1990) 653.

\bibitem{Berends:1994xn}
  F.~A.~Berends, R.~Pittau and R.~Kleiss,
  Comput.\ Phys.\ Commun.\  {\bf 85} (1995) 437
  [hep-ph/9409326];\\
%
%
M.~Skrzypek, S.~Jadach, W.~P\l aczek and Z.~W\c{a}s,
  Comput.\ Phys.\ Commun.\  {\bf 94} (1996) 216;\\
%
   G.~Passarino,
   Comput.\ Phys.\ Commun.\  {\bf 97} (1996) 261
   [hep-ph/9602302];\\
%
   E.~Accomando and A.~Ballestrero,
   Comput.\ Phys.\ Commun.\  {\bf 99} (1997) 270 
   [hep-ph/9607317];\\
%
   J.~Fujimoto {\it et al.},
   Comput.\ Phys.\ Commun.\  {\bf 100} (1997) 128 
   [hep-ph/9605312];\\
%
   D.~Y.~Bardin {\it et al.},
   Comput.\ Phys.\ Commun.\  {\bf 104} (1997) 161
   [hep-ph/9612409];\\
%
   S.~Jadach  {\it et al.},
   Comput.\ Phys.\ Commun.\  {\bf 119} (1999) 272 
   [hep-ph/9906277];\\
%
%
   F.~A.~Berends, C.~G.~Papadopoulos and R.~Pittau,
   Comput.\ Phys.\ Commun.\  {\bf 136} (2001) 148
   [hep-ph/0011031];\\
%
   E.~Accomando, A.~Ballestrero and E.~Maina,
   Comput.\ Phys.\ Commun.\  {\bf 150} (2003) 166 
   [hep-ph/0204052].
%
%

\bibitem{Beenakker:1996kt}
W.~Beenakker {\it et al.},
in {\sl Physics at LEP2}, eds.\ G.~Altarelli, T.~Sj\"o\-strand and
F.~Zwirner (CERN 96-01, Geneva, 1996), Vol.~1, p.~79
[hep-ph/9602351].

\bibitem{Bardin:1997gc}
  D.~Y.~Bardin {\it et al.},
in {\sl Physics at LEP2}, eds.\ G.~Altarelli, T.~Sj\"o\-strand and
F.~Zwirner (CERN 96-01, Geneva, 1996), Vol.~2, p.~3
[hep-ph/9709270].

\bibitem{Grunewald:2000ju}
M.~W.~Gr\"unewald {\it et al.},
in {\it Reports of the Working Groups on Precision Calculations
for LEP2 Physics}, eds.\ S.~Jadach, G.~Passarino and R.~Pittau
(CERN 2000-009, Geneva, 2000), p.~1
[hep-ph/0005309].

\bibitem{Fadin:1993kg}
  V.~S.~Fadin, V.~A.~Khoze and A.~D.~Martin,
  Phys.\ Lett.\ B {\bf 311} (1993) 311;\\
%
  D.~Y.~Bardin, W.~Beenakker and A.~Denner,
  Phys.\ Lett.\ B {\bf 317} (1993) 213.

\bibitem{Beenakker:1998gr}
W.~Beenakker, F.~A.~Berends and A.~P.~Chapovsky,
Nucl.\ Phys.\ B {\bf 548} (1999) 3
[hep-ph/9811481].

\bibitem{Jadach:1998tz}
S.~Jadach {\it et al.},
Phys.\ Rev.\ D {\bf 61} (2000) 113010
[hep-ph/9907436];
%
Comput.\ Phys.\ Commun.\  {\bf 140} (2001) 432
[hep-ph/0103163];
Comput.\ Phys.\ Commun.\  {\bf 140} (2001) 475
[hep-ph/0104049];
%
Phys.\ Rev.\ D {\bf 65} (2002) 093010
[hep-ph/0007012].

\bibitem{Denner:2000kn}
A.~Denner, S.~Dittmaier, M.~Roth and D.~Wackeroth,
Phys.\ Lett.\ B {\bf 475} (2000) 127
[hep-ph/9912261];
%
Eur.\ Phys.\ J.\ direct C {\bf 2} (2000) 4
[hep-ph/9912447];
%
in {\it Proc. of the 5th International Symposium on Radiative Corrections (RADCOR 2000)}, ed. H.~E. Haber,
hep-ph/0101257;
%
Comput.\ Phys.\ Commun.\  {\bf 153} (2003) 462
[hep-ph/0209330].

\bibitem{Denner:2000bj}
A.~Denner, S.~Dittmaier, M.~Roth and D.~Wackeroth,
Nucl.\ Phys.\ B {\bf 587} (2000) 67
[hep-ph/0006307].

\bibitem{Kurihara:2001um}
Y.~Kurihara, M.~Kuroda and D.~Schildknecht,
Phys.\ Lett.\ B {\bf 509} (2001) 87
[hep-ph/0104201].

\bibitem{Denner:1999gp}
A.~Denner, S.~Dittmaier, M.~Roth and D.~Wackeroth,
Nucl.\ Phys.\ B {\bf 560} (1999) 33
[hep-ph/9904472].


\bibitem{Jadach:2001cz}
S.~Jadach {\it et al.},
Phys.\ Lett.\ B {\bf 523} (2001) 117
[hep-ph/0109072];\\
%
  F.~Cossutti,
  DELPHI note 2004-050 PHYS 944, hep-ph/0505232;\\
%
R.~Bruneli\`ere {\it et al.},
Phys.\ Lett.\ B {\bf 533} (2002) 75
[hep-ph/0201304].

\bibitem{Vicini:1998iy}
A.~Vicini,
Acta Phys.\ Polon.\ B {\bf 29} (1998) 2847.

\bibitem{Berends:1969nt}
  F.~A.~Berends and G.~B.~West,
  Phys.\ Rev.\ D {\bf 1} (1970) 122;\\
%
  Y.~Kurihara, D.~Perret-Gallix and Y.~Shimizu,
  Phys.\ Lett.\ B {\bf 349} (1995) 367
  [hep-ph/9412215].

\bibitem{Argyres:1995ym}
  E.~N.~Argyres {\it et al.},
  Phys.\ Lett.\ B {\bf 358}(1995) 339 
  [hep-ph/9507216].

\bibitem{Aeppli:1993cb}
  A.~Aeppli, F.~Cuypers and G.~J.~van Oldenborgh,
  Phys.\ Lett.\ B {\bf 314} (1993) 413
  [hep-ph/9303236].

\bibitem{Beenakker:1996kn}
  W.~Beenakker {\it et al.},
  Nucl.\ Phys.\ B {\bf 500} (1997) 255
  [hep-ph/9612260].

\bibitem{Stuart:1991xk}
  R.~G.~Stuart,
  Phys.\ Lett.\ B {\bf 262} (1991) 113.

\bibitem{Aeppli:1993rs}
  A.~Aeppli, G.~J.~van Oldenborgh and D.~Wyler,
  Nucl.\ Phys.\ B {\bf 428} (1994) 126
  [hep-ph/9312212].

\bibitem{Baur:1995aa}
  U.~Baur and D.~Zeppenfeld,
  Phys.\ Rev.\ Lett.\  {\bf 75} (1995) 1002
  [hep-ph/9503344].

\bibitem{Passarino:1999zh}
  G.~Passarino,
  Nucl.\ Phys.\ B {\bf 574} (2000) 451
  [hep-ph/9911482];\\
%
  E.~Accomando, A.~Ballestrero and E.~Maina,
  Phys.\ Lett.\ B {\bf 479} (2000) 209
  [hep-ph/9911489].

\bibitem{Beenakker:1999hi}
  W.~Beenakker, F.~A.~Berends and A.~P.~Chapovsky,
  Nucl.\ Phys.\ B {\bf 573} (2000) 503
  [hep-ph/9909472];\\
%
  W.~Beenakker {\it et al.},
  Nucl.\ Phys.\ B {\bf 667} (2003) 359
  [hep-ph/0303105].

\bibitem{Beneke:2003xh}
M.~Beneke, A.~P.~Chapovsky, A.~Signer and G.~Zanderighi,
Phys.\ Rev.\ Lett.\  {\bf 93} (2004) 011602
[hep-ph/0312331] and
%
Nucl.\ Phys.\ B {\bf 686} (2004) 205
[hep-ph/0401002].

\bibitem{Boudjema:2004id}
F.~Boudjema {\it et al.},
Nucl.\ Phys.\ Proc.\ Suppl.\  {\bf 135} (2004) 323
[hep-ph/0407079].

\bibitem{Denner:2005es}
  A.~Denner, S.~Dittmaier, M.~Roth and L.~H.~Wieders,
  Phys.\ Lett.\ B {\bf 612} (2005) 223
  [hep-ph/0502063].

\bibitem{Dittmaier:2000mb}
S.~Dittmaier,
Nucl.\ Phys.\ B {\bf 565} (2000) 69
[hep-ph/9904440];\\
%
M.~Roth,
PhD thesis, ETH Z\"urich No. 13363 (1999),
hep-ph/0008033.

\bibitem{Hollik:1988ii}
W.~F.~L.~Hollik,
Fortschr.\ Phys.\  {\bf 38} (1990) 165.

\bibitem{Denner:1994xt}
A.~Denner, S.~Dittmaier and G.~Weiglein,
Nucl.\ Phys.\ B {\bf 440} (1995) 95
[hep-ph/9410338].

\bibitem{Kublbeck:1990xc}
J.~K\"ublbeck, M.~B\"ohm and A.~Denner,
Comput.\ Phys.\ Commun.\  {\bf 60} (1990) 165;\\
H.~Eck and J.~K\"ublbeck, {\it Guide to FeynArts 1.0\/},
University of W\"urzburg, 1992.

\bibitem{Hahn:2000kx}
T.~Hahn,
Comput.\ Phys.\ Commun.\  {\bf 140} (2001) 418
[hep-ph/0012260].

\bibitem{Hahn:1998yk}
T.~Hahn and M.~Perez-Victoria,
Comput.\ Phys.\ Commun.\  {\bf 118} (1999) 153
[hep-ph/9807565];\\
%
T.~Hahn,
Nucl.\ Phys.\ Proc.\ Suppl.\  {\bf 89} (2000) 231
[hep-ph/0005029].

\bibitem{Denner:1993kt}
A.~Denner,
Fortsch.\ Phys.\  {\bf 41} (1993) 307.

\bibitem{Me65} D.~B.~Melrose, {Nuovo Cimento}~{\bf XL~A} (1965) 181.

\bibitem{Denner:2002ii}
A.~Denner and S.~Dittmaier,
Nucl.\ Phys.\ B {\bf 658} (2003) 175
[hep-ph/0212259].

\bibitem{Passarino:1979jh}
G.~Passarino and M.~Veltman,
Nucl.\ Phys.\ B {\bf 160} (1979) 151.

\bibitem{'tHooft:1979xw}
G.~'t Hooft and M.~Veltman,
Nucl.\ Phys.\ B {\bf 153} (1979) 365.

\bibitem{Beenakker:1990jr}
W.~Beenakker and A.~Denner,
Nucl.\ Phys.\ B {\bf 338} (1990) 349.

\bibitem{Denner:1991qq}
A.~Denner, U.~Nierste and R.~Scharf,
Nucl.\ Phys.\ B {\bf 367} (1991) 637.

\bibitem{Vicini:2001pd}
  A.~Vicini,
  Phys.\ Lett.\ B {\bf 531} (2002) 83
  [hep-ph/0104164].

\bibitem{Sirlin:1981pi}
A.~Sirlin,
Nucl.\ Phys.\ B {\bf 192} (1981) 93.

\bibitem{Denner:2003iy}
  A.~Denner, S.~Dittmaier, M.~Roth and M.~M.~Weber,
  Nucl.\ Phys.\ B {\bf 660} (2003) 289
  [hep-ph/0302198].

\bibitem{Denner:2003zp}
  A.~Denner, S.~Dittmaier, M.~Roth and M.~M.~Weber,
  Nucl.\ Phys.\ B {\bf 680} (2004) 85
  [hep-ph/0309274].

\bibitem{Bohm:2004zi}
  A.~R.~Bohm and Y.~Sato,
  Phys.\ Rev.\ D {\bf 71} (2005) 085018
  [hep-ph/0412106].

\bibitem{Cutkosky:1960sp}
  R.~E.~Cutkosky,
  J.\ Math.\ Phys.\  {\bf 1} (1960) 429.

\bibitem{Veltman:1963th}
  M.~J.~G.~Veltman,
  Physica {\bf 29} (1963) 186.

\bibitem{Aoki:1980ix}
K.~I.~Aoki {\it et al.},
  Prog.\ Theor.\ Phys.\  {\bf 65} (1981) 1001;
%
  Prog.\ Theor.\ Phys.\ Suppl.\  {\bf 73} (1982) 1.

\bibitem{Stuart:1990vk}
  R.~G.~Stuart,
in Proceedings of the XXVth Rencontre de Moriond,
{\it $\PZ^0$ physics}, ed.\  J.~Tr\^an
Thanh V\^an, (Editions Fronti\`eres, Gif-sur-Yvette, 1990), p.~41.

\bibitem{Sirlin:1991fd}
  A.~Sirlin,
  Phys.\ Rev.\ Lett.\  {\bf 67} (1991) 2127;
%
  Phys.\ Lett.\ B {\bf 267} (1991) 240;\\
%
%
  R.~G.~Stuart,
  Phys.\ Rev.\ Lett.\  {\bf 70} (1993) 3193;\\
%
  M.~Passera and A.~Sirlin,
  Phys.\ Rev.\ Lett.\  {\bf 77} (1996) 4146
  [hep-ph/9607253];\\
%
  P.~Gambino and P.~A.~Grassi,
  Phys.\ Rev.\ D {\bf 62} (2000) 076002
  [hep-ph/9907254];\\
%
  P.~A.~Grassi, B.~A.~Kniehl and A.~Sirlin,
  Phys.\ Rev.\ D {\bf 65} (2002) 085001
  [hep-ph/0109228];\\
%
  A.~Freitas, W.~Hollik, W.~Walter and G.~Weiglein,
  Nucl.\ Phys.\ B {\bf 632} (2002) 189
  [Erratum-ibid.\ B {\bf 666} (2003) 305]
  [hep-ph/0202131].

\bibitem{Denner:1997kq}
  A.~Denner and T.~Hahn,
  Nucl.\ Phys.\ B {\bf 525} (1998) 27
  [hep-ph/9711302]; \\
A.~Bredenstein, S.~Dittmaier and M.~Roth, hep-ph/0506005, to appear in
Eur. Phys.~J.~C.

\bibitem{Eidelman:2004wy}
S.~Eidelman {\it et al.}  [Particle Data Group Collaboration],
Phys.\ Lett.\ B {\bf 592} (2004) 1.

\bibitem{Azzi:2004rc}
P.~Azzi {\it et al.}  [CDF and D0 Collaborations, 
and Tevatron Electroweak Working Group],
hep-ex/0404010.

\bibitem{Jegerlehner:2001ca}
F.~Jegerlehner,
DESY 01-029, LC-TH-2001-035, hep-ph/0105283.

\bibitem{Dittmaier:1998nn}
  S.~Dittmaier,
  Phys.\ Rev.\ D {\bf 59} (1999) 016007
  [hep-ph/9805445].

\bibitem{Denner:2010tr}
  A.~Denner, S.~Dittmaier,
  Nucl.\ Phys.\  {\bf B844 } (2011)  199
  [arXiv:1005.2076 [hep-ph]].

\end{thebibliography}
\end{document}